\documentclass[aps,reprint,nofootinbib]{revtex4-1}
\usepackage{graphicx,amsfonts,amsmath,amssymb}
\usepackage{bm}

\begin{document}
\title{Minkowski Spacetime and QED from Ontology of Time}
\date{\today}
\author{C. Baumgarten}
\affiliation{5244 Birrhard, Switzerland}
\email{christian-baumgarten@gmx.net}

\def\begeq{\begin{equation}}
\def\endeq{\end{equation}}
\def\begary{\begeq\begin{array}}
\def\endary{\end{array}\endeq}
\def\bmtx{\left(\begin{array}}
\def\emtx{\end{array}\right)}
\def\d{\partial}
\def\e{\eta}
\def\w{\omega}
\def\W{\Omega}
\def\s{\sigma}
\def\eps{\varepsilon}
\def\a{\alpha}
\def\b{\beta}
\def\g{\gamma}
\def\y{\gamma}
\def\d{\partial}
\def\U{{\cal E}}
\def\P{{\cal P}}
\def\S{{\cal S}}

\def\leftD#1{\overset{\leftarrow}{#1}}
\def\rightD#1{\overset{\rightarrow}{#1}}

\begin{abstract}
Classical mechanics, relativity, electrodynamics and quantum mechanics are 
often depicted as separate realms of physics, each with its own formalism and
notion. This remains unsatisfactory with respect to the unity of nature and
to the necessary number of postulates.
We uncover the intrinsic connection of these areas of physics and describe 
them using a common symplectic Hamiltonian formalism. Our approach is based 
on a proper distinction between variables and constants,  i.e. on a 
basic but rigorous ontology of time. We link these concepts with the obvious
conditions for the possibility of measurements. The derived consequences put 
the measurement problem of quantum mechanics and the Copenhagen interpretation 
of the quantum mechanical wavefunction into perspective.
According to our (onto-) logic we find that spacetime can not be fundamental.
We argue that a geometric interpretation of symplectic dynamics emerges from 
the isomorphism between the corresponding Lie algebra and the representation of a
Clifford algebra. Within this conceptional framework we derive the
dimensionality of spacetime, the form of Lorentz transformations and of the 
Lorentz force and fundamental laws of physics as the Planck-Einstein relation, 
the Maxwell equations and finally the Dirac equation. 
\end{abstract}
\maketitle

\section{Introduction}

\subsection{Spacetime vs. Proper Time}
\label{sec_opening}

Schr\"odinger once wrote that ``In Einstein's theory of gravitation matter and
its dynamical interaction are based on the notion of an intrinsic geometric
structure of the space-time continuum''\cite{ES1950}. What we will discuss in 
this article suggests to conjecture the reverse statement, i.e. that the 
intrinsic geometric structure of spacetime is based on the very notion of matter 
and its dynamical interaction. The idea that spacetime is not fundamental but 
emergent has been proposed in the past by several authors~\cite{KW1985,Pir03,Pir04,
Pir05,ARM,VBF,ACFKGS,Smilga,Hiley,Seiberg,KnuthB}. Some discussed the 
relation between spacetime and quantum communication~\cite{HandM}. 
Our conjecture results from a different, almost classical, notion of quantum mechanics, 
closely connected to the phase space picture of classical 
statistical mechanics. A significant number of publications support our
direction of thought~\cite{Stueckelberg,RalstonQM,Buric,BGH,AGust1,Elze,Briggs1,Briggs2,Skinner,CGE,BPR}. 

We shall start with the distinction of variables and constants, i.e. from an 
(onto-)logic of time. Consider the basic quantummechanical relationship
\begeq
i\hbar\,\partial_t\psi=E\,\psi\,.
\label{eq_change}
\endeq
The left side is the rate of change of a wavefunction $\psi$ and the equation
expresses that this rate of change is equal to the energy of the system. 
``Energy'' is probably the most fundamental concept in physics. 
The conservation of energy has no serious exception and physics assigns to the 
energy the role of substance. Any entity that falls under the notion of ``object'' 
is ``charged'' with a certain amount of energy and is therefore subject
to change with a frequency $\omega=E/\hbar$. The rate of change is what
quantifies the ``passage of time''. This is the meaning of saying that time
and energy are conjugate quantities. The passage of time is measured by clocks, i.e. by the 
rate of change of a reference device. And any system that can be described by 
Eq.~\ref{eq_change} is a clock in itself. Metaphorically we say it exists {\it in time}.

Seen by light, Eq.~\ref{eq_change} is nothing but the equation of motion (EQOM)
of an harmonic oscillator. If we write the real and imaginary part of the so-called
``wave-function'' separately $\psi=X+i\,Y$, then we obtain with $E/\hbar=\omega$:
\begary{rcl}
\dot\psi&=&\dot X+i\,\dot Y=-i\omega(X+iY)\\
\dot X&=&\omega\,Y\\
\dot Y&=&-\omega\,X\,,
\label{eq_change2}
\endary
where the dot indicates the temporal derivative. In matrix form this
reads\footnote{
The equality of Eq.~\ref{eq_change} and Eq.~\ref{eq_change3} is known for long
(see for instance Ref.~(\cite{RalstonQM})), but the way of understanding and
teaching quantum mechanics has not changed.}:
\begeq
\bmtx{c}\dot X\\\dot Y\emtx=\omega\bmtx{cc}
0&1\\
-1&0\\
\emtx\,\bmtx{c}X\\Y\emtx\,.
\label{eq_change3}
\endeq
The interpretation of the use of the unit imaginary in Eq.~\ref{eq_change} 
seems to be the litmus test of our attitude towards quantum mechanics.
It is as often presented as a necessary ingredient as its necessity is strictly
denied. We believe that the unit imaginary is nothing mysterious or magical
that distinguishes quantum from classical mechanics. It is just a compact 
form of writing Hamilton's equations of motion (EQOM) of a classical harmonic 
oscillator (CHO). However, Eq.~\ref{eq_change3} is the {\it normal form} of
an algebraically more general equation and it is in this respect an unmotivated
limitation of the EQOM - as we are going to show in this essay.

The difference between the two interpretations is in some correspondence with 
two conflicting attitudes towards the wave function. Some scientist believe 
it is a mere mathematical tool while others tend to interpret its components 
as the true dynamical variables. We will argue that  both attitudes miss the
point. It is the {\it pure form} of physical theories - i.e. the equation of 
motion - that requires the definition of some fundamental variables. And it
is a {\it deformation} to believe that we can directly assign physical meaning
to these variables. Hence the ``classical'' form of physical theories is - 
though mathematically sound - conceptually inconsistent or at least incomplete.
Let us briefly explain why this is so.

Eq.~\ref{eq_change} tells us that the wavefunction of a system with energy 
{\it changes} at every time. Instead of {\it postulating} the validity of 
Eq.~\ref{eq_change}, we again reverse the argument: We postulate that
{\it change} is immanent to physical reality~\cite{Radovan}. The permanent
change is the physical mode of existence. Again metaphorically we say that 
material objects {\it exist in time}. Any fundamental physical model of 
reality has to represent this continuous change. This is the essence 
of Eq.~\ref{eq_change}: Variables $\psi$ that represent {\it existing} 
fundamental quantities have to vary continuously. We show in the following
that Eq.~\ref{eq_change} can be derived from this postulate.  But before doing
so, let us briefly describe why classical physics is conceptionally
incomplete. The most trivial flaw is the inability of classical physics to
give an account of its measurement standards. The most trivial being the
length. Classically we take the existence of a solid rod for granted. 
Einstein made an enlightening remark: ``One should always be aware that the 
presupposition of the existence in principle of rigid rods is a presupposition 
suggested by approximate experience but is, in principle, arbitrary''\cite{EinsteinBio}.
Why is this so? Because classical physics taken as classical mechanics,
electrodynamics and relativity can not explain the existence of a finite
and fixed atomic radius and hence can not explain the existence of a measuring
rod. Bohr's orbital theory tried to establish fixed radii by an ad-hoc
postulate - but the idea of definite orbitals conflicts with electrodynamics.
We will not derive a finite radius in the following, instead we will make
the presumption explicit and incorporate it as a general principle in what
follows. 

On the fundamental level existence implies continuous change. But though 
all fundamental quantities continuously change, physics as an experimental 
science requires the constancy of measurement standards. Without constant 
reference standards we could not test physical models. These two apparently 
trivial facts are the starting point of the ontology of time.
Our line of reasoning is in contrast to concepts that postulate 
spacetime to be fundamental. Eq.~\ref{eq_change} does not require 
a concept of spacetime. If, in the following, we speak of time then it 
is always understood as a rate of change~\footnote{In accelerator physics this
is called {\it phase advance}.}. However we shall argue in this 
essay that and how the conjugate concept of spacetime - namely the 
energy-momentum relationship - and Maxwell equations emerge in an
algebraically simple and elegant way from a reinterpretation of 
Eq.~\ref{eq_change}, i.e. from a proper distinction between those quantities 
that change and those that stay constant.
 
\subsection{The Form of Physics}
\label{sec_intro}

Usually textbooks on (classical) physics begin with the equations of motion
of point-masses, the definition of positions, velocities and accelerations
etc. This means that the fundamental variables of the theory are usually 
directly charged with physical meaning, a meaning that is supposed to have
its origin in our ``classical'' macroscopic experience. 
However this ``method'' implies that we have to presuppose a considerable 
number of mechanical concepts. A proper definition of these concepts however 
relies on and refers to an elaborated theory. 
Like Baron Munchhausen, who claimed to have pulled himself by the hair 
out of the swamp, the theory is based upon notions that become meaningful 
only in the context of the spelled-out physical theory~\cite{Ronde}.
It is the fate of human existence to be thrown into an already existing world 
and it seems to be the fate of physics students to be thrown right into a
framework of concepts without the chance for a stepwise systematic and logical
{\it (re-)construction}. The classical theory implicitly claims that these
concepts are in some way ad-hoc derivatives from everyday experience - though
seen by light Hamiltonian and Lagrangian mechanics can rarely be applied 
directly to everyday empirical evidence.

We invite the reader to follow a different path and to put on hold with the 
conceptions of mass, point particles, wave functions and also spacetime. 
A renewed analysis of spacetime is required as soon as we understand that the 
rod of constant length - and with it spacetime - is a mere postulate. 
But if quantum mechanics is required to explain the existence of rods of 
constant length, then the same is true for the concept of spacetime:
It requires an explanation - in the optimal case in form of a derivation~\footnote{
There are more good physical reasons to reconsider the concept of spacetime.
The general theory of relativity (GTR) claims that the geometry of spacetime
is determined by the distribution of the masses ``in it''. If this is true, then 
matter and spacetime are not separable entities. Secondly, it is a
well-established part of quantum mechanics that the wavefunction can either 
be defined as a function of space and time or equivalently as a function of 
energy and momentum. Both representations of the wavefunction are related to
each other via a Fourier transformation. If the most fundamental theories
suggest that spacetime can not be consistently understood independent from 
matter and its dynamics, then we should not ignore this fact.}.

Physics is an ``empirical'' science, based on objective measurements. 
A measurement is the comparison of {\it variable} properties of objects 
(measurands) with the corresponding constant properties of reference 
objects. Before we are able to measure distances, we need to have a ruler, 
for instance a solid rod. Certainly we can {\it think} about (the concept of) 
length without having a ruler, i.e. we can develop geometry.
But we can not perform a measurement nor predict a measurement outcome 
without the ruler.
The material {\it object} that is used as a ruler (the rod) is the fundament of 
{\it objectivity} of measurement. Only with the possibility in principle of 
the existence of reference objects like the prototype meter it becomes
physically meaningful to speak about the length of an object or the distance
between objects, i.e. space. In practice we say almost unreflectedly that the 
length {\bf L} is {\bf x} inch, meter or lightyears. However the {\it production}
of a ruler is the first thing we have to do before we can measure length.
{\it It is not sufficient to define} it - it has to be physically made. 
It is a necessary though not a sufficient condition for a measurement. 
The {\it handbook of metrology and testing} describes a measurement as follows:
``Measurement begins with the definition of the measurand,
the quantity intended to be measured. The specification of a measurand 
requires knowledge of the kind of quantity and a description of the object 
carrying the quantity. When the measurand is defined, it
must be related to a measurement standard, the realization
of the definition of the quantity to be measured.''~\cite{metrology}.
Theory should follow practice: Before we postulate spacetime, we should
explain how a measurement of distances is at all possible. 

This article is not about metrology. But what is important about the measurement
standard is that it must be realized and {\it may not change with time}. 
The measurement of a property of a physical entity requires the existence of 
material entities where the respective property stays constant (the rulers) 
and other entities - {\it with the same type of properties} - (the measurands) 
where this property varies or at least {\it can} vary. If the measurand changes 
with time, it is the purpose 
of a physical model to {\it predict} the time-dependence of the measurand. 
When we say that the ruler must not change with time, then we include other 
invariances as well. Wigner pointed this out by referring to {\it invariance} 
as a fundamental principle~\cite{Wigner}.
Hence we can say that physics is irresolvably committed to time, i.e. to 
continuous change {\it and} to constancy. This is not postulate, it is a 
{\it conditio sine qua non}\footnote{ 
Sir Hamilton was not only aware of the intimate connection between physics 
and time - he even had the vision to develop algebra as the science 
of pure time~\cite{Hamilton} and it was his deep belief that 
``that the intuition of time is more deep-seated in the human mind than 
the intuition of space''~\cite{Ohrstrom}.}. 

The unity of nature is next major premise: To be part of the physical world
implies the possibility in principle of (direct or indirect) interaction with 
all other things that are part of this world and therefore we (have to)
presume the existence of a common denominator, a {\it fundamental
  level}\footnote{The basis of this believe is much the same as most theorists 
believe that it should finally be possible to describe all forces of nature 
within a grand unified theory.}. Physical models are based on a quantitative 
description of reality. 
Whatever a physical theory considers to be fundamental must essentially be 
representable by quantities. We call these quantities ``variables''. The 
mathematical model of these fundamental entities is hence based on a list of 
variables $\psi$ of (yet) unknown dimension, which all continuously change at 
all times. But if the components of $\psi$ change at all times and if 
they are fundamental, then there is no way to define a measurement standard 
for them. Nevertheless there might be functions of the fundamental variables 
that are constant in time: such constants are called {\it constants of 
motion (COMs)}. These functions may not include linear terms - otherwise
it is easy to show that the functions can not be constants~\footnote{
Dragt {\it et al} have shown in Ref.~(\cite{Dragt}) that there are no non-zero
first order moments in linear Hamiltonian systems.}. 
Hence the dimension (unit) of these functions is different from the dimension 
(unit) of the fundamental variables. And therefore they can not serve as rulers 
for the considered fundamental variables. The logical consequence is that 
a direct measurement of the fundamental variables of $\psi$ is not possible.
The above considerations explain why the components of the quantum mechanical (QM)
wave function have to vary at all times, why the quantum mechanical wave
function implies an interpretational problem and why QM has to postulate that 
measurement devices must be ``classical'' or ``macroscopic'': Due to the
absence of constant rulers, the fundamental variables (the components of the 
wavefunction) can not be directly measured~\footnote{
Or with the words of  S. Fortin and O. Lombardi: ``The difficulties can be
overcome once it is recognized that {\it classicality is a property of the
observables}''~\cite{FortinLombardi}.
}. 

The definition of general and abstract quantities like the fundamental
variables and their evolutions in time will not suffice to make up a
physical theory. At some stage we need an interpretation that maps quantities 
of the model to measurable quantities in the world in order to obtain physical
meaning. The physical meaning can then only be induced from the mathematical
structure of the physical quantity. The interpretation can only refer to already 
known physical laws in order to identify possible interpretations 
of the quantities and the relations between them. We can only recognize
fundamental physical meaning of certain algebraic relationships because an 
elaborated theory of elementary particles that has been shown to be 
experimentally successful in the description of fundamental physical phenomena 
is already available.

\subsection{Outline}

The paper is organized as follows: In Sec.~\ref{sec_setup} we describe our 
understanding of fundamental physical entities. In Sec.~\ref{sec_constants} we 
derive Hamilton's equations of motion (EQOM) and the basic properties of the
symplectic unit matrix. In Sec.~\ref{sec_sigma} constants of motion are 
introduced with the help of Lax Pairs. We review important algebraic 
relations that indicate the construction principles of spacetime: 
The required congruence of the algebraic structure of (skew-) Hamiltonian 
matrices with the basic elements (i.e. generators) of Clifford algebras. 
We describe the basic measurable entities in a world based on time: 
(Second) moments of fundamental variables.
Since second moments can be represented by expectation values of matrix 
operators, the relations between these matrices are the relations between 
the observables (i.e. the laws of physics).

In Sec.~\ref{sec_geom} we derive the conditions for the emergence of
a geometric space from symplectic dynamics. Observables are expectation 
values of matrix operators and hence spacetime geometry should be representable
by a system of basic matrices with certain properties. Such matrix systems 
indeed exist and are known as (representations of) Clifford algebras. 

In Sec.~\ref{sec_structure} we describe symplectic transformations 
as structure preserving transformations. 
These transformations are isomorphic to Lorentz transformations and are
the basis for the apparent geometry of spacetime.
We explore the meaning of structure defining transformations.

In Sec.~\ref{sec_clifford} we give a short overview over the basic
properties of Clifford algebras in general and especially of 
$Cl_{N-1,1}(\mathbb{R})$. We derive conditions that limit the possible
dimensionality for an emergent spacetime. 
Further analysis of the properties of these Clifford algebras allows to 
restrict spacetime to 3+1 dimensions. Hence the appropriate 
algebra is the algebra of the real Dirac matrices.

In Sec.~\ref{sec_relativity} we describe Lorentz transformations - boosts and
rotations - as structure preserving (symplectic) transformations. 
We present an interpretation of the Dirac matrix system called the
``electromechanical equivalence'' (EMEQ). Guided by the EMEQ we derive the Lorentz 
force and in Sec.~\ref{sec_method} we derive (Quantum-) electrodynamics. 
We give arguments why momentum and energy should be related to spatial
and temporal derivatives and sketch the path towards the Dirac equation,
a representation of massive spinors and the describe the significance of
CPT-transformations in our approach. 

Sec.~\ref{sec_outlook} finalizes the discussion and a summary is given in
Sec.~\ref{sec_summary}.

\section{Obligations of a Fundamental Theory }
\label{sec_setup}

The basic variables of a fundamental theory are continuously varying quantities~\footnote{
The physical constants ($\varepsilon_0$, 
$\mu_0$, $c$, $\hbar$ ...) as listed in handbooks are not {\it material} 
but rather {\it theorectical} objects. Some of these are merely conversion
factors for units. If we refer to ``constants'', then we address properties of 
material objects that are potentially useful as measurement standards. For
instance the orbital radii of hydrogen atoms.} - they are dynamical variables.
Since quantum theory postulates that the wavefunction is fundamental, it is nearby
to assume that we finally have to identify the components of the wavefunction with 
(parts of) our list of ``fundamental variable(s)''. The Copenhagen interpretation 
avoids the reference problem by assigning a mere probabilistic meaning to the
wavefunction, i.e. by postulating that a) the wave function has only a probabilistic
meaning and b) the wavefunction gives a complete description of reality, i.e. is
fundamental. Most discussions of the Copenhagen interpretation focus on
the question, if the probability interpretation is correct, or whether the
probability is ``classical'' or ``quantum'' in nature and what the ontological
status of the wavefunction is. We do not address these questions. Instead our claim is, 
that {\it if} the wavefunction is truly fundamental, then a meaningful {\it physical} 
interpretation of the wavefunction is impossible. This does not imply that it 
is {\it less real} than the macroscopic quantities derived from it. 
However a meaningful answer to the question {\it what} varies requires the 
existence of some constant entity for comparison, i.e. a reference for a 
measurement. If we say that an entity represented by a variable in our 
equation is a length then we need to have, in principle, a unit (for instance 
the meter) to refer to and we need to have an idea of spacetime. 
Not only that this reference has to be of the same type 
(i.e. dimension), but it also has to be macroscopic as ``meaning'' can only be 
derived from everyday (macroscopic) experience. Physical meaning and objectivity 
depend on a constant quantity of the same dimension that can be used as a reference.
We have an idea and a concept of length as we are surrounded by objects with this
property. The lack of reference is according to our premises a logical consequence 
of fundamentality. With respect to the fundamental variables there is no
meaningful answer to the question {\it what} varies, but only {\it how} they
vary\footnote{In other words: ``The wave function does not describe matter, it describes how
  matter behaves''~\cite{Allori}.}. 

This is a logical limit in physics: A fundamental physical theory can describe 
how entities change but it can not tell what the entities are that are changing.
Hence the only possible {\it physically real} rulers, the only possible 
measurement standards are constants of motion: the prototype meter is an object 
composed of atoms and molecules which consist of ``elementary particles''. 
The motion of these components is stable and the length of the reference 
object ``meter'' is a constant of motion of these particles. But at the 
level of these particles - or at the level of their constituents - at some 
level we face the pure variation of abstract fundamental variables. 
We conclude that if it is possible to directly measure (the value
of) a variable, it can not be fundamental.

There is no doubt that we can measure distances and angles {\it since we have
rods and goniometers}. Hence space (or spacetime, respectively) is not 
fundamental. Spacetime is a construction for our theoretical needs - it is 
not itself a ``thing'' but a formalism to express the relations
between things. This is not to say that spacetime does not ``exist''. 
If things exist and relations between things exist, then we can 
speak of the space that exists. However it is {\it not fundamental}, which
basically means that it can be {\it derived}. To say this implies the obligation
to derive the dimensionality of spacetime, the Lorentz transformations
and the role played by electrodynamics, i.e. the speed of light. 
We shall argue that spacetime is based on the isomorphism of a Lie algebra 
with (reps of) real Clifford algebras $Cl(N-1,1)$. 
The ``perception'' of spacetime is a result of pattern recognition in loose 
analogy to the perception of acoustic signals in terms of sound and music~\footnote{
We should be aware that ``physically'' there are just wavelike density 
fluctuations in air. It is our mind that {\it processes} the perceived data 
and interprets them as sound, language or music. This interpretation is 
indeed adequate. Nevertheless it remains an interpretation.}.
Our ansatz is based on a list of fundamental and abstract variables, the values 
of which can not be directly measured. Such a measurement would require 
constants of motion from ``more'' fundamental entities. But then the more 
fundamental variables could not be measured, etc. Hence it is impossible 
to directly assign physical meaning to the fundamental dynamical variables. 
In this respect fundamental variables are not identical to degrees of 
freedom (DOFs) in classical mechanics as DOFs are usually assumed to be 
measurable at least in principle. 
But in most other aspects fundamental variables are similar to (DOFs) in 
classical mechanics and at first sight there is no reason to refuse the 
possibility of a classical description of the dynamics of these variables.

E.T. Jaynes wrote that ``Because of their empirical origins, QM and QED are
not physical theories at all. In contrast, Newtonian celestial mechanics, 
Relativity, and Mendelian genetics are physical theories, because their
mathematics was developed by reasoning out the consequences of clearly
stated physical principles from which constraint the possibilities''.
And he continues ``To this day we have no constraining principle from which
one can deduce the mathematics of QM and QED; [...] In other words, the 
mathematical system of the present quantum theory is [...] unconstrained by
any physical principle''~\cite{Jaynes}.
We agree with this statement insofar as we think that it is important to
constraint possibilities and to develop a theory according to well-defined
principles. The introduction of the unit imaginary into quantum mechanics 
is not of this kind. No physical principle has ever been formulated that
explains the necessity for the use of the unit imaginary in QM. There are 
just ad-hoc postulates. We believe that the use of the unit imaginary as 
it is usually done, is a mistake. It introduces a structure without clear 
and explicite motivation. 
Hence we will avoid or at least delay the introduction of the unit imaginary 
up to the point, when its use becomes {\it reasonable} and legitimate. 
In the meantime we restrict our considerations to the reals. Or more precisely: 
we demand that all fundamental variables are of the same type, as we consider it 
scientifically not legitimate to introduce an asymmetry like the asymmetry between the 
reals and the imaginary numbers {\it without reason}. Hence all fundamental 
variables are either all real or all imaginary. The latter might be preferable 
to indicate their unmeasurability, however as long as there is no need 
to introduce the distinction between real and imaginary variables, 
Ockham's razor demands to stay with reals~\footnote{For discussions about the use of 
real and/or complex numbers generally in physics and specifically in 
quantum mechanics, see also Refs.~\cite{Stueckelberg,imag,imag2,Gibbons}.}.

It follows from this conception that fundamental ``material objects'' 
like for instance electrons have to be described by their structure. Since the
location in spacetime and hence a continuous trajectory is according to our 
principles not fundamental and may hence not serve to define or verify {\it
 sameness}, elementary particles have no material identity, but only a 
structural identity. The only structure at hand is the structure of the 
variations of the variables - the structure of
their dynamics. The fundamental objects (elementary particles) are not directly 
represented by the fundamental variables, but rather by the dynamical
structure of the variations, by the patterns of motion. 
Physics can be described as a method to analyze patterns of motion, it
is to some degree identical to pattern recognition~\footnote{Recall that 
the birth of modern physics is connected with names like Kepler, who 
recognized the ``true'' pattern of planetary motion.}.
If ``objects'' do not change their structure (i.e. their ``identity'') in some 
interaction, then the involved dynamical processes must be structure preserving: 
If a material entity like an elementary particle is defined and understood by its
structure, then its continuous existence requires - besides a continuous variation 
of its constituents (the variables) - that the dynamical structure must be preserved. 
However if objects can only be identified by a certain dynamical structure then two 
objects with the same structure can not be distinguished. If we could distinguish 
particles experimentally that are indistinguishable in our physical model, then the model 
would be incomplete. Thus, if quantum mechanics is indeed fundamental, then it must 
logically include a concept of identical particles - as particles in a 
fundamental physical theory have to be represented by structures.

\section{Variables, Constants and the Hamiltonian}
\label{sec_constants}

\subsection{Variables}

We suggest the following principles for an ontologically proper basis of physics:
\begin{enumerate}
\item Existence happens in time. The ``time'' that is meant here does not have 
to be identical to the time that an observer would measure using a clock, but 
we insist that there is a bijective functional relationship~\footnote{We
  explicitly include the possibility of regularizing transformations~\cite{KS1,KS2}.
}.
\item Existence in time is manifest by variation. All quantifiable properties
  of all fundamental physical entities continuously change and are representable by 
  real quantities that continuously vary, i.e. by ``variables''.
\item Measurements require constant references (rulers), i.e. a physical model requires 
constants.
\end{enumerate}
From these axioms it follows that
\begin{enumerate}
\item There are no other (physical) constants than constants of motion (COMs). 
\item The fundamental variables have no measurement standard, i.e. can not be directly measured.
\end{enumerate}

The value of a single variable varies. Hence a single variable can not generate 
constants of motion. Therefore we start with an arbitrary number $k>1$ of dynamical 
variables~\footnote{Ockham's razor commits us to determine the minimum number of variables
required to generate objects with spacetime properties.}.
According to the axioms there exist one (or a set of) constant function(s)
${\cal H}(\psi_1, \psi_2,\dots \psi_k)=\mathrm{const}$ of the dynamical variable 
list ${\bf \psi}$. Hence we may write (with the presumption ${\cal H}=\mathrm{const}$ 
we imply here ${\partial{\cal H}\over\partial t}=0$):
\begeq
{d{\cal H}\over dt}={\d{\cal H}\over \d \psi_1}\,\dot \psi_1+{\d{\cal H}\over \d \psi_2}\,\dot \psi_2+\dots+{\d{\cal H}\over \d \psi_k}\,\dot \psi_k=0\,,
\label{eq_constH}
\endeq
or in vector notation:
\begeq
{d{\cal H}\over dt}=(\nabla_\psi\,{\cal H})^T\cdot\dot {\bf\psi}=0\,.
\label{eq_constHvec}
\endeq
The (simplest) general solution is given by
\begeq
\dot {\bf \psi}={\bf S}\,(\nabla_\psi\,{\cal H})\,,
\label{eq_Heqom0}
\endeq
where ${\bf S}$ is a $k\times k$ non-singular skew-symmetric real matrix,
i.e. ${\bf S}^T=-{\bf S}$. The skew-symmetry of ${\bf S}$ is sufficent to
solve Eq.~\ref{eq_Heqom0} so that ${\cal H}$ is a constant of motion. 
According to linear algebra there exists a non-singular matrix ${\bf Q}$ such
that~\footnote{See for instance Ref.~(\cite{MHO1}) and App.~(\ref{sec_vectortime},\ref{sec_nsym_dirac})}:
\begeq
{\bf Q}^T\,{\bf S}\,{\bf Q}=\textrm{diag}(\eta_0,\eta_0,\eta_0,\dots\,,0,0,0)
\label{eq_strucdef}
\endeq
where 
\begeq
\eta_0=\bmtx{cc}0&1\\-1&0\emtx\,,
\endeq
is the basic form of the symplectic unit matrix.
According to the axioms the matrix must have full rank $2\,n\le k$,
i.e. all variables vary and hence there are no constants in the 
state vector such, i.e.
\begeq
{\bf Q}^T\,{\bf S}\,{\bf Q}=\textrm{diag}(\eta_0,\eta_0,\eta_0,\dots\,,\eta_0)\equiv\y_0\,.
\label{eq_gamma0}
\endeq
The transformation Eq.~\ref{eq_gamma0} is a linear change of variables which
is used to find (or recognise or define) the ``natural'' or {\it normal} variables. 
We call such transformations {\it structure defining} (see below and App.~(\ref{sec_nonsym}))
and rewrite Eq.~\ref{eq_constHvec} accordingly
\begary{rcl}
{d{\cal H}\over dt}&=&({\bf Q}^{-1}\,\nabla\,{\cal H})^T\,{\bf Q}^T\,{\bf S}\,{\bf Q}({\bf Q}^{-1}\,\nabla\,{\cal H})=0\\
&=&({\bf Q}^{-1}\,\nabla\,{\cal H})^T\,\y_0\,({\bf Q}^{-1}\,\nabla\,{\cal H})=0\,.
\endary
In the following we assume that the transformation into the normal variables
has been done such that the normal variables are given by $\psi$ and that ${\bf
  S}$ has between transformed into $\y_0$ so that
\begeq
\dot\psi=\y_0\,\nabla_\psi\,{\cal H}\,.
\label{eq_Heqom2}
\endeq
The matrix $\y_0$ has the even dimension $2n\times 2n$: in any time-like
physical world the dynamical variables come in pairs~\footnote{ 
The unit imaginary has indeed significance in quantum mechanics as
it ensures an even number of fundamental variables and a continuous variation
of these variables due to Eq.~\ref{eq_change}. However our approach contradicts 
the frequently expressed opinion that QM inherently requires the use of complex 
numbers in the sense that it could not be formulated without the unit imaginary.}.
It is therefore sensible to refer to a pair of variables when we speak of a degree of freedom. 
We call $\psi_{2j}=q_j$ the j-th {\it canonical coordinate} and $\psi_{2j+1}=p_j$ the j-th 
{\it canonical momentum}, i.e. ${\bf\psi}=(q_1,p_1,\dots,q_n,p_n)^T$,
but this nomenclature is purely formal as long as the variables $\psi$ are
fundamental. It is just the structure of the matrix $\y_0$ that leads to this
distinction. The matrix $\y_0$ is called the {\it symplectic
  unit matrix} and the equations of motion (EQOM) have Hamilton's form:
\begary{rcl}
\dot q_k&=&{\d{\cal H}\over\d p_k}\\
\dot p_k&=&-{\d{\cal H}\over\d q_k}\,.
\label{eq_Heqom1}
\endary
Whenever we have constants (of motion) of the dynamical variables, we can 
derive Hamilton's EQOM in some way. We can interpret the abstract and
intrinsically unmeasurable basic variable list $\psi$ as ``spinors'', i.e. 
as ``objects'' in a phase space. 
Just formally we call the components ``coordinates'' and ``momenta''. But in 
fact every $\psi$ is a point in an abstract $2n$-dimensional phase space~\footnote{
In classical statistical mechanics this kind of space is called $\Gamma$-space~\cite{Becker}
but the intimate relation to quantum mechanics is known as for instance described
by Kim and Noz~\cite{KimNoz}.}.
However the Hamiltonian formalism itself does not require that the variables are 
fundamental and therefore this formalism can be applied to any system with dynamical
constants. The difference to the ``classical'' Hamiltonian formalism
is solely that the classical state vector $\psi$ describes the (average) 
properties of systems as the position of the center of mass or the average
momentum and does not consist of fundamental variables. 
Thus the components of a classical state vector $\psi$ can - at least in
principle - be measured. 
But if the state vector represents fundamental quantities then the components 
can not be directly and individually measured. Before we are able to present 
an interpretation of $\psi$ we first have to construct constants of motion 
and the corresponding observables. One constant of motion has been introduced 
already: The Hamiltonian function ${\cal H}$. In classical physics this function 
most often represents the energy of a physical system. We will suggest a
similar interpretation in what follows.

\subsection{The Hamiltonian}
\label{sec_hamiltonian}

We introduced the ``Hamiltonian'' ${\cal H}$ as (an arbitrary) constant of
motion. Typically there are several constants of motion and hence the function 
${\cal H}$ is not yet well-defined. In the following we assume that ${\cal H}$
is positive (semi-) definite with respect to the variables $\psi$. This
restriction is neither arbitrary nor weak: If the constructed
constant of motion ${\cal H}$ is a reference of existing measurable things, 
then there must be a Hamiltonian function ${\cal H}$ that reflects the amount of
 something, for instance the amount of substance. In a physical theory that
is free of ad-hoc postulates the constant existence of a substance must 
be represented by a positive (semi-) definite Hamiltonian {\it as positivity
is immanent to the notion of substance}. 

We write the ``Hamiltonian'' ${\cal H}$ as a Taylor series of the $2\,n$
variables $\psi_k$:
\begeq
{\cal H}(\psi)={\cal H}_0+\eps^T\,\psi+\frac{1}{2!}\,\psi^T\,{\bf A}\,\psi+\frac{1}{3!}\,{\bf B}_{ijk}\,\psi_i\,\psi_j\,\psi_k+\dots\,,
\label{eq_hamiltonian_general}
\endeq
where $\eps$ is a $2\,n$-dimensional vector and ${\bf A}$ is a symmetric $2\,n\times 2\,n$ matrix and
${\bf B}$ is a tensor that is symmetric in all indices. We assume that ${\cal H}$ has a local minimum
{\it somewhere}. Further we may set ${\cal H}_0=0$ as it has no influence on
the equations of motion.
A typical method in physics is to stepwise study the solutions of Eq.~\ref{eq_hamiltonian_general}, starting
with low amplitudes first. Hence the first step is to neglect higher order terms and to focus on
small amplitude solutions, i.e. to write
\begeq
{\cal H}(\psi)\approx\eps^T\,\psi+\frac{1}{2!}\,\psi^T\,{\bf A}\,\psi\,.
\label{eq_hamiltonian2}
\endeq
After the truncation to second order, an offset $\tilde\psi=\psi-\psi_0$ of size
$\psi_0=-{\bf A}^{-1}\,\eps$ ($\tilde\psi=\psi-\psi_0$) enables to get rid of the linear term, so that:
\begeq
{\cal H}(\psi)\approx\frac{1}{2}\,\psi^T\,{\bf A}\,\psi
\label{eq_Hamiltonian}
\endeq
The Hamiltonian EQOM~(\ref{eq_Heqom2}) can then be written as a product of a 
matrix ${\bf F}$ and the vector $\psi$:
\begeq
\dot\psi=\y_0\,\nabla_\psi\,{\cal H}={\bf F}\,\psi\,,
\label{eq_Heqom}
\endeq
where we defined the matrix 
\begeq
{\bf F}\equiv\y_0\,{\bf A}\,.
\label{eq_Fdef}
\endeq

\section{Constants of Motion}
\label{sec_sigma}

We define the dyad $\Sigma_{ij}\equiv \psi_i\,\psi_j$ which can be understood
as a matrix of all possible quadratic forms. If we optionally consider some
type of averaging, then this matrix becomes a matrix of second moments $\Sigma$:
\begeq
\Sigma=\langle\psi\psi^T\rangle\,,
\label{eq_sigma0}
\endeq
where the angles $\langle\rangle$ indicate some sort of averaging. 
We define the matrix ${\bf S}$ by ${\bf S}\equiv\Sigma\,\y_0$, so that:
\begary{rcl}
\dot{\bf S}&=&\dot\Sigma\,\y_0=\dot\psi\,\psi^T\,\y_0+\psi\,\dot\psi^T\,\y_0\\
           &=&{\bf F}\,\psi\,\psi^T\,\y_0+\psi\,\psi^T\,{\bf F}^T\,\y_0\\
           &=&{\bf F}\,{\bf S}-{\bf S}\,\y_0\,{\bf F}^T\,\y_0\,,
\label{eq_env}
\endary
where we use the fact that $\y_0^2=-{\bf 1}$ and $\y_0=-\y_0^T$.
From Eq.~\ref{eq_Fdef} it follows that
\begeq
{\bf F}^T={\bf A}^T\,\y_0^T=(\y_0)^2\,{\bf A}\,\y_0=\y_0\,{\bf F}\,\y_0\,,
\label{eq_symplexdef}
\endeq
where we used the fact that ${\bf A}$ is symmetric. We then obtain
\begeq
\dot{\bf S}={\bf F}\,{\bf S}-{\bf S}\,{\bf F}\,.
\label{eq_eqom_meas}
\endeq
That is, if ${\bf F}$ and ${\bf S}$ commute, then the matrix ${\bf S}$
is constant, while $\psi$ varies. Operators ${\bf S}$ and ${\bf F}$ that fulfill 
Eq.~(\ref{eq_eqom_meas}) form a so-called ``Lax Pair''. For such pairs it can be shown that
\begeq
Tr({\bf S}^k)=\mathrm{const}\,,
\label{eq_COM}
\endeq 
holds for any natural number $k$~\cite{Lax,Lax1,Lax2}. Hence the bilinear
form~(\ref{eq_sigma0}) is the basis for a set of constants as given by Eq.~(\ref{eq_COM}). 
Note that the equation of motion (Eq.~\ref{eq_eqom_meas}) of the second
moments fullfills the requirement to represent observable properties of a 
physical system. If ${\bf S}$ and ${\bf F}$ are similar, i.e. if they
{\it share a system of eigenvectors}, then ${\bf S}$ is constant. If this is
not the case, then we still have the constants of motion given by
Eq.~\ref{eq_COM} as reference quantity. 

The average of the quadratic form equals a matrix of second moments, if summed over 
an ensemble of $N$ spinors:
\begeq
\Sigma={1\over N}\,\sum\limits_{k=1}^N\,\psi_k\,\psi_k^T\,,
\label{eq_sigma1}
\endeq
or - which is a complementary description -  if we use a ``density'' $\rho(\psi)$ to describe
the distribution of states in $\Gamma$-space. In this case the matrix of second moments
is given by
\begeq
\Sigma={\int\,\rho(\psi)\,\psi\,\psi^T\,d^{2n}\psi\over\int\,\rho(\psi)\,d^{2n}\psi}\,.
\label{eq_sigma2}
\endeq
Eq.~(\ref{eq_eqom_meas}) holds also for a single $\Gamma$-space trajectory $\psi(t)$, but
for a single trajectory one finds for all $k$ that $Tr({\bf S}^k)=0$, so that the constancy
is trivial: The eigenvalues of a matrix of second moments $\Sigma$ depend on 
whether they are computed for a single vector $\psi$ (which results in a matrix $\Sigma$
with vanishing determinant and vanishing eigenvalues), for two linear independent 
vectors $(\psi_1,\psi_2)$ ($\Sigma$ has vanishing determinant, but two non-zero 
eigenvalues) or $\ge 2\,n$ linear independent vectors $\psi_k$ as in Eq.~(\ref{eq_sigma1}). 
Only in the latter case, $\Sigma$ is non-singular~\footnote{In accelerator physics the eigenvalues
of ${\bf S}=\Sigma\,\y_0$ are the ``emittances'' (times the unit imaginary). It is evident,
that the position of a single particle at time $t$ defines a point but not an area in
phase space~\cite{rdm_paper,geo_paper}.}. This implies that our ansatz unfolds to
full generality only if we consider ensembles. Or more precise, the
observables are averaged over ensembles. 

Since ${\bf S}$ (and ${\bf F}$) are by definition the product of a symmetric matrix $\Sigma$ and the 
skew-symmetric matrix $\y_0$, it follows that 
\begeq
Tr({\bf S})=Tr({\bf F})=0\,.
\endeq 
so that the simplest meaningful constants of motion are given as
\begeq
Tr({\bf S}^2)=\mathrm{const}\,.
\endeq 
Eq.~\ref{eq_Hamiltonian} describes a $n$-dimensional harmonic oscillator.
The truncation of the Hamiltonian to second order is not arbitrary - it
guarantees stability. It is well-known, that non-linearities yield (in many or 
even most cases) unstable or chaotic behavior. Therefore in order to
establish a system of stable references linear systems are preferable. 
This restriction is {\it per se} not problematic, since the 
equations that we will derive are also linear~\footnote{
We refer to the well-known regularization methods that allows to map the Kepler problem and a large
variety of other central potential problems to the harmonic oscillator~\cite{Belenkii}.}.
We might also interpret this restriction as a low-energy approximation as in Ref.~(\cite{KW1985}).

The constants of motion that can be used as references are quadratic forms (\ref{eq_COM}).
If a reference is based on a quadratic form, the variable compared with the reference (the measured 
variable) also must be a quadratic form. Therefore the ``dynamical variables'' $q_i$ and $p_i$ cannot 
be measured directly, but only (functions of) quadratic forms based on these 
variables. We call these (functions of) quadratic forms observables. 

Matrices that fulfill Eq.~\ref{eq_symplexdef} are called ``Hamiltonian'' or sometimes ``infinitesimally symplectic''. 
In our opinion these names are misleading, the former mainly because the matrix ${\bf F}$ does not appear in the 
Hamiltonian and the latter since ${\bf F}$ is neither symplectic nor infinitesimal. Therefore we use the
term {\it symplex} (plural {\it symplices})~\cite{rdm_paper,geo_paper}. 

A {\it cosymplex} (or ``antisymplex'' or ``skew-Hamiltonian'' matrix) 
is a matrix ${\bf C}$ that holds~\footnote{The nomenclature of (co-) symplices
  combines the terms ``symplectic'', ``(co-) sine'' and ``matrix''. 
The connection to (co-) sine will become obvious in Sec.~\ref{sec_symtrans}. Furthermore, 
in geometry a ``1-simplex'' is a line, a ``2-simplex'' is a triangle and a ``3-simplex'' is a 
volume (tetrahedron). Equivalently the elements of a Clifford algebra (for instance the Dirac algebra) 
include vectors (vertices), bi-vectors (edges) and tensors (faces). 
Pascal's triangle can be used for both - the number of k-simplices of an n-simplex ($k<=n$) 
and to the number of traceless k-(co-) symplices of an $N$-dimensional Clifford algebra. 
Hence the Clifford algebra $Cl_{3,1}$ represented by the real Dirac-matrices can be 
compared to a regular 3-simplex (tetrahedron). See Fig.~\ref{fig_geom} in the appendix.}:
\begeq
{\bf C}^T=-\y_0\,{\bf C}\,\y_0\,.
\label{eq_cosymplexdef}
\endeq
A cosymplex ${\bf C}$ can always be written as a product of $\y_0$ and a skew-symmetric matrix. 
The sums of (co-) symplices are (co-) symplices, i.e. the superposition principle holds. 
Hence (co-) symplices form a linear vector space and any (co-) symplex can be written as a linear 
combination of ``basic'' (co-) symplices. The algebra of (co-) symplices is the Lie algebra $sp(2\,n)$.

Denoting symplices by ${\bf S}$ and cosymplices by ${\bf C}$ (optionally with subscript) it is easily 
shown that the anticommutator of two symplices is a cosymplex:
\begary{rcl}
({\bf S}_1\,{\bf S}_2+{\bf S}_2\,{\bf S}_1)^T&=&{\bf S}_2^T\,{\bf S}_1^T+{\bf S}_1^T\,{\bf S}_2^T\\
&=&\y_0\,{\bf S}_2\,\y_0\,\y_0\,{\bf S}_1\,\y_0+\y_0\,{\bf S}_1\,\y_0\,\y_0\,{\bf S}_2\,\y_0\\
&=&-\y_0\,({\bf S}_2\,{\bf S}_1+{\bf S}_1\,{\bf S}_2)\,\y_0\,.
\endary
The following general rules can be derived:
\begary{ccc}
\left.\begin{array}{c}
{\bf S}_1\,{\bf S}_2-{\bf S}_2\,{\bf S}_1\\
{\bf C}_1\,{\bf C}_2-{\bf C}_2\,{\bf C}_1\\
{\bf C}\,{\bf S}+{\bf S}\,{\bf C}\\
{\bf S}^{2\,n+1}\\
\end{array}\right\} & \Rightarrow & \mathrm{symplex}\\&&\\
\left.\begin{array}{c}
{\bf S}_1\,{\bf S}_2+{\bf S}_2\,{\bf S}_1\\
{\bf C}_1\,{\bf C}_2+{\bf C}_2\,{\bf C}_1\\
{\bf C}\,{\bf S}-{\bf S}\,{\bf C}\\
{\bf S}^{2\,n}\\
{\bf C}^n\\
\end{array}\right\} & \Rightarrow & \mathrm{cosymplex}\\
\label{eq_cosy_algebra}
\endary
The relations Eq.~(\ref{eq_cosy_algebra}) define the structure of a specific algebra, called the 
Lie algebra $sp(2\,n)$ of the symplectic group $Sp(2\,n)$. 
As the matrix of generators ${\bf F}$, also the matrix of observables ${\bf S}$ of Eq.~(\ref{eq_COM}) 
is a symplex. According to Eq.~(\ref{eq_cosy_algebra}) odd powers of ${\bf S}$ are symplices with 
zero trace. Hence the non-zero (i.e. useful) constants of motion given by Eq.~(\ref{eq_COM}) are 
of even order.

Since the matrix ${\bf A}$ is symmetric (Eq.~\ref{eq_Hamiltonian}), the maximal number $\nu$ 
of free parameters of a symplex for $n$ degrees of freedom (DOF) is
\begeq
\nu_s={2\,n\,(2\,n+1)\over 2}\,.
\label{eq_nu_s}
\endeq
If the matrix ${\bf F}$ can be expressed as a sum of basic matrices, 
the basis $\nu_s$ elements. A basis for all real $2n\times 2n$ matrices is completed 
by $\nu_c$ cosymplices:
\begeq
\nu_c={2\,n\,(2\,n-1)\over 2}\,.
\label{eq_nu_c}
\endeq

\subsection{Measurable Quantities: Observables}
\label{sec_observables}

Usually the second moments are assumed to be average values of $N$ identical systems 
as in Eq.~(\ref{eq_sigma1}), but the equations hold equivalently for a single system.
However it should be noted that for a vector of $2\,n$ fundamental variables, 
$2\,n$ linear independent ``samples'' are required, if the matrix ${\bf
  S}=\Sigma\,\y_0$ is supposed to have a non-vanishing determinant.
Later we will come back to this point.
 
We can define the $\Sigma$-matrix as well using a (normalized) probability density $\rho$:
\begeq
\Sigma=\int\,\rho(\psi)\,\psi\,\psi^T\,d^{2n}\psi\equiv \langle\psi\,\psi^T\rangle\,,
\label{eq_rhodef}
\endeq
with the normalization according to:
\begeq
1=\int\,\rho(\psi)\,d^{2n}\psi\,,
\endeq
i.e. the density $\rho(\psi)$ is defined as a function of the {\it fundamental variables}
or {\it phase space} variables. 
 
We introduced the (``S-matrix'') ${\bf S}=\Sigma\y_0$ and derived Eq.~(\ref{eq_eqom_meas}).
Let the {\it adjunct} vector $\bar\psi$ be defined by
\begeq
\bar\psi\equiv\psi^T\,\y_0\,,
\endeq 
so that ${\bf S}=\vert\psi\rangle\langle\bar\psi\vert$ and the expectation value of an operator ${\bf O}$:
\begeq
\langle{\bf O}\rangle\equiv\langle\bar\psi\,{\bf O}\,\psi\rangle\,.
\endeq
Now consider that ${\bf O}$ is a cosymplex, i.e. can be written as $\y_0\,{\bf B}$ with some 
skew-symmetric matrix ${\bf B}$, then the expectation value of ${\bf O}=\y_0\,{\bf B}$ vanishes:
\begary{rcl}
\langle{\bf O}\rangle&=&\langle\psi^T\,\y_0^2{\bf B}\,\psi\rangle\\
                     &=&-\langle\psi^T\,{\bf B}\,\psi\rangle=0\,.
\label{eq_zerocosy}
\endary
Hence only symplices represent measurable properties of (closed) systems. Cosymplices ${\bf C}$ 
only yield a non-vanishing result when they appear between different (i.e. linearily independent) states:
\begeq
\bar\phi\,{\bf C}\,\psi\ne 0\Rightarrow \phi\ne\,\lambda\,\psi\,,
\endeq
with some arbitrary (non-zero) factor $\lambda$. Vice versa - if two (normalized) states $\phi$ and $\psi$ 
may not ``generate'' non-vanishing expectation values of the cosymplex type,
then is follows that:
\begeq
\phi=\pm\psi\,.
\endeq
Since we assume that physical reality can be described by the totality of 
all fundamental variables, the coefficients of the matrix ${\bf F}$ are functions
of (internal or external) measurable quantities, i.e. of some symplex ${\bf S}$ 
generated from fundamental variables. And since cosymplices have a vanishing 
expectation value, they can not contribute. 
We will use this important result in Sec.~\ref{sec_method} in the derivation 
of MWEQ. Furthermore, if a structure that represents a ``particle'' is preserved
in a physical process, then the involved driving terms must be symplices. 
Otherwise the evolution in time would not be symplectic, i.e. 
{\it structure preserving}. And vice versa: 
If the particle structure (particle type) is transformed in a process, then  
there must be a contribution from cosymplices as for instance {\it axial
vector currents} as in the V-A theory of weak interaction.

This means that {\it if we focus on a closed system}, then (the expectation values of) 
all cosymplices have to vanish. 

The properties of ``open'' and ``closed'' have a relative meaning which depends on the context. 
If we consider two separate ``particles'' described by state vectors $\psi$ and $\phi$ and let them get in 
contact, then we would for instance combine the state vectors into a larger state vector $\Psi=(\psi,\phi)^T$. 
In App.~(\ref{sec_rpm}) we give the explicit form of general $2\,n\times
2\,n$-symplices. It is shown that non-diagonal sub-blocks may {\it per se} 
have arbitrary terms. Hence the question, if a (sub-) matrix is
a symplex or a cosymplex depends on the context.
However {\it all} quantities - including the matrix ${\bf F}$ - must by some
means result from ``fundamental variables'', i.e. from state vectors. Also a 
static electromagnetic field has its origin in charges and currents of
``particles'' that are described by state vectors.
If reality is described by the totality of the fundamental variables, then also 
the symplex ${\bf F}$ must somehow be generated from $\psi$. Since the matrix 
of observables ${\bf S}=\Sigma\,\y_0$ is a symplex as well, the functional 
relationship might be simple: consider (for instance) a Hamiltonian of the form
\begeq
{\cal H}\propto\psi^T\,\Sigma^{-1}\,\psi\,,
\label{eq_dual}
\endeq
which implies ${\bf F}\propto\,\y_0\,\Sigma^{-1}$. In this special case a
Boltzmann distribution $\rho(\psi)\propto\exp{(-{{\cal H}\over k\,T})}$ is 
identical to a multivariate Gaussian density distribution~\cite{stat_paper}. 
In general the functional relationship between ${\bf F}$ and ${\bf S}$ might
be more involved, but obviously there is a kind of {\it duality} between the 
matrix of observables ${\bf S}=\Sigma\,\y_0$ and the matrix of generators 
${\bf F}=\y_0\,{\bf A}$ that can be illustrated by Eq.~(\ref{eq_dual}).
This duality is well-known in classical mechanics. The Hamiltonian is
an observable and the generator of translations in time (evolution in time),
the angular momentum is an observable and the generator of rotations, the
momentum is the generator of spatial translations~\cite{Grgin}.

The time derivative of an expectation value is (assuming the observable ${\bf O}$ does not explicitly depend on time):
\begary{rcl}
\langle{\bf\dot O}\rangle&=&\langle\dot{\bar\psi}\,{\bf O}\,\psi\rangle+\langle\bar\psi\,{\bf O}\,\dot\psi\rangle\\
&=&\langle\dot\psi^T\,\y_0\,{\bf O}\,\psi\rangle+\langle\bar\psi\,{\bf O}\,{\bf F}\,\psi\rangle\\
&=&\langle\psi^T\,{\bf F}^T\,\y_0\,{\bf O}\,\psi\rangle+\langle\bar\psi\,{\bf O}\,{\bf F}\,\psi\rangle\\
&=&\langle\psi^T\,\y_0\,{\bf F}\,\y_0^2\,{\bf O}\,\psi\rangle+\langle\bar\psi\,{\bf O}\,{\bf F}\,\psi\rangle\\
&=&\langle\bar\psi\,({\bf O}\,{\bf F}-{\bf F}\,{\bf O})\,\psi\rangle\\
\label{eq_eqom}
\endary
This general law connects the commutator of an operator ${\bf O}$ and ${\bf F}$ with 
its time derivative. This is the linear form of the Poisson bracket and it is purely
classical. One conclusion from Eq.~\ref{eq_eqom} in combination with 
Eq.~\ref{eq_cosy_algebra} is that the time evolution is such that (co-) symplices stay
(co-) symplices. They belong to different domains.

\section{From Lie Algebra to Spacetime Geometry}
\label{sec_geom}

From the (anti-) commutation properties described by Eq.~\ref{eq_cosy_algebra}
it follows that any fundamental system of (co-) symplices requires that
the basic (co-) symplices either commute or anticommute: Only in this case the 
product is a pure symplex or cosymplex, i.e. either observable or not.  
If we consider pure (co-) symplices ${\bf A}$ and ${\bf B}$, then
\begeq
({\bf A}\,{\bf B})^T={\bf B}^T\,{\bf A}^T=\pm\,\y_0\,{\bf B}\,{\bf A}\,\y_0\,.
\endeq
It follows that the product of pure (co-) symplices is again a pure (co-) symplex,
iff ${\bf A}\,{\bf B}=\pm {\bf B}\,{\bf A}$. This implies that all (anti-)commutators
of pure (co-)symplices are again pure (co-)symplices. Hence - if there are
systems of (anti-)commuting matrices, they are expected to play a fundamental
role.

In the following we will show how spacetime structures emerge from
the algebra of (co-)symplices (Eq.~\ref{eq_cosy_algebra}),
i.e. we will explain why and how the spinors $\psi$ - which are composed 
of fundamental variables - are able to produce the relations between 
observables that represent the spacetime structures.
In order to do this we have to identify algebraic objects that represent 
(unit) vectors for the space and time directions (or - equivalently - the energy and
momentum directions). For this purpose we need a definition of ``length'' (a norm) and of 
orthogonality. So far we only introduced time and a single symplex: $\y_0$.
Apparently $\y_0$ is the natural choice of the ``unit vector'' in the ``time direction''. 

Usually vector norms in Euclidean spaces represent the length of a vector,
i.e. a property that is invariant under rotations and translations. Hence 
we prefer an invariant quantity for the desired definition of a norm. 
Besides the Hamiltonian the only invariants we introduced so far are given by 
Eq.~(\ref{eq_COM}). The simplest non-trivial $k$-norm is given for $k=2$, i.e.: 
\begeq
\Vert {\bf A}\Vert^2\equiv\frac{1}{2\,n}\,\mathrm{Tr}({\bf A}^2)\,.
\label{eq_norm}
\endeq
where the division by $2\,n$ is introduced to let the unit matrix have unit norm.
The matrix $\y_0$ hence has a norm of $-1$, i.e. the norm (\ref{eq_norm}) is not
positive definite.

Two elements of an algebra are said to be orthogonal, if their inner product vanishes.
Since we have no inner product defined (yet), we use an alternative 
{\it geometric} definition which we consider to be equivalent in {\it flat
  spacetime} - the {\it Pythagorean theorem}. 
Two vectors ${\vec v}$ and ${\vec w}$ are orthogonal, iff~\cite{napier}
\begeq
\Vert {\vec v}+{\vec w}\Vert^2=\Vert {\vec v}\Vert^2+\Vert {\vec w}\Vert^2\,,
\endeq
which implies (in conventional vector notation) that the inner product vanishes:
\begeq
({\vec v}+{\vec w})^2={\vec v}^2+{\vec w}^2+2\,\vec v\cdot\vec w\,.
\endeq
For two elements ${\bf A}$ and ${\bf B}$ of the Lie algebra $sp(2\,n)$ the 
use of Eq.~\ref{eq_norm} then yields
\begary{rcl}
\Vert {\bf A}+{\bf B}\Vert^2&=&\frac{1}{2\,n}\,\mathrm{Tr}\left[({\bf A}+{\bf B})^2\right]\\
&=&\frac{1}{2\,n}\,\mathrm{Tr}({\bf A}^2+{\bf B}^2+{\bf A}{\bf B}+{\bf B}{\bf A})\\
&=&\Vert {\bf A}\Vert^2+\Vert{\bf B}\Vert^2+\frac{1}{2\,n}\,\mathrm{Tr}({\bf A}{\bf B}+{\bf B}{\bf A})\,.
\endary
As we argued above, all elements of the desired algebra either pairwise commute or anticommute.
Two anticommuting elements ${\bf A}$ and ${\bf B}$ with
\begeq
\mathrm{Tr}\left({\bf A}\,{\bf B}+{\bf B}\,{\bf A}\right)=0\,.
\label{eq_ortho}
\endeq
are then orthogonal in the sense that they obey the Pythagorean theorem:
\begeq
\Vert {\bf A}+{\bf B}\Vert^2=\Vert {\bf A}\Vert^2+\Vert{\bf B}\Vert^2\,.
\endeq
Hence the (trace of the) anticommutator is structurally equal to the inner product~\footnote{
Since the anticommutator does not necessarily result a scalar, one may
distinguish between inner product and scalar product. The scalar product  
can then be defined by the trace of the inner product:
\begeq
({\bf A}\cdot{\bf B})_S\equiv {1\over 2\,n}\mathrm{Tr}({{\bf A}\,{\bf B}+{\bf B}\,{\bf A}\over 2})\,,
\endeq
where the index ``S'' indicates the scalar part.
}:
\begeq
{\bf A}\cdot{\bf B}\equiv {{\bf A}\,{\bf B}+{\bf B}\,{\bf A}\over 2}\,.
\label{eq_innerprod}
\endeq
The above arguments and the fact that $\y_0$ seems to have the form of a unit
vector suggests to search for other unit vectors $\y_k$ for $k\in[1\dots
  N-1]$, especially other unit vectors that anticommute with $\y_0$. For 
such unit vectors we find by the use of $\y_0^2=-1$:
\begary{rcl}
\y_k\,\y_0+\y_0\,\y_k&=&0\\
\y_k\,\y_0^2+\y_0\,\y_k\,\y_0&=&0\\
\y_k&=&\y_0\,\y_k\,\y_0\,,
\label{eq_onetimedim}
\endary
so that in accordance with Eqs.~\ref{eq_symplexdef} and \ref{eq_cosymplexdef} $\y_k$ ($k\ne 0$) 
is a symplex, iff it is symmetric. It is a cosymplex, iff it is
skew-symmetric. This fact is remarkable. Since up to now we just defined time,
it is nearby to interpret $\y_0$ as the unit vector ``in the direction of
time''. Then from Eq.~\ref{eq_onetimedim} we can take, that all other unit
vectors which are perpendicular to $\y_0$ and that correspond to driving terms 
of the Hamiltonian, are symmetric and hence formally different from ``time''.
In other words - there is only one observable direction of time. Using the
algebraic identity 
$$\mathrm{Tr}({\bf A}\,{\bf A}^T)=\sum\limits_{i,j}\,a_{ij}^2$$ 
- where $a_{ij}$ are the elements of the matrix ${\bf A}$ - it follows, that
(with an appropriate normalization) symmetric unit ``vectors'' ${\bf A}$
square to $+{\bf 1}$ and skew-symmetric ${\bf B}$ square to $-{\bf 1}$:
\begary{rcl}
\mathrm{Tr}({\bf A}^2)&=&\mathrm{Tr}({\bf A}\,{\bf A}^T)=\sum\limits_{i,j}\,a_{ij}^2\ge 0\\
\mathrm{Tr}({\bf B}^2)&=&-\mathrm{Tr}({\bf B}\,{\bf B}^T)=-\sum\limits_{i,j}\,b_{ij}^2\le 0\,.
\endary
Therefore there are two types of measurable unit vectors: $\y_0$ that 
represents time and squares to $-{\bf 1}$ and a number of symmetric symplices $\y_k$ 
(anticommuting with $\y_0$ and with each other) that square to $+{\bf 1}$. 
We interpret the latter (if existent) as the unit vectors in space-like 
directions. Skew-symmetric elements that are orthogonal to $\y_0$ are cosymplices 
and have a vanishing expectation value.

In summary: if a (flat) spacetime emerges from dynamics, all symplices that
represent orthogonal directions have to {\it anticommute}. 
Therefore the obvious choice of the {\it inner product} or {\it scalar product} 
of two elements of a dynamically generated geometric algebra is the anticommutator.
The unit vector in the time-direction squares to $-{\bf 1}$ and all vector-type 
symplices $\y_x$ which are orthogonal to $\y_0$ are symmetric and square to $+{\bf 1}$.

Given that we can find a (sub-) set of $N$ pairwise orthogonal symplices $\y_k$ with 
$k\,\in\,[0,1,\dots,N-1]$ in $sp(2\,n)$, then the product of two of them holds:
\begary{rcl}
\y_i\,\y_j&=&{\y_i\,\y_j+\y_j\,\y_i\over 2}+{\y_i\,\y_j-\y_j\,\y_i\over 2}\\
          &=&{\y_i\,\y_j-\y_j\,\y_i\over 2}\,,
\endary
for $i\ne j$. It is easy to prove that the product $\y_i\,\y_j$ with $i\ne j$
is orthogonal to both - $\y_i$ and $\y_j$. Hence we identify the {\it exterior product} 
with the commutator of ${\bf A}$ and ${\bf B}$:
\begeq
{\bf A}\,\wedge {\bf B}={{\bf A}\,{\bf B}-{\bf B}\,{\bf A}\over 2}\,.
\label{eq_ext}
\endeq
A system of $N$ pairwise anti-commuting real matrices (that square 
to $\pm{\bf 1}$) represents a Clifford algebra $Cl_{p,q}$, $N=p+q$. 
$p$ is the number of unit vectors that square to $+{\bf 1}$ and $q$ the
number of unit vectors that square to $-{\bf 1}$. As only a single
time-direction is possible we can fix $q=1$.
If (as we presumed) space-time emerges from the dynamics of fundamental
variables, then the apparent geometry of spacetime is described by the 
(representation of) a Clifford algebra $Cl_{N-1,1}$.
The unit vectors are called {\it generators} of the Clifford algebra.

There are $\left({N\atop k}\right)$ possible combinations (without
repetition) of $k$ elements from a set of $N$ generators. 
Then there are
$\left({N\atop 2}\right)$ products of 2 basic matrices (named {\it bivectors}), 
$\left({N\atop 3}\right)$ products of 3 basic matrices (named {\it trivectors}) and so on.
The product of all $N$ basic matrices is called {\it pseudoscalar}.
The $N$ anti-commuting generators are the (unit) {\it vectors}.
The total number of all k-vectors then is~\footnote{The case $k=0$ can be identified with 
the unit matrix.}:
\begeq
\sum\limits_{k=0}^N\,\left({N\atop k}\right)=2^N\,.
\endeq
As we have no reason to allow more elements in the algebra than $2^N$, 
the vector space $sp(2\,n)$ and $Cl_{N-1,1}$ have the same dimension:
\begeq
2^N=(2n)^2\,,
\label{eq_complete}
\endeq
which - if fulfilled - is nothing but the condition for {\it completeness}.

If such a basis of anticommuting unit matrices exists, we know that
\begary{rcll}
\y_k^2&=&\pm\,{\bf 1}&\textrm{ for all }k\,\in\,[0,\dots,\,4n^2-1]\\
\y_k&=&\pm\,\y_k^T&\textrm{ for all }k\,\in\,[0,\dots,\,4n^2-1]\\
\y_k\,\y_j&=&\pm\,\y_j\,\y_k&\textrm{ for all }j\ne k,\,\in\,[0,\dots,\,4n^2-2]\,.
\label{eq_basic_gamma_def}
\endary
From Eqs.~\ref{eq_basic_gamma_def} above we readily conclude:
\begary{rcl}
\y_0\,\y_k\,\y_0&=&\pm\,\y_k^T\\
\y_k\,\y_0\,\y_k^T&=&\pm\,\y_0\,,
\label{eq_discrete}
\endary
so that all $\y_k$ are either symplices or cosymplices and they are either symplectic ($\y_k\,\y_0\,\y_k^T=\y_0$) 
or ``cosymplectic'' ($\y_k\,\y_0\,\y_k^T=-\y_0$).
Since $\mathrm{Det}({\bf A}{\bf B})=\mathrm{Det}({\bf A})\mathrm{Det}({\bf B})$
and since $\y_k^2=\pm\,{\bf 1}$ it follows that all basic matrices have a determinant of $\pm 1$.

Given such a matrix system exists for some $n$ and $N$, then it has remarkable
symmetry properties. Note that we derived these properties without fixing 
the size of the matrices. 
If (as we demand) Eq.~(\ref{eq_complete}) holds, then the system is a complete {\it basis} of the vector space of real 
$2n\times 2n$-matrices and therefore these matrices ${\bf M}$ can be written as a linear combination of 
the $\y_k$ according to:
\begeq
{\bf M}=\sum\limits_{k=0}^{(2n)^2-1}\,m_k\,\y_k\,,
\endeq
so that with the signature $s_j$ defined by $s_j\equiv \mathrm{Tr}(\y_j^2)/(2n)$ we have
\begary{rcl}
\frac{1}{4n}\,\mathrm{Tr}({\bf M}\,\y_j+\y_j\,{\bf M})&=&\frac{1}{4n}\,\mathrm{Tr}\left(\sum\limits_{k=0}^{2n-1}\,m_k\,(\y_j\,\y_k+\y_k\,\y_j)\right)\\
&=&s_j\,m_j\,.
\label{eq_scalprod}
\endary

Given that $\y_k$ with $k\,\in\,[0,\dots,N-1]$ are the generating elements of the algebra, 
then the {\it vector} ${\bf A}=\sum\limits_{k=0}^{N-1}\,a_k\,\y_k$ squares to a scalar:
\begeq
{\bf A}^2=\sum\limits_{k=0}^{N-1}\,s_k\,a_k^2=\sum\limits_{k=1}^{N-1}\,a_k^2-a_0^2\,.
\endeq
Hence the algebra of a dynamical system generates a flat Minkowski spacetime with proper 
observables if the corresponding real matrix algebra is a representation of a Clifford 
algebra in which all generating elements are symplices. The idea to relate Clifford algebras 
and symplectic spaces is not new (see for instance Ref.~(\cite{BGB,BB}) and references therein), 
however (to the knowledge of the author) it has not been shown yet that the concept 
of a geometric (Clifford) algebra can be derived from the concept of a dynamically emergent 
spacetime.

\section{Structure Preservation}
\label{sec_structure}

\subsection{Transfer Matrizes and Symplectic Transformations}
\label{sec_symtrans}

The solution of the Hamiltonian EQOM (Eq.~\ref{eq_Heqom}) for the linearized
Hamiltonian (with constant matrix ${\bf A}$) are given by the matrix exponential:
\begeq
\psi(t)=\exp{({\bf F}\,(t-t_0))}\,\psi(t_0)={\bf M}(t,t_0)\,\psi(t_0)\,.
\label{eq_solution}
\endeq
The matrix ${\bf M}(t,t_0)=\exp{({\bf F}\,(t-t_0))}$ is usually called {\it transfer matrix} 
and it is well-known to be symplectic if ${\bf F}$ is a symplex (``Hamiltonian''), that is, ${\bf M}$
fulfills the following (equivalent) equations~\cite{MHO}:
\begary{rcl}
{\bf M}\,\y_0\,{\bf M}^T&=&\y_0\\
{\bf M}^{-1}&=&-\y_0\,{\bf M}^T\,\y_0\\
({\bf M}^{-1})^T&=&-\y_0\,{\bf M}\,\y_0\\
{\bf M}^T&=&-\y_0\,{\bf M}^{-1}\,\y_0\,.
\label{eq_symplectic}
\endary

Since the matrix exponential of ${\bf F}$ is quite complicated in general, it is more 
convenient to consider transformations generated by basic symplices $\y_j$, since all 
$\y_j$ square to $\pm{\bf 1}$:
\begary{rcl}
{\bf R}_j&=&\exp{(\y_j\,\eps)}\\
         &=&\left\{
\begin{array}{lcl}
{\bf 1}\,\cos{(\eps)}+\y_j\,\sin{(\eps)}&\mathrm{for}&\y_j^2=-1\\
{\bf 1}\,\cosh{(\eps)}+\y_j\,\sinh{(\eps)}&\mathrm{for}&\y_j^2=1\\
\end{array}\right.\\
{\bf R}_j^{-1}&=&\exp{(-\y_j\,\eps)}\\
       &=&\left\{
\begin{array}{lcl}
{\bf 1}\,\cos{(\eps)}-\y_j\,\sin{(\eps)}&\mathrm{for}&\y_j^2=-1\\
{\bf 1}\,\cosh{(\eps)}-\y_j\,\sinh{(\eps)}&\mathrm{for}&\y_j^2=1\\
\end{array}\right.\\
\label{eq_sine_cosine}
\endary
The solutions are the sum of the unit matrix and $\y_j$ with (hyperbolic) (co-) sine coefficients. 
From Eqs.~\ref{eq_sine_cosine} we find that skew-symmetric symplices (those that square to $-{\bf 1}$) are
generators of orthogonal transformations, since 
\begary{rcl}
{\bf R}_j&=&{\bf 1}\,\cos{(\eps)}+\y_j\,\sin{(\eps)}\\
{\bf R}_j^T&=&{\bf 1}\,\cos{(\eps)}-\y_j\,\sin{(\eps)}\\
\Rightarrow&=&{\bf R}_j\,{\bf R}_j^T={\bf 1}\,.
\endary
Hence skew-symmetric symplices are generators of (symplectic) {\it rotations}. The symplectic transformations
generated by symmetric symplices are essentially {\it boosts}. 
As we will show in more detail in Sec.~\ref{sec_emeq}, it is legitimate to call the 
structure preserving property of symplectic transformations the {\it principle of (special) relativity} 
since Lorentz transformations of ``spinors'' $\psi$ in Minkowski spacetime are a subset of 
the symplectic transformations for two DOFs. 
 
The structure of the matrix exponential of a symplex ${\bf F}$ and hence the
structure of a symplectic matrix is given by:
\begeq
{\bf M}(t)=\exp{({\bf F}\,t)}={\bf C}+{\bf S}\,,
\endeq
where the (co-) symplex ${\bf S}$ (${\bf C}$) is given by:
\begary{rcl}
{\bf S}&=&\sinh{({\bf F}\,t)}\\
{\bf C}&=&\cosh{({\bf F}\,t)}\,,
\endary
since (the linear combination of) all odd powers of a symplex are symplices and all even powers are cosymplices.
The inverse transfer matrix ${\bf M}^{-1}(t)$ is then given by:
\begeq
{\bf M}^{-1}(t)={\bf M}(-t)={\bf C}-{\bf S}\,.
\endeq

For the matrix of second moments $\Sigma$ and ${\bf S}=\Sigma\,\y_0$
Eq.~(\ref{eq_solution}) and Eq.~(\ref{eq_symplectic}) yield:
\begary{rcl}
\Sigma(t)&=&\psi(t)\,\psi^T(t)={\bf M}(t,t_0)\,\psi(t_0)\,\psi(t_0)^T\,{\bf M}^T(t,t_0)\\
\Sigma(t)\,\y_0&=&-{\bf M}(t,t_0)\,\Sigma(t_0)\,\y_0^2\,{\bf M}^T(t,t_0)\,\y_0\\
{\bf S}(t)&=&{\bf M}(t,t_0)\,{\bf S}(t_0)\,{\bf M}^{-1}(t,t_0)\,,
\label{eq_similarity}
\endary
so that the evolution of ${\bf S}$ in time is a {\it symplectic similarity transformation}.
Hence all {\it eigenvalues} of ${\bf S}$ are constants.

From Eq.~\ref{eq_solution} it is obvious that the column vectors of the transfer matrix
are the solutions for initial conditions $\psi(0)$ that are equal to unit vectors.
The first column $\psi_1(t)$ of ${\bf M}(t,t_0)$ is given by:
\begeq
\psi_1(t)={\bf M}(t,t_0)\,(1,0,0,0)^T\,.
\label{eq_unitvec}
\endeq
Furthermore we note that if a symplex ${\bf F}$ is symplectically similar to
$\y_0$, then it follows from Eq.~(\ref{eq_symplectic})
\begeq
{\bf F}\propto{\bf M}\,\y_0\,{\bf M}^{-1}={\bf M}\,{\bf M}^T\,\y_0\,,
\label{eq_gamma0st}
\endeq
so that ${\bf F}$ equals the matrix ${\bf S}=\Sigma\,\y_0$ if the second
moments are the average over the orthogonal unit vectors as exemplified by 
Eq.~(\ref{eq_unitvec}):
\begeq
\Sigma={\bf M}\,{\bf M}^T=\sum_k\,\psi_k(t)\psi_k^T(t)\,.
\endeq
This is another example for the mentioned duality between observables and generators 
(see Eq.~(\ref{eq_dual}) and Ref.~(\cite{Grgin})).

\subsection{Time-Dependent Hamiltonian}

We started out from a Hamiltonian that has no explicit time dependence.
Let us see, if we can still use the same concepts if the driving term in
Eq.~\ref{eq_Hamiltonian} is time-dependent:
\begary{rcl}
{d{\cal H}\over dt}&=&{\d{\cal H}\over\d t}+\nabla_{\psi}{\cal H}\,\dot\psi=0\\
                  0&=&\frac{1}{2}\psi^T\dot{\bf A}\psi+\psi^T\,{\bf
                    A}\,\dot\psi=0\\
\label{eq_Hoft}
\endary
We use the Ansatz:
\begeq
\dot\psi=(\y_0\,{\bf A}+{\bf G})\,\psi\,,
\endeq
with a general symplex ${\bf G}$ and plug it into Eq.~\ref{eq_Hoft}
\begary{rcl}
0&=&\frac{1}{2}\psi^T\dot{\bf A}\psi+\psi^T\,{\bf A}\,(\y_0\,{\bf A}+{\bf G})\,\psi\\
0&=&\psi^T\left(\frac{1}{2}\dot{\bf A}+{\bf A}\,\y_0\,{\bf A}+{\bf A}\,{\bf G}\right)\,\psi\\
\endary
As we know that $\psi^T{\bf A}\,\y_0\,{\bf A}\psi=0$ due to the skew-symmetry
of ${\bf A}\,\y_0\,{\bf A}$, the remaining matrix has to vanish separately,
i.e. has to be skew-symmetric. Furthermore we have to assume that $\dot{\bf
  A}=\dot{\bf A}^T$ as the symmetry is logically required in the Hamiltonian
for a single system:
\begary{rcl}
0&=&(\frac{1}{2}\dot{\bf A}+{\bf A}\,{\bf G})^T+\frac{1}{2}\dot{\bf A}+{\bf A}\,{\bf G}\\
0&=&\dot{\bf A}+{\bf G}^T\,{\bf A}+{\bf A}\,{\bf G}\\
0&=&\dot{\bf A}+\y_0\,{\bf G}\,\y_0\,{\bf A}+{\bf A}\,{\bf G}\\
0&=&\y_0\,\dot{\bf A}-{\bf G}\,\y_0\,{\bf A}+\y_0\,{\bf A}\,{\bf G}\\
\dot{\bf F}&=&{\bf G}\,{\bf F}-{\bf F}\,{\bf G}\\
\label{eq_foo1}
\endary
This shows us that - as long as $\dot{\bf A}=\dot{\bf A}^T$ holds, then ${\cal
  H}$ remains constant if the time-dependence of ${\bf F}$ is the result of
symplectic motion. Since ${\cal H}$ is positive definite, we may write ${\bf
  F}$ in the general form
\begary{rcl}
{\bf F}&\equiv&{\bf B}\,{\bf B}^T\,\y_0\\
\dot{\bf F}&=&{\bf\dot B}\,{\bf B}^T\,\y_0+{\bf B}\,{\bf\dot B}^T\,\y_0\\
\dot{\bf F}&=&{\bf\dot B}\,{\bf B}^{-1}\,{\bf B}\,{\bf B}^T\,\y_0+{\bf  B}\,{\bf B}^T\,({\bf B}^T)^{-1}{\bf\dot B}^T\,\y_0\\
\dot{\bf F}&=&({\bf\dot B}\,{\bf B}^{-1})\,{\bf F}-{\bf F}\,(\y_0\,({\bf B}^T)^{-1}{\bf\dot B}^T\,\y_0)\\
\label{eq_foo2}
\endary
The last lines of Eq.~\ref{eq_foo1} and Eq.~\ref{eq_foo2} are equivalent, if
\begary{rcl}
{\bf G}&=&\y_0\,({\bf B}^T)^{-1}{\bf\dot B}^T\,\y_0\\
\y_0\,{\bf G}\,\y_0&=&({\bf B}^T)^{-1}{\bf\dot B}^T\\
{\bf G}^T&=&({\bf B}^T)^{-1}{\bf\dot B}^T\\
{\bf G}&=&{\bf\dot B}\,{\bf B}^{-1}\\
{\bf\dot B}&=&{\bf G}\,{\bf B}\\
\endary
Hence we again find a spinor-type of linear equation with a symplex ${\bf G}$ 
as driving term: 
\begeq
{\bf\dot B}={\bf G}\,{\bf B}\,,
\label{eq_foo3}
\endeq
which tells us that symplectic motion can be regarded as a kind of ``closed
system'', if all spinors change according to Eq.~\ref{eq_foo3} and all
symplices change according to Eq.~\ref{eq_foo2}. According to classical
statistical mechanics, the phase space density $\rho(\psi)$ is a constant, 
if it is a function of the Hamiltonian (or - more generally - of the constants
of motion) only. In our case we can expect a system in thermodynamical
equilibrium, if the density is a function of $\psi^T\,{\bf A}\,\psi$,
i.e. a function of the bilinear forms that we identified as measurable
quantities~\cite{Becker}. In Sec.~\ref{sec_emeq} we will show that the 
constants of motion can be interpreted by the {\it mass} of the system.

\subsection{Cosymplices as Generators of Transformations}

The definition of what we consider to be a symplex or a co-symplex depends on
the choice of the symplectic unit matrix $\y_0$. Eq.~(\ref{eq_sine_cosine})
holds also for (skew-) symmetric cosymplices as it holds for symplices.
Consider a skew-symmetric cosymplex $\y_j$ (which we know to anticommute with $\y_0$), 
then the matrix exponential is an orthogonal similarity transformation, so that the
transformed matrix $\y_0$ is skew-symmetric and squares to the negative unit matrix:
\begary{rcl}
\tilde\y_0^T&=&({\bf R}_j\,\y_0\,{\bf R}_j^T)^T={\bf R}_j\,\y_0^T\,{\bf R}_j^T\\
&=&-{\bf R}_j\,\y_0\,{\bf R}_j^T=-\tilde\y_0\\
\tilde\y_0^2&=&({\bf R}_j\,\y_0\,{\bf R}_j^T)^2={\bf R}_j\,\y_0^2\,{\bf R}_j^T\\
&=&-{\bf R}_j\,{\bf R}_j^T=-{\bf 1}\\
\endary
The transformed matrix {\it differs} from $\y_0$, but has the same properties.

If we consider on the other hand a symmetric cosymplex $\y_j$ (which we know to commute with $\y_0$), 
then the matrix exponential is also symmetric and commutes with $\y_0$ so that
\begary{rcl}
({\bf R}_j\,\y_0\,{\bf R}_j^T)^T&=&{\bf R}_j\,\y_0^T\,{\bf R}_j^T=-{\bf R}_j^2\,\y_0\\
({\bf R}_j\,\y_0\,{\bf R}_j^T)^2&=&{\bf R}_j^4\,\y_0^2=-{\bf R}_j^4\,{\bf 1}\\
\endary
The transformed matrix is still antisymmetric but (in general) it does not square 
to the (negative) unit matrix.

One may conclude that skew-symmetric cosymplices are generators of time-like rotations:
They change the ``orientation'' of $\y_0$, the system is rotated towards a new ``time direction''
$\tilde\y_0={\bf R}\,\y_0\,{\bf R}^T$, but remains structurally similar
(though not the same). 

The {\it physical meaning} of such non-symplectic transformations can be
related to a variety of phenomena. In Ref.~(\cite{newlook}) for instance
non-symplectic transformations have been shown to be related to the appearance
of vector potentials and to the transformation into curvilinear coordinate
systems. In App.~\ref{sec_nonsym} we give an example which is directly related
to the structure of a system. Furthermore we explicitly give a matrix that
transforms between the six possible choices for $\y_0$ of the Dirac algebra
in App.~\ref{sec_nsym_dirac}.

\section{The Dimensionality of Spacetime}
\label{sec_clifford}

\subsection{Clifford Algebras}

Sets of anticommuting nonsingular matrices are well-known as 
generators of {\it Clifford algebras}. The inner products of the basic 
elements are the elements of the {\it metric tensor}.
The vector space constructed in this way is Euclidean only for diagonal metric
tensors with signatures $s_1=s_2=\dots=s_N=1$, i.e. Euclidean unit vectors 
square to $+1$. Therefore generators that square to $+1$ are {\it spatial} 
or {\it space-like} and generators that square to $-1$ {\it time-like}~\footnote{
This is only meaningful for {\it real} representations, since the case of 
complex representations allows to switch the sign of matrices by the 
multiplication with the unit imaginary.}.
A Clifford algebra with $p$ space-like and $q$ time-like generator matrices 
is indicated by $Cl_{p,q}$.

From Eq.~\ref{eq_complete} it follows that such matrix systems can only be
complete for even $N=2\,M$~\footnote{
A Clifford algebra also can be defined for odd $N$, but these algebras are not 
isomorphic to a representation in form of $2n\times 2n$ real matrices.}. Then:
\begeq
n=2^{M-1}\,
\endeq
Any real matrix system that fulfills these requirements is based on a number 
of DOFs which is a power of $2$. The simplest system hence is the one for which 
$M=1$ ($N=2$, $n=1$) holds, i.e. it consists of $2\times 2$-matrices.

As mentioned in Sec.~\ref{sec_sigma} only those elements $\y_k$ that are symplices 
have a non-vanishing expectation value $\bar\psi\,\y_k\,\psi$ and are therefore 
{\it observables}. Hence we restrict ourselves to Clifford algebras with generators 
that can be represented by symplices. This includes $\y_0$ as the direction of time. 
Since any generator that anticommutes with $\y_0$ is symmetric and therefore 
squares to $+1$, we have $q=1$, i.e. the Clifford algebra that represents
spacetime is of the type $Cl_{p,1}(\mathbb{R})$.  

If $\y_0,\y_1,\dots,\y_{N-1}$ are the generators of a Clifford algebra 
$Cl_{N-1,1}$ with metric 
\begeq
\y_\mu\,\y_\nu+\y_\nu\,\y_\mu=2\,\textrm{Diag}(-1,1,\dots,1)\,,
\endeq 
then the bivector-elements $\y_0\,\y_k$ with $k=1,2,\dots\,N-1$ are symmetric
and therefore generators of boosts (``boosters'') while the bivector-elements
$\y_j\,\y_k$ with $j\ne k$ and $j,k\ne 0$ are the skewsymmetric 
generators of spatial rotations (``rotators'' or ``gyrators''). We come back to this 
in the next subsection.

\subsection{The Number of Space Dimensions}
\label{sec_space_dimensions}

We have shown that it is possible to construct spacetime geometry on the basis of
(Hamiltonian) dynamics, i.e. on the geometric {\it interpretation} of the matrix 
operators that represent the observables. The possible numbers of spacetime 
dimensions $N=p+q$ in this approach is given by the representation theory of 
real matrices~\cite{Okubo}:
\begeq
p-q=0\textrm{ or }2\textrm{ mod }8\,.
\endeq
As we have derived above, there is only one single skew-symmetric (time-like) 
symplex-generator of the Clifford algebra (i.e. $\y_0$), i.e. we have to set 
$q=1$ and therefore there are $p=N-1$ space dimensions so that:
\begeq
N-2=0\textrm{ or }2\textrm{ mod }8\,.
\label{eq_dim}
\endeq
Hence the dimensionality of dynamically emergent spacetime geometries are algebraically 
limited to $1+1$, $3+1$, $9+1$, $11+1$, $17+1$, $19+1$, $25+1$, $27+1$,
$\dots$ dimensions, corresponding to symplices of size 
$2\times 2$, $4\times 4$, $32\times 32$, $64\times 64$, $\dots$~\cite{Gibbons,Herbut}.

Symplices of size $2\times 2$ have {\it either} real {\it or} pure imaginary
eigenvalues and are therefore not able to represent the general 
(i.e. complex) case. But there is another argument, why one degree of freedom
is not sufficient. A single degree of freedom does not enable to describe
interaction. We can not expect to derive physics based on non-interaction.  
Especially spacetime can only be explained based on interaction.
Furthermore one oscillating degree of freedom implies turning points, 
where the momentum (or kinetic energy) is zero. At these points there is a 
moment without change. An hypothetical observer could not 
experimentally decide, whether the direction of time has changed at these 
points or not. Only in case of two (or more) dimensions there is an additional 
constant of motion - the angular momentum - that (if it is non-zero) ensures 
that there {\it is change at all times}. In case of two dimensions and spin, 
time reversal is always connected to a non-zero {\it action}. Therefore the 
$2\times 2$ case can not be generalized and is rejected as too simple.
 
Certainly there are several different lines of argumentation, why spacetime
has $3+1$ dimensions. 
However the algebraic framework described here has the advantage that it
is under certain constraints maximal abstract and does not require to
postulate and establish an elaborated physical framework based on which the 
arguments achieve explanatory power. Precisely speaking, our considerations
are less concerned with ``the world'' but instead with the most simple general 
interpretation of clusters of variables as objects of ``a world''. 
\begin{enumerate}
\item The eigenvalues of $2n\times 2n$ matrices over the reals are (for $n>1$) pairs of complex conjugate
eigenvalues. Since $N$ is even, all representations for $N>1$ have matrix dimensions that are multiples 
of $4$ and can therefore always be block-diagonalized to $4\times 4$-blocks. Therefore higher dimensional
cases based on $4^M\times 4^M$-matrices can be split into $M$ objects with representation 
$4\times 4$, i.e. into $M$ objects in $Cl_{3,1}$. 
\item 
The number of variables of the spinor should be equal to the dimension of the geometry
constructed from it. The Pauli and the Dirac algebra are the only cases where
the state vector $\psi$ has the same dimension as the spacetime constructed
from it, i.e. the only case where $2\,n=N$ so that
\begeq
2^N=N^2\,.
\label{eq_equaldim}
\endeq
This criterium is extremely strong and restricts $N$ to $N=2$ or $N=4$. 
We will discuss this point in Sec.~\ref{sec_mass}.
\item Since $N$ is even in all possible dynamically generated spacetime
  dimensions, the dimension of the spinor $2\,n$ is always a multiple of
  $4$. Therefore all higher-dimensional spinors (and hence spacetime
  geometries) can be re-interpreted as a number of interacting 4-spinors.   
\item In 3 space dimensions the number of rotation axes equals the number of spatial directions since 
one can form 3 rotator pairs $\y_j\,\y_k$ (both space-like vectors) without repetition from 3
spatial directions. Both - rotators and boosters - mutually anti-commute with each other 
and hence generate an orthogonal space of the same dimension. In more than 3 space dimensions this 
changes: The number of boosts stays proportional to the number of spatial
dimensions, but the number of (symplectic, i.e. binary) rotators grows
quadratically with the number of spatial directions $N-1$ since one can select 
$\left( N-1\atop 2\right)=(N-1)(N-2)/2$ rotators (see Tab.~\ref{tab_obs}). 
For $N-1=4$ space dimensions this gives $6$ rotators and not all of them are orthogonal to each other.
In more than $3$ space dimensions some rotators commute, for instance in a 4-dimensional space
the rotators $\y_1\y_2$ and $\y_3\y_4$ commute~\cite{Nibart}. This means that (dynamically generated) 
spacetimes with more than 3 spatial dimensions are topologically not homogeneous. More precisely:
$D=3$ is the only space dimension in which all rotational axis have the same
relation to all others.
\end{enumerate}
Hence the simplest possible generally useful dimensionality of ``objects'' is 
$3+1$~\footnote{Other authors also considered
it worthwhile to analyze aspects of the dimensionality of space-time from a
Lie-algebraic point of view~\cite{ChoKong}.
}. 

In Sec.~\ref{sec_sigma} we raised the question of measurability. 
We emphasized that all unit vectors have to be symplices in order to be measurable.
Since all basic elements (unit vectors) anticommute, also the bi-vectors - i.e. the
products of two unit vectors - are symplices.
Geometrically bi-vectors are ``oriented faces''. However products of 3 orthogonal symplices 
- which would correspond to a volume - are cosymplices and therefore have vanishing expectation 
values (see also App.~\ref{app_highdim})). 
It follows that in a dynamically emerging spacetime ``objects'' can have
positions (or directions), but they have no measurable volume. 
I.e. fundamental (measurable) objects have - interpreted as objects in
spacetime - no volume, they are point-like particles.

\subsection{The Real Dirac Matrices}
\label{sec_rdm}

In the previous section we have shown that the case of   
$4\times 4$-matrices that represent a geometric (Clifford) algebra is unique and fundamental.
It is the simplest and most general algebra that allows for construction of 
spacetime. And it is the only algebra that generates a spacetime {\it as we experience it}.
With respect to the basic variables $\psi$ it represents systems with 2 DOFs. 
It can also be regarded as ``fundamental'' as it describes the smallest possible 
oscillatory system with (internal) coupling or ``interaction''~\footnote{
The role of the Dirac matrices as generators of the symplectic group $Sp(4,R)$ has been 
described before, see for instance Ref.~(\cite{DGL}). It has been noted already in Ref.~\cite{Dirac}.}.

An appropriate choice of the basic matrices of the Clifford algebra $Cl_{3,1}\mathbb(R)$
are the well known Dirac matrices. Ettore Majorana was the first who described a system
of Dirac matrices that contains exclusively imaginary (or exclusively real) values~\cite{Majorana}.
We call these matrices the ``real Dirac matrices'' (RDMs) and since we are -
on the fundamental level - committed to the reals,
we will use the RDMs instead of the conventional system~\footnote{
According to the fundamental theorem of the Dirac matrices by Pauli, we are free to choose 
any system of Dirac matrices, since all systems are algebraically similar, i.e. can
be mapped onto each other via similarity transformations~\cite{Pauli}. 
The transformation to the conventional system is given in App.~\ref{sec_diracconv}.
The isomorphism of this matrix group to the 3+2-dimensional de Sitter group has been
pointed out by Dirac in Ref.~\cite{Dirac}.
}.

Often the term ``Dirac matrices'' is used in more restrictive way and
designates only four matrices, i.e. the generators of the Clifford algebra~\footnote{
Note that we define $\y_{14}=\y_0\,\y_1\,\y_2\,\y_3$, which is labeled $\y_5$ in QED.}.
The RDMs as we define them here include the complete system of $16$ real
matrices and include all elements of the Clifford algebra generated by $\y_\mu$:
\begary{cccp{5mm}ccc}
\y_{14}&=&\y_0\,\y_1\,\y_2\,\y_3;&&\y_{15}&=&{\bf 1}\\
\y_4&=&\y_0\,\y_1;&&\y_7&=&\y_{14}\,\y_0\,\y_1=\y_2\,\y_3\\
\y_5&=&\y_0\,\y_2;&&\y_8&=&\y_{14}\,\y_0\,\y_2=\y_3\,\y_1\\
\y_6&=&\y_0\,\y_3;&&\y_9&=&\y_{14}\,\y_0\,\y_3=\y_1\,\y_2\\
\y_{10}&=&\y_{14}\,\y_0&=&\y_1\,\y_2\,\y_3&&\\
\y_{11}&=&\y_{14}\,\y_1&=&\y_0\,\y_2\,\y_3&&\\
\y_{12}&=&\y_{14}\,\y_2&=&\y_0\,\y_3\,\y_1&&\\
\y_{13}&=&\y_{14}\,\y_3&=&\y_0\,\y_1\,\y_2&&\\
\label{eq_gamma_def}
\endary
There are $2\,n!=24$ possible permutations of the variables in the state vector $\psi$,
but since a swap of the two DOFs does not change the form of $\y_0$, there
are $12$ possible basic systems of RDMs~\cite{rdm_paper}. Since $\y_0$ is antisymmetric, 
there are only six possible choices for $\y_0$. Each of them allows to choose between two 
different sets of the ``spatial'' matrices~\footnote{
Given one system is known as $\y_\mu$ then the alternative system is $\y_0,\,\y_0\y_1,\,\y_0\y_2,\,\y_0\y_3$.}.
The explicit form of the RDMs is given in Ref.~(\cite{rdm_paper,geo_paper}).
The four basic RDMs are anti-commuting
\begeq
\y_\mu\,\y_\nu+\y_\nu\,\y_\mu=2\,g_{\mu\nu}\,{\bf 1}\,,
\label{eq_metric}
\endeq 
and the ``metric tensor'' $g_{\mu\nu}$ has the form~\footnote{
The convention of QED ($g_{\mu\nu}=\mathrm{Diag}(1,-1,-1,-1)$)
cannot be represented with real Dirac matrices~\cite{Okubo,Scharnhorst}.}
\begeq
g_{\mu\nu}=\mathrm{Diag}(-1,1,1,1)\,.
\endeq
Any real $4\times 4$ matrix ${\bf A}$ can be written as a linear combination
of RDMs according to
\begeq
{\bf A}=\sum\limits_{k=0}^{15}\,a_k\,\y_k\,,
\label{eq_RDM_LC}
\endeq
where a quarter of the trace $\mathrm{Tr}({\bf A})$ equals the {\it scalar component}
$a_{15}$ - since all other RDMs have zero trace.
The RDM-coefficients $a_k$ are given by the {\it scalar product}
\begeq
a_k={\bf A}\,\cdot\,\y_k={\mathrm{Tr}(\y_k^2)\over 4}\,\mathrm{Tr}({{\bf A}\,\y_k+\y_k\,{\bf A}\over 8})\,.
\label{eq_scalarprod_def}
\endeq
Any real $4\times 4$ Hamiltonian matrix (symplex) ${\bf F}$ can be written as a linear combination in this basis
 \begeq
{\bf F}=\sum\limits_{k=0}^{9}\,f_k\,\y_k\,.
\label{eq_symplex_LC}
\endeq
If Minkowski spacetime is interpreted as a physical reality without intrinsic connection to quantum mechanics,
then it appears to be a jolly contingency that the (real) Dirac matrices have the geometric properties that 
exactly fit to the needs of spacetime geometry~\cite{Hestenes,JCA}.
In Ref.~\cite{JCA} the authors write: ``Historically, there have been many attempts to produce the 
appropriate mathematical formalism for modeling the nature of physical space, such as Euclid's geometry, 
Descartes' system of Cartesian coordinates, the Argand plane, Hamilton's quaternions, Gibbs' vector system 
using the dot and cross products. We illustrate however, that Clifford's geometric algebra (GA) provides 
the most elegant description of physical space.'' 
If spacetime emerges from dynamics then there is deeper reason for this elegance. 
Geometric algebra (GA) - if introduced by postulating abstract mathematical anticommuting objects -
leaves a mark of ``artificial construction'' that has few explanatory
power. But GA is way more than an elegant way to write down equations:
if {\it spacetime emerges from dynamics} in the form of the Dirac algebra, then we have the 
{\it missing link}, the {\it physical reason} for the congruence of GA with $3+1$-dimensional
Minkowski spacetime. Then the relativistic covariance of the Dirac algebra is
the only logical explanation for the form of the Lorentz transformations.
This implies a reversal of common understanding. Neither the experimentally
found or postulated properties of spacetime nor the postulated constancy
of the speed of - yet undefined - light waves are the proper reason for
the form of the Lorentz transformations, but the algebraic structure of
Hamiltonian dynamics of the fundamental variables.
It is a consequence of the physical restriction that constants of
motion are required as references for measurements. Measurable physical
properties of physical objects are those properties that appear or can be
related to (functions of) constants of motion - in some (other) physical
system. Non-measurable (boolean) properties are typically due to symmetry 
properties - again determined by the properties of the Dirac algebra.
We will show this in detail in the next section.

\section{Lorentz Transformations are Symplectic Transformations}
\label{sec_relativity}

{\it Structure preservation} refers to Eq.~(\ref{eq_strucdef}). That is: the transformation
of the skew-symmetric matrix ${\bf S}$ to $\y_0$ is the exact contrary of structure preserving -
it is {\it structure defining}. Only transformations that preserve the form of $\y_0$
are called structure preserving. And since $\y_0$ represents the form of Hamilton's equations, 
structure preserving transformations are {\it canonical}. {\it Linear} structure preserving 
transformations are called {\it symplectic}.

However, the form of $\y_0$ - as a result of Eq.~\ref{eq_strucdef} - is not unique in the
Dirac algebra. {\it Any} skew-symmetric $\y$-matrix squares to $-{\bf 1}$ and is equally 
useable as symplectic unit matrix and can equally well represent the time direction. 
The use of the other skew-symmetric matrices ($\y_{10}$,$\y_{14}$,$\y_{7}$,$\y_{8}$,$\y_{9}$)
implies a permutation of the variables in $\psi$ and a permutation of the indices of the other 
matrices. There are six different equally legitimate systems~\cite{rdm_paper}. If we consider 
interactions between spinors, we have to take the possibility into account that not all (sub-)
systems are described by the same structure defining transformation according to
Eq.~\ref{eq_strucdef} - though we will not discuss this in detail here.

Using Eq.~\ref{eq_symplectic} it is easy to show that symplectic transformations are 
{\it structure preserving}, i.e. a transformation of the dynamical variables $\psi$ 
that does not change the form of the equations, neither it changes the form
of $\y_0$. For the matrix of observables ${\bf S}=\Sigma\y_0$ we expect a
similarity transformation. The linear transformation ${\bf R}$ is given as
\begeq
\tilde\psi={\bf R}\,\psi\,.
\label{eq_trans0}
\endeq
Hence 
\begary{rcl}
{\bf S}&=&\langle\psi\psi^T\rangle\,\y_0\\
{\bf\tilde S}&=&\langle\tilde\psi\tilde\psi^T\rangle\,\y_0\\
             &=&\langle{\bf R}\,\psi\psi^T\,{\bf R}^T\rangle\,\y_0\\
             &=&-{\bf R}\,\langle\psi\psi^T\rangle\,\y_0\,\y_0\,{\bf R}^T\y_0\\
             &=&-{\bf R}\,{\bf S}\,\y_0\,{\bf R}^T\y_0\,,
\endary
so that the conditions are fulfilled if ${\bf R}$ is symplectic (fulfills Eq.~\ref{eq_symplectic}):
\begary{rcl}
-\y_0\,{\bf R}^T\y_0&=&{\bf R}^{-1}\\
{\bf R}\,\y_0\,{\bf R}^T&=&\y_0\,.
\endary
Then it is easy to show that symplectic transformations are {\it structure preserving}, i.e. a 
symplex ${\bf F}$ transformed by a symplectic similarity transformation ${\bf R}$ remains a symplex:
\begary{rcl}
{\bf\tilde F}&=&{\bf R}\,{\bf F}\,{\bf R}^{-1}\\
\Rightarrow {\bf\tilde F}^T&=&({\bf R}^{-1})^T\,{\bf F}^T\,{\bf R}^T\\
                           &=&(-\y_0\,{\bf R}\,\y_0)\,(\y_0\,{\bf F}\,\y_0)\,(-\y_0\,{\bf R}^{-1}\,\y_0)\\
                           &=&\y_0\,{\bf R}\,{\bf F}\,{\bf R}^{-1}\,\y_0\\
                           &=&\y_0\,{\bf\tilde F}\y_0\,.
\endary
In combination with Eq.~(\ref{eq_similarity}) one finds that a transformation 
is symplectic if it is possible to find a (constant) Hamiltonian from which 
the transformation can be derived: a transformation is symplectic, if it can 
be expressed as the result of a {\it possible evolution in time}. This implies
that not all similarity transformations are possible evolutions in time.

As we show in the following, it is legitimate to call the structure preserving
property of symplectic transformations the {\it principle of (special) relativity} 
since Lorentz transformations (LTs) of spinors in Minkowski spacetime are a subset of 
the possible symplectic transformations for two DOFs. 
It is well-known that the matrix exponential of a ``Hamiltonian matrix'' (i.e. a symplex)
is a symplectic transformation. It remains to be shown which symplices generate LTs.
Consider we transform a matrix ${\bf X}={\cal E}\,\y_0+P_x\,\y_1+P_y\,\y_2+P_z\,\y_3$ using 
${\bf R}_4=\exp{(\y_4\,\eps/2)}$,
then we obtain after decomposition of the transformed matrix ${\bf X}'$ into the RDM-coefficients:
\begary{rcl}
{\bf X}'&=&{\bf R}_4\,{\bf X}\,{\bf R}_4^{-1}\\
{\bf X}'&=&{\cal E}'\,\y_0+P_x'\,\y_1+P_y'\,\y_2+P_z'\,\y_3\\
{\cal E}'&=&{\cal E}\,\cosh{(\eps)}+P_x\,\sinh{(\eps)}\\
P_x'&=&P_x\,\cosh{(\eps)}+{\cal E}\,\sinh{(\eps)}\\
P_y'&=&P_y\\
P_z'&=&P_z\\
\label{eq_lt1}
\endary
With the usual parametrization where $\eps$ is the ``rapidity''
($\beta=\tanh{(\eps)}$, $\y=\cosh{(\eps)}$, $\beta\y=\sinh{(\eps)}$), one finds
\begary{rcl}
{\cal E}'&=&\y\,{\cal E}+\beta\,\y\,P_x\\
P_x'&=&\y\,P_x+\beta\,\y\,{\cal E}\,,
\endary
which is a Lorentz boost along $x$.
Same transformation, but ${\bf X}=E_x\,\y_4+E_y\,\y_5+E_z\,\y_6+B_x\,\y_7+B_y\,\y_8+B_z\,\y_9$:
\begary{rcl}
{\bf X}'&=&{\bf R}\,{\bf X}\,{\bf R}^{-1}\\
E_x'&=&E_x\\
B_x'&=&B_x\\
E_y'&=&E_y\,\cosh{(\eps)}+B_z\,\sinh{(\eps)}\\
E_z'&=&E_z\,\cosh{(\eps)}-B_y\,\sinh{(\eps)}\\
B_y'&=&B_y\,\cosh{(\eps)}-E_z\,\sinh{(\eps)}\\
B_z'&=&B_z\,\cosh{(\eps)}+E_y\,\sinh{(\eps)}\\
\label{eq_lt2}
\endary
which is again a Lorentz boost along $x$, but now it shows the transformation behavior
of electromagnetic fields. Obviously the algebraic structure of the Dirac matrices
in combination with symplectic dynamics results directly in Lorentz transformations. 
Indeed the transformations above are identical to the Lorentz transformations of the 
Dirac equation\footnote{Most textbooks on relativistic QM contain inadequately
short and/or incomprehensible descriptions of the covariance of the Dirac
equation. The best depiction found by the author is given in Ref.~(\cite{Schmueser}).}. 

Our approach does not refer to the speed of light, to time dilation
or with alike. N.D. Mermin once wrote: ``Relativity is not a branch of 
electromagnetism''~\cite{Mermin}. Our approach goes beyond this statement. 
We did not directly introduce spacetime. But we introduced Lorentz
transformations in direct combination with the electromagnetic fields. 
Lorentz transformations, electrodynamics, spinors, and the mass-energy 
relation all together are facets of the same phenomenon. Space-time is
generated by transformation operators, which also represent the
electromagnetic interaction.
The conventional approach derives LTs from the ``constancy of the
speed of light''. This is a phenomenological Ansatz which certainly has a 
legitimization in the history of physics, but is {\it logically misleading}: 
the ``constancy'' of the speed of light is neither the cause nor the reason
for, but just one aspect of (special) relativity. {\it Relativity} as such
is nothing but structure preservation, i.e. it corresponds to the linear
canonical transformations of classical mechanics. Singh continued the work of
Mermin in the (algebraic) direction~\cite{Singh}. At about the same time Penrose and 
Rindler wrote in Ref.~(\cite{PR1}): 
``The formalism most commonly used for the mathematical treatment of manifolds and
their metrics is, of course, the tensor calculus (or such essentially equivalent
alternatives as Cartan's calculus of moving frames). But in the specific
case of four dimensions and Lorentzian metric there happens to exist - by
accident or providence - another formalism which is in many ways more
appropriate, and that is the formalism of 2-spinors.''
What Rindler formulates in terms of complex 2-spinors, can (more ``classically'')
be expressed by real 4-spinors as we present it here. In this way it becomes
more evident that the emergence of spacetime is no accident, but the simplest 
method of pattern recognition of abstract and general symplectic dynamics.
Furthermore, by the use of real 4-spinors we discover that the electromagnetic
fields are intrinsically connected to the emergent spacetime geometry. This
is the reason why the speed of light is related to the geometry of spacetime.
This is explained in more detail in Sec.~\ref{sec_emeq}.

The possibility to derive electrodynamics from relativity alone has been 
questioned~\cite{FLS}. But in the context of symplectic dynamics as suggested here
there are additional restrictions and MWEQs can indeed be derived. It is an 
experimental fact that electromagnetic interactions are structure preserving. 
No electromagnetic process known to the author is able to transform a fermion 
into another. There is pair production and particle-antiparticle annihilation,
but it is conventional wisdom to identify the antiparticle with the particle
going ``backwards in time''~\cite{Schmueser}. 
The fact that Lorentz transformations of the Dirac spinor are 
indeed linear canonical transformations, is long known~\cite{Dirac,Jauch}. 
But it is rarely mentioned in discussions on relativity - likely because 
relativity is almost exclusively depicted as a theory of space and time.
However - as Edward J. Gillis remarked - ``relativity is not about spacetime''~\cite{Gillis}.

\subsection{The Electromechanical Equivalence (EMEQ)}
\label{sec_emeq}

We started our considerations with abstract entities. It is therefore unavoidable
to present an {\it interpretation} of the terms (i.e. the observables) derived 
from these entities. The interpretations themselves can neither be ``derived'' 
nor be proven to be ``correct''. We can only show that they are consistent and
meaningful and that they have explanatory power. Interpretations make sense or 
they don't - but they do not exclude other interpretations.

The transformation properties (\ref{eq_lt1}) and (\ref{eq_lt2}) 
suggest the introduction of the electromechanical equivalence (EMEQ), as presented in 
Refs.~(\cite{rdm_paper},\cite{geo_paper}).
In those papers the EMEQ was merely used as a {\it formal tool} that allows to
obtain a descriptive interpretation of symplectic transformations.
Here we argue substantially and present a derivation of MWEQs.

In Sec.~\ref{sec_symtrans} we have shown that the fundamental solution of the EQOM 
is given by a symplectic transfer matrix. In order to produce {\it constants of motion},
the transfer matrix should represent a strongly stable system, i.e. all of its eigenvalues lie
on the unit circle in the complex plane. In this case the symplex ${\bf F}$ has exclusively 
purely imaginary eigenvalues. We analyzed the structure of ${\bf F}$ by introducing 
the RDMs which are a representation of the Clifford algebra $Cl_{3,1}$. Therefore the
list of unit symplices includes the four basic elements $\y_0\dots\y_3$ and
the six bi-vectors. The other $6$ members of the group of the RDMs are cosymplices.
The analysis of the transformation properties indicates a possible physical
interpretation of the coefficients of the algebra: we associate the $f_k$ 
(Eq.~\ref{eq_symplex_LC}) as follows:
\begary{rclr}
f_0&\equiv&{\cal E}&\textrm{Energy}\\
(f_1,f_2,f_3)^T&\equiv&{\vec P}&\textrm{Momentum}\\
(f_4,f_5,f_6)^T&\equiv&{\vec E}&\textrm{electric field}\\
(f_7,f_8,f_9)^T&\equiv&{\vec B}&\textrm{magnetic field}\\
\label{eq_emeq}
\endary
This association is not arbitrary as the transformation behavior under rotations and 
Lorentz boosts of the $f_k$ has to fit to the corresponding physical quantities. However
it is not unique. Besides the electromagnetic fields, any second rank tensor would fulfill
the formal requirements. If we would not be interested in electrodynamics but for instance
in hydrodynamics or the motion of rigid bodies, we would have to use a different scheme of
interpretation. This ambiguity is unavoidable since our starting point was the abstract
structure of Hamiltonian symplectic dynamics~\footnote{And it tells us that major parts of the laws
of physics are the result of pattern recognition in the same sense as the interpretation of 
acoustic signals in terms of sound and music is. Because, that is what we do
here: We analyze the patterns of fundamental variables in Hamiltonian motion.}.

However our choice is legitimate if the EMEQ is consistent, meaningful and if it helps in
structural analysis as we will show in what follows. Dimensional problems
should not appear as long as we can find physical constants that allow to
translate energy, momentum, electric and magnetic fields into frequencies (the
dimension of ${\bf F}$ is a frequency. For the energy the required factor is $\hbar$:
\begary{rcl}
{\cal E}/\hbar&=&\mathrm{frequency}\\
\vert\vec P\vert\,c/\hbar&=&\mathrm{frequency}\\
{e\over m_e\,c}\,\vert\vec E\vert&=&\mathrm{frequency}\\
{e\over m_e}\,\vert\vec B\vert&=&\mathrm{frequency}\\
\endary
We assume that in our unit system all these units have the numerical value of
unity, i.e. $m_e=e=\hbar=c=1$.

With this nomenclature the eigenvalues $\pm\,i\,\w_1,\,\pm\,i\,\w_2$ of ${\bf F}$ are given as:
\begary{rcl}
\omega_1 &=&\sqrt{K_1+2\,\sqrt{K_2}}\\
\omega_2 &=&\sqrt{K_1-2\,\sqrt{K_2}}\\
K_1&=&-\mathrm{Tr}({\bf F}^2)/4={\cal E}^2+\vec B^2-\vec E^2-\vec P^2\\
K_2&=&\mathrm{Tr}({\bf F}^4)/16-K_1^2/4\\
&=&({\cal E}\,\vec B-\vec P\times\vec E)^2-(\vec E\cdot\vec B)^2-(\vec P\cdot\vec B)^2\\
\mathrm{Det}({\bf F})&=&\w_1^2\,\w_2^2=K_1^2-4\,K_2\\
\label{eq_eigenfreq}
\endary
Bi-vectors form a so-called {\it even subalgebra} of the Dirac algebra\footnote{
We define elements to be {\it even}, if they do not change sign when all {\it basic elements} 
$\y_0,\dots,\y_3$ change sign. They are odd otherwise. I.e. bi-vectors are even, vectors
and tri-vectors are odd.}. We have shown that any Clifford algebra $Cl_{p,q}$ which
is able to represent spacetime has an even number of generators $N$. This holds also
for the Dirac algebra. Even dimensional algebras have {\it even subalgebras}, which means that
even elements generate exclusively even elements, whereas odd elements can be used to generate 
the full algebra: Fermions can generate electromagnetic fields, but photons
can not generate fermions.
Consequently there should be (differential-) equations that allow to derive the bi-vector fields 
$\vec E$ and $\vec B$ from vectors, but not vice versa. This means, that the description of a 
free particle without external fields should not refer to bi-vectors. 
Hence it is natural to distinguish the vector and bi-vector components and for vanishing 
bi-vectors ($\vec E=\vec B=0$) one finds:
\begary{rcl}
K_1&=&{\cal E}^2-\vec P^2\\
K_2&=&0\\
\omega_1 &=&\omega_2 =\sqrt{K_1}=\sqrt{{\cal E}^2-\vec P^2}\,,
\label{eq_noEB}
\endary
which are the relativistic invariants of matter-fields.
At the end of Sec.~\ref{sec_sigma} we argued that a constant measurement reference can be 
constructed as a quadratic form. 
Here we have an example, since $\sqrt{K_1}=m=\sqrt{{\cal E}^2-\vec P^2}$:
{\it The mass is a constant of motion and has the physical meaning and unit of frequency}.
This is an important result and a significant step towards quantum mechanics:
It is the combination of the Planck-Einstein relation $E=\hbar\,\omega$ and
Einstein's $E=m\,c^2$. It follows that the ``time'' parameter as we introduced
it in Eqs.~(\ref{eq_Heqom1}), corresponds to the {\it eigentime} or {\it
  proper time}. The eigenfrequency that defines the scale of the eigentime is the
mass and hence a massive particle is described by an inertial frame of reference in combination 
with an oscillator of constant frequency (i.e. a clock). This is the deeper meaning of 
Einstein's clock attached to all inertial frames. 

For vanishing vector components (${\cal E}=0=\vec P$) we have
\begary{rcl}
K_1&=&\vec B^2-\vec E^2\\
K_2&=&-(\vec E\cdot\vec B)^2\,,
\label{eq_noEP}
\endary
which are the well-known relativistic invariants of the electromagnetic field.
It follows that the frequencies are only real-valued for $\vec E\cdot\vec B\ge 0$
and $\vec B^2-\vec E^2\ge 0$.
Furthermore, the eigenfrequencies of a pure electromagnetic wave vanish since
$K_1=K_2=0$. This implies that we can not transform by whatever means into
a system of a pure e.m.-wave: The {\it ideal} electromagnetic wave
has no inertial system, no eigenfrequency and no intrinsic phase advance.
An analysis of the transformations generated by the bi-vectors yields that the
elements $\y_4, \y_5$ and $\y_6$ are responsible for Lorentz boosts, while 
$\y_7, \y_8$ and $\y_9$ are generators of (spatial) rotations. This corresponds to 
our physical intuition since (charged) particles are {\it accelerated} by electric 
fields but their ``trajectories'' are bended (i.e. rotated) in magnetic fields. 
It is therefore no surprise that - using the EMEQ - the derivation of the Lorentz 
force is straightforward (see Sec.~\ref{sec_lorentzforce}).

Furthermore we note that there are vector-elements associated with ${\cal E}$ and $\vec P$, but no
elements associated directly with the spacetime coordinates. For a representation of a single elementary
particle this makes sense insofar as the dynamics of a single particle may not refer to absolute
positions (unless via the bi-vector fields). Spacetime coordinates do not refer to particle properties,
but to the relativ position (relation) of particles. Hence as long as we refer to a single object
based on fundamental variables, we observe the momentum and energy {\it of} a particle as it behaves
in an {\it external electromagnetic field}. 

As we already mentioned in Sec.~(\ref{sec_sigma}), the symplex 
${\bf S}=\Sigma\,\y_0$ is singular, if we use only a single real spinor $\psi$ to define 
\begeq
\Sigma=\langle\,\psi\,\psi^T\,\y_0\,\rangle\,,
\endeq
where the angles imply some (unspecified) sort of average.
Since the eigenvalues of the symplex ${\bf S}$ yield the ``frequency'' or ``mass''
$m=\sqrt{{\cal E}^2-\vec P^2}$ of the structure which we identify with a
particle, we need at least $N\ge 4$ linear independent spinors to obtain a
massive fermion. In the classical Dirac theory, the electron spinor is a
linear combination of two complex spinors, i.e. also requires 4 real spinors.
From this point of view {\it inertial mass} might be interpreted statistically -
as a property that can only be derived using averaging - either over samples or in time.
The idea that mass might be a statistical phenomenon is neither new nor
extraordinarily exotic, see for instance Ref.~(\cite{Verlinde,MvR1}) and 
references therein. But here we do not refer to entropy but rather to 
dynamical or algebraic properties of real spinors.

\subsection{The Lorentz Force}
\label{sec_lorentzforce}

The 4-momentum of a particle (fermion) is defined by vectors. 
The vector components of ${\bf S}$ are associated with the 4-momentum:
\begeq
{\bf P}={\cal E}\,\y_0+P_x\,\y_1+P_y\,\y_2+P_z\,\y_3={\cal E}\,\y_0+\vec P\,\vec\y\,,
\endeq
where ${\cal E}$ is the energy and $\vec P$ the momentum. Accordingly the fields are 
associated with the bi-vectors of the ``force'' matrix ${\bf F}$.
Then Eq.~(\ref{eq_env}) is (in appropriate units) the Lorentz force equations
\begeq
{d{\bf P}\over d\tau}=\dot{\bf P}={q\over 2\,m}\left({\bf F}\,{\bf P}-{\bf P}\,{\bf F}\right)\,,
\label{eq_lorentzforce}
\endeq
where $\tau$ is the proper time and ${q\over 2 m}$ is a relative scaling factor.
In the lab frame time $dt=\y\,d\tau$ EQ.~\ref{eq_lorentzforce} yields (setting $c=1$):
\begary{rcl}
{d{\cal E}\over d\tau}&=&{q\over m}\,\vec P\,\vec E\\
{d\vec P\over d\tau}&=&{q\over m}\,\left({\cal E}\,\vec E+\vec P\times\vec B\right)\\
\y\,{d{\cal E}\over dt}&=&q\,\y\,\vec v\,\vec E\\
\y\,{d\vec P\over dt}&=&{q\over m}\,\left(m\,\y\,\vec E+m\,\y\,\vec v\times\vec B\right)\\
{d E\over dt}&=&q\,\vec v\,\vec E\\
{d\vec P\over dt}&=&q\,\left(\vec E+\vec v\times\vec B\right)\,.
\endary
(Note that Eq.~\ref{eq_lorentzforce} allows to add arbitrary multiples of ${\bf P}$ to ${\bf F}$
without an effect on ${\bf \dot P}$.)

\section{Electrodynamics from Symplectic Spacetime}
\label{sec_method}

The electromagnetic terms of the symplex ${\bf F}$ are bi-vectors. 
It has been pointed out in Sec.~\ref{sec_emeq} that bi-vectors form
an {\it even subalgebra} which implies that they are generated by the
product of two vectors. Up to now only a single type of 4-vector 
has been introduced, namely the energy-momentum 4-vector of a single 
system (``particle'') of interest. Using a pair of 4-vectors one can
construct a bi-vector by the use of the commutator of two vectors. 
However, in order to have a non-zero bivector, these two vectors must
be different. This could either be the energy-momentum 4-vector of a 
second system or a 4-vector representing something different than energy 
and momentum. 
Firstly we can argue that a second particle did - in contrast to the 
bi-vectors - not (yet) appear in our considerations and therefore 
it is unavoidable to introduce a new 4-vector. And secondly, the EMEQ
introduces momenta without the corresponding coordinates. Hence
we write this 4-vector as 
\begeq
{\bf X}=\y_0\,t+\vec\y\cdot\vec x\,,
\endeq
and express the fields $\vec E$ and $\vec B$ - which were 
originally understood as functions of the eigentime $\tau$ -
by $\vec x$ and $t$:
\begary{rcl}
\vec E(\tau)&\to&\vec E(\vec x,t)\\
\vec B(\tau)&\to&\vec B(\vec x,t)\,.
\label{eq_construct0}
\endary
This means that we extend the dynamics of (single) observables to the dynamics 
of ``fields'', depending on the new vector type parameters $t$ and $\vec x$. 
For now we keep open the question, how these vectors are related to $({\cal E},\vec P)$. 
But we can't resist to cite Einstein here:
``Spacetime does not claim existence on its own, but only as a structural 
quality of the field''~\cite{Einstein}.

There have been several attempts to ``derive'' MWEQs in the 
past~\cite{Krefetz,Kobe1,Kobe2,Dyson,Gersten,DWGS}. None of these succeeded in finding
general acceptance. Most textbooks treat just the Lorentz covariance of 
MWEQs. A derivation of MWEQs from the covariance condition alone is considered 
impossible~\cite{FLS}. 
But our ansatz is based on an algebraic framework with additional (i.e. symplectic) 
constraints. Symplectic constraints are well-known in different context and 
led (for instance) to the non-squeezing theorem, i.e. to the metapher of the 
``symplectic egg''~\cite{Gromov,Gosson}. 
We have shown that Lorentz transformations as well as the action of the Lorentz 
force are symplectic. The appearance of cosymplices as generators is excluded 
for the structure-preserving electromagnetic theory.

In order to keep track of the transformation properties with respect to
symplectic transformations the bi-vectors fields can only
be expressed (for instance in form of a Taylor series) by algebraic
expressions that respect the appropriate transformation properties. 
According to Eq.~(\ref{eq_emeq}) the field matrix is given by
\begeq
{\bf F}=E_x\,\y_4+ E_y\,\y_5+ E_z\,\y_6+ B_x\,\y_7+ B_y\,\y_8+ B_z\,\y_9\,.
\endeq
If we write the fields as a Taylor series, the first term {\it must} have the 
following form in order to be linear in ${\bf X}$ {\it and} to yield a bi-vector:
\begeq
{\bf F}={\bf F}_0+({\cal D}{\bf F})\,{\bf X}-{\bf X}({\cal D}{\bf F})+\cdots\,,
\label{eq_MWEQ0}
\endeq
where ${\cal D}{\bf F}$ is an appropriate derivative taken at ${\bf X}=0$
and {\it must} be a vector.
Alternative forms of the linear term either include an ``axial'' vector (cosymplex) 
${\bf V}$ or a pseudoscalar, both having a vanishing expectation value:
\begary{rcl}
{\bf V}&=&V_t\,\y_{10}+ V_x\,\y_{11}+ V_y\,\y_{12}+ V_z\,\y_{13}\\
{\bf F}_1&=&({\bf V}\,{\bf X}+{\bf X}{\bf V})/2\\
\endary
Since we can exclude the appearance of cosymplices in structure preserving 
interactions, the first order term has the form of Eq.~\ref{eq_MWEQ0}. 

In the following we analyze first and second order expressions that can be 
constructed with respect to the transformation properties. We skip constant 
terms for the moment and analyze the relations of the partial derivatives. 
Linear terms can exclusively be expressed by the commutator of
two vector quantities, i.e. ${\bf X}$ and another (constant) 4-vector ${\bf J}=\rho_0\,\y_0+\vec j_0\cdot\vec\y$:
\begeq
{\bf F}_1=(\y_0\,\vec E+\y_{14}\,\y_0\vec B)\cdot\vec\y={4\pi\over 3}\,\frac{1}{2}\,({\bf X}\,{\bf J}-{\bf J}\,{\bf X})\,,
\endeq
so that we obtain
\begary{rcl}
\vec E&=&{4\pi\over 3}\,(\vec j_0\,t-\rho_0\,\vec x)\\
\vec B&=&{4\pi\over 3}\,(\vec x\times\vec j_0)\\
\endary
and find that these linear terms fulfill MWEQs:
\begary{rcl}
\vec\nabla\cdot\vec E&=&4\pi\,\rho_0\\
\vec\nabla\cdot\vec B&=&0\\
\vec\nabla\times\vec E+\partial_t\vec B&=&0\\
\vec\nabla\times\vec B-\partial_t\vec E&=&{4\pi\vec j_0}\\
\endary
We continue with a second order term~\footnote{In first order the two partials of the induction law vanish separately, 
i.e $\vec\nabla\times\vec E=0=\partial_t\vec B$.}:
\begeq
{\bf F}_2=\frac{1}{4}\,({\bf X}\,{\bf J}\,{\bf X}\,{\bf J}-{\bf J}\,{\bf X}\,{\bf J}\,{\bf X})\,,
\endeq
and it also fulfills MWEQs:
\begary{rcl}
\vec\nabla\cdot\vec E&=&4\,\pi\,\rho_1\\
\vec\nabla\cdot\vec B&=&0\\
\vec\nabla\times\vec E+\partial_t\vec B&=&0\\
\vec\nabla\times\vec B-\partial_t\vec E&=&4\,\pi\,\vec j_1\,,
\endary
where 
\begary{rcl}
\rho_1&=&-\vec j_0^2\,t-3\,t\,\rho_0^2+4\,(\vec j_0\cdot\vec x)\,\rho_0\\
\vec j_1&=&4\,(\vec j_0\cdot\vec x-\rho_0\,t)\vec j_0+(\rho_0^2-\vec j_0^2)\,\vec x\,.
\endary
Both orders fulfill the continuity equation:
\begeq
\partial_t\,\rho+\vec\nabla\cdot\vec j=0\,.
\endeq
So far this is not a derivation of MWEQs, but it shows that the transformation
properties of bi-vectors and the construction of spacetime imply MWEQs. 
There is a method to express these ideas mathematically by the use of differential operators. 
Since the aim is the construction of spacetime, the appropriate generalized
derivative must by of vector-type, i.e. we define the covariant derivative by
\begeq
\d\equiv -\d_t\,\y_0+\d_x\,\y_1+\d_y\,\y_2+\d_z\,\y_3\,.
\label{eq_deriv}
\endeq
The non-abelian nature of matrix multiplication requires to distinguish differential operators 
acting to the right and to the left, i.e. we have $\d$ as defined in Eq.~\ref{eq_deriv}, $\rightD{\d}$ and
$\leftD{\d}$ which is written to the right of the operand (thus indicating the order of the matrix multiplication) 
so that 
\begary{rcl}
{\bf F}\leftD{\d}&\equiv& -\d_t\,{\bf F}\,\y_0+\d_x\,{\bf F}\,\y_1+\d_y\,{\bf F}\,\y_2+\d_z\,{\bf F}\,\y_3\\
\rightD{\d}{\bf F}&\equiv& -\y_0\,\d_t\,{\bf F}+\y_1\,\d_x\,{\bf F}+\y_2\,\d_y\,{\bf F}+\y_3\,\d_z\,{\bf F}\,.
\endary
In order to keep the transformation properties of derivative expressions transparent, we distinguish
between the {\it commutative} 
\begeq
\d\wedge{\bf A}\equiv\frac{1}{2}\,\left(\rightD{\d}{\bf A}-{\bf A}\leftD{\d}\right)
\endeq
and the {\it anti-commutative} 
\begeq
\d\cdot{\bf A}\equiv\frac{1}{2}\,\left(\rightD{\d}{\bf A}+{\bf A}\leftD{\d}\right)
\endeq
derivative. Then we find:
\begary{rcl}
\frac{1}{2}\,\left(\rightD{\d}\textrm{ vector }-\textrm{ vector }\leftD{\d}\right)&\Rightarrow&\textrm{ bi-vector}\\
\frac{1}{2}\,\left(\rightD{\d}\textrm{ vector }+\textrm{ vector }\leftD{\d}\right)&\Rightarrow&\textrm{ scalar}=0\\
\frac{1}{2}\,\left(\rightD{\d}\textrm{ bi-vector }-\textrm{ bi-vector }\leftD{\d}\right)&\Rightarrow&\textrm{ vector}\\
\frac{1}{2}\,\left(\rightD{\d}\textrm{ bi-vector }+\textrm{ bi-vector }\leftD{\d}\right)&\Rightarrow&\textrm{ axial vector}=0\\
\label{eq_difftypes}
\endary
Since (in case of the Dirac algebra) any symplex is the sum of a vector and a bivector, we find (for this special case):
\begary{rcl}
\frac{1}{2}\,\left(\rightD{\d}\textrm{ symplex }+\textrm{ symplex }\leftD{\d}\right)&=&\d\cdot\textrm{ symplex }=0\\
\label{eq_difftypes2}
\endary

Now we return to Eq.~\ref{eq_MWEQ0} and find that ${\cal D}{\bf F}$ must be a vector and 
therefore has (in ``first order'') the form
\begeq
{\cal D}{\bf F}=\frac{1}{2}\,\left(\rightD{\d}{\bf F}-{\bf F}\leftD{\d}\right)=4\,\pi\,{\bf J}\,,
\label{eq_MWinhomo}
\endeq
which is nothing but a {\it definition} of the {\it vector current}
\begeq
{\bf J}=\rho\,\y_0+j_x\,\y_1+j_y\,\y_2+j_z\,\y_3\,.
\endeq
Written explicitly in components, Eq.~\ref{eq_MWinhomo} is given by 
\begary{rcl}
\vec\nabla\cdot\vec E&=&4\,\pi\,\rho\\
\vec\nabla\times\vec B-\d_t\vec E&=&4\,\pi\,\vec j\\
\endary

We define a vector-potential ${\bf A}=\phi\,\y_0+A_x\,\y_1+A_y\,\y_2+A_z\,\y_3$ 
which has (to first order) the form~\footnote{
The commutative derivative of the trivial first order form ${\bf A}\propto{\bf X}$ vanishes and is therefore
useless in this context.}:
\begeq
{\bf A}_1=\frac{1}{4}\,\left({\bf X}\,{\bf F}_0-{\bf F}_0\,{\bf X}\right)\,,
\label{eq_vecpot}
\endeq
where ${\bf F}_0$ is the constant term of Eq.~\ref{eq_MWEQ0}. 
It is easily verified that ${\bf F}_0=\frac{1}{2}\,\left(\rightD{\d}{\bf A}_0-{\bf A}_0\leftD{\d}\right)$ and hence
the electromagnetic (bi-vector-) fields of the general symplex ${\bf F}$ is given by
\begeq
{\bf F}=\frac{1}{2}\,\left(\rightD{\d}{\bf A}-{\bf A}\leftD{\d}\right)\,,
\label{eq_pot_grad}
\endeq
or explicitly in components:
\begary{rcl}
\vec E&=&-\vec\nabla\phi-\d_t\vec A\\
\vec B&=&\vec\nabla\times\vec A\,.
\endary

It is well-known that the homogeneous MWEQs are a direct consequence of Eq.~\ref{eq_pot_grad}:
\begary{rcl}
\rightD{\d}{\bf F}+{\bf F}\leftD{\d}&=&\frac{1}{2}\,\left(\rightD{\d}^2\,{\bf A}-{\bf A}\leftD{\d}^2\right)=0\\
\vec\nabla\cdot\vec B&=&0\\
\vec\nabla\times\vec E+\d_t\vec B&=&0\\
\label{eq_MWhomo}
\endary
since the squared operators are scalars and commute with ${\bf A}$. 
Accordingly the continuity equation is a direct consequence of Eq.~\ref{eq_MWinhomo}:
\begary{rcl}
\rightD{\d}{\bf J}+{\bf J}\leftD{\d}&=&\frac{1}{8\pi}\,\left(
\rightD{\d}^2{\bf F}-\rightD{\d}{\bf F}\leftD{\d}+\rightD{\d}{\bf F}\leftD{\d}-{\bf F}\leftD{\d}^2\right)\\
&=&\d_t\rho+\vec\nabla\vec j=0\,.
\label{eq_continuity}
\endary
The validity of the Lorentz gauge follows from Eq.~\ref{eq_difftypes}
\begeq
\frac{1}{2}\,\left(\rightD{\d}{\bf A}+{\bf A}\leftD{\d}\right)=\d_t\,\phi+\vec\nabla\vec A=0\,,
\label{eq_Lgauge}
\endeq
and one obtains the wave equation of the vector potential ${\bf A}$:
\begary{rcl}
4\,\pi\,{\bf J}&=&-\rightD{\d}^2{\bf A}=(\d_t^2-\vec\nabla^2)\,{\bf A}\,.
\label{eq_waveeq}
\endary
In a ``current free region'' the fields have to fulfill the homogeneous wave equation:
\begeq
\rightD{\d}^2\,{\bf F}=(\vec\nabla^2-\d_t^2){\bf F}=0\,.
\label{eq_emwave}
\endeq
It follows that plane electromagnetic fields ``in vacuum'' hold $\w^2=\vec
k^2$ so that the fields have no ``eigenfrequency'' and no dispersion.

Note that Eqs.~(\ref{eq_MWhomo}) and~(\ref{eq_continuity}) are of the same form.
The continuity equation (\ref{eq_continuity}) describes charge as something with 
the properties of a substance in spacetime. 

In summary: We have shown that Lorentz transformations can indeed be
understood as symplectic transformations.
The generators of these transformations are electromagnetic fields.
Since electrodynamics is a structure preserving interaction, the description
of electromagnetic fields requires that exclusively structure preserving terms
appear.
With these ``additional conditions'' we were able to derive MWEQs 
if we interpret the bi-vectors of the matrix ${\bf F}$ as functions of vector-type 
spacetime coordinates of $Cl_{3,1}(\mathbb{R})$. The bi-vector fields 
in ${\bf F}$ fulfill a wave equation that describes the propagation of electromagnetic 
waves, i.e. of light. One finds that electromagnetic waves ``in vacuum'' fulfill the 
relation $\w^2=\vec k^2$. From the constant group velocity
${\partial\w\over\partial k}=1$ we obtain the constancy of the speed of light.
Therefore we did not derive MWEQs from special relativity, but both theories are the
result of the construction of spacetime from symplectic dynamics.

Above equations are not restricted to $3$ space dimensions.
However the connections between commutators and anti-commutators are way more
complex in $10$ or $12$ dimensions. We had to consider additional
``channels'': In 3-dimensional space, the symplex ${\bf F}$ is composed
exclusively of components of the vector and bi-vector type. 
In higher-dimensional space (10- or 12-dimensional spacetime), the general symplex ${\bf F}$ 
consists (besides vectors and bi-vectors) of n-vectors with $n\in [5,6,9,10,\dots]$ 
(see App.~\ref{app_highdim}). Accordingly we had to include higher-order terms
into the Hamiltonian and likely we would have trouble with the stability of
the described objects.

\subsection{The Density}
\label{sec_density}

From the combination of Eq.~\ref{eq_construct0} and Eq.~\ref{eq_waveeq} it follows that also
the (current-) density has also to be constructed as a function of space and time:
\begary{rcl}
\rho&=&\rho(\vec x,t)\\
\vec j&=&\vec j(\vec x,t)\,.
\endary
and it is clear that we have to normalize the density by
\begeq
\int\,\rho(\vec x,t)\,d^3x=const=Q\,,
\endeq
where $Q$ is the ``charge''.
Especially the relation between the density $\rho(\vec x,t)$ and the ``phase space density''
$\tilde\rho(\psi)$ according to Eq.~\ref{eq_rhodef} requires some attention. It is clear that the 
naive assumption 
\begeq
\rho(\vec x,t)=\tilde\rho(\psi(\vec x,t))
\endeq
can not be applied directly. According to Eq.~\ref{eq_rhodef} we have
\begeq
1=\int\,\tilde\rho(\psi)\,d^4\psi=\int\,\tilde\rho(\psi(\vec x,t))\,\sqrt{g}\,d^3x\,,
\label{eq_normalization}
\endeq
where $g$ is the appropriate Gramian determinant. We assume in the following that the normalization
has been adjusted accordingly.

If spacetime would be fundamental, we could be sure that the wave function $\psi(\vec x,t)$ was 
well-defined and single valued for any coordinate of Minkowski spacetime. However, if spacetime 
emerges from dynamics, then spacetime coordinates might be functions of the phase space position 
$t(\psi),\vec x(\psi)$. This includes the possibility that a) different phase space positions are 
mapped to the same spacetime-position and b) that space and time coordinates are not unique, i.e. 
particles appear ``instantaneously'' at different ``locations'' of spacetime. Finally this might 
indeed be the reason why $\psi$ has to be interpreted as a ``probablity density''.

\subsection{The Wave Equations}
\label{sec_waveeq}

According to Eq.~\ref{eq_waveeq} the vector potential fulfills a wave equation - and also 
the (electromagnetic) fields can be described by waves. Solutions of wave equations are usually 
analyzed with the help of the Fourier transformation, i.e.
the solutions can be written as superpositions of plane waves
\begary{rcl}
\phi(\vec x,t)&=&\int\,\tilde\phi(\vec k,\w)\,e^{i(\vec k\cdot\vec x-\w t)}\,d^3k\\
\tilde\phi(\vec k,\w)&=&\int\,\phi(\vec x,t)\,e^{-i(\vec k\cdot\vec x-\w t)}\,d^3x\,,
\endary
where we skipped normalization constants for simplicity.
One important feature of the Fourier transform is the replacement of nabla operator $\vec\nabla$
with the wave vector $i\vec k$ and of the time derivative with the frequency $\d_t\to -i\,\w$. 
The Fourier transformed MWEQs are algebraic conditions for wave functions (we skip 
the tilde as it is usually clear from the context, if we refer to the fields or their Fourier transform):
\begary{rcl}
i\,\vec k\cdot\vec B&=&0\\
i\,\vec k\times\vec E-i\,\w\vec B&=&0\\
i\,\vec k\cdot\vec E&=&4\pi\,\rho\\
i\,\vec k\times\vec B+i\,\w\vec E&=&{4\pi\vec j}\\
\label{eq_maxwellfourier}
\endary
The second of Eq.~\ref{eq_maxwellfourier} implies that all (single) solutions also
fulfill $\vec E\cdot\vec B=0$. The homogeneous parts of Eq.~\ref{eq_maxwellfourier} are 
structurally an exact copy of the expressions that we obtain from Eq.~\ref{eq_eigenfreq} 
with the condition $K_2=0$: 
\begary{rcl}
{\cal E}\,\vec B-\vec P\times\vec E&=&0\\
\vec P\cdot\vec B&=&0\\
\vec E\cdot\vec B&=&0\\
\endary

In Ref.~\cite{geo_paper} we made use of the condition $\vec P\cdot\vec B=0$ 
and $\vec E\cdot\vec B=0$ which have to be fulfilled in order to block-diagonalize 
the matrix ${\bf F}$. But block-diagonalization (``decoupling'') did not require 
the third term ${\cal E}\,\vec B-\vec P\times\vec E$ to vanish.
However, if the corresponding MWEQ would not yield zero, then we had to deal 
with a non-zero but divergence-free magnetic current density ${4\pi\vec j_m}=\d_t\,\vec B+\vec\nabla\times\vec E$ 
- which has not been found to date. There is another more abstract reason,
why we may restrict the solutions to the case of $K_2=0$: Only with this condition
being fulfilled the symplex ${\bf F}$ is a symplectic similarity transformation of 
the pure time direction as described by Eq.~(\ref{eq_gamma0st})~\footnote{
For $K_2>0$ there are two different frequencies while for $K_2=0$ the eigenfrequencies
are identical as in case of $\y_0$.}. 
The third scalar product 
$\vec P\cdot\vec E$ (and the corresponding field equation $\vec\nabla\cdot\vec E$) is in the
general case non-zero and we emphasize that this is another correspondence to the symplectic
decoupling formalism (Ref.~\cite{geo_paper}). It is also remarkable, that the so-called duality 
rotations~\footnote{
Concerning the electro-magnetic duality rotation see for instance the short theoretical review 
article of J.A. Mignaco~\cite{Duality1} and the latest experimental results~\cite{Duality2}.} of the 
electromagnetic field do not fit into this approach~\cite{rdm_paper}: Electric and magnetic
fields can not be exchanged or mixed. Though both are bi-vector fields (forming a ``second rank'' 
tensor), the special role of the time coordinate breaks the suspected symmetry between $\vec E$ 
and $\vec B$. Furthermore the only generator that might be used for such a rotation is the 
pseudo-scalar, i.e. a cosymplex. Insofar the symplectic foundation of electrodynamics has a higher 
explanatory power than the ``conventional'' formalism.

If we summarize (and extend) the stability conditions $K_2\ge 0$ and $K_1\ge 2\,\sqrt{K_2}$ 
(Ref.~\cite{geo_paper}) by the condition $K_2=0$, then it is appropriate (or even mathematically 
inevitable) to use this striking structural similarity and to {\it postulate} the equivalence 
of $\vec k$ and $\vec P$ ($\w$ and ${\cal E}$). That is - we claim that $\vec P\propto\vec k$ 
(${\cal E}\propto\w$) and hence that the momentum and energy equal via the Fourier 
transform the spatial and time derivatives, respectively. There is still an optional proportionality
factor. But for the same reason that we did not refer to the speed of light, we also skip the 
proportionality factor $\hbar$, since it has no physical significance, but depends on the choice 
of units~\cite{Ralston,Hsu,Duff}:
\begary{rcl}
{\cal E}&=&\w=i\,\d_t\\
{\vec P}&=&\vec k=-i\,\vec\nabla\\
\label{eq_QMop}
\endary
This means that the constant $K_2$ in Eq.~\ref{eq_eigenfreq} vanishes for every plane
wave solution since all single terms forming $K_2$ vanish separately:
\begary{rcl}
K_2&=&({\cal E}\,\vec B-\vec P\times\vec E)^2-(\vec E\cdot\vec B)^2-(\vec P\cdot\vec B)^2\\
{\cal E}\,\vec B-\vec P\times\vec E&\to&\w\,\vec B-\vec k\times\vec E=0\\
\vec P\cdot\vec B&\to&\vec k\cdot\vec B=0\\
\vec E\cdot\vec B&=&0\\
\endary
It is easy to show that ``in vacuum'' we also have $\vec E^2-\vec B^2=0$ so
that also $K_1=0$ and there are no eigenfrequencies of electromagnetic waves. 
Furthermore they do not generate inertial frames of reference.

As we projected the bivector elements (fields) of the general symplex ${\bf F}$ into spacetime, 
we need to do the same with the phase space variables $\psi$ and thus obtain the ``wave function'' 
or spinor $\psi(\vec x,t)$. A normalization might be written as
\begeq
\int\,\psi(\vec x,t)^T\psi(\vec x,t)\,d^3x=1\,.
\endeq
The phase space density function $\rho$ (and optionally the Gramian, see Eq.~\ref{eq_normalization}) 
might be ``included'' into the spinor function, such that
\begeq
\tilde\psi=\sqrt{\rho}\,\psi\,.
\endeq
Since the electric current must be a vector, the simplest Ansatz for the 
current density of a ``particle'' described by $\psi$ is~\footnote{
This definition looks quite similar to the definition of the momentum by the EMEQ: 
Mass and charge density are proportional in case of ``point particles''.}
\begary{rcl}
\rho&=&-\bar\psi\,\y_0\,\psi\\
j_x&=&\bar\psi\,\y_1\,\psi\\
j_y&=&\bar\psi\,\y_2\,\psi\\
j_z&=&\bar\psi\,\y_3\,\psi\\
\label{eq_fourcurrent}
\endary
The EQOM of $\psi(\vec x,t)$ must result in a current density (Eq.~\ref{eq_fourcurrent}) that
fulfills the continuity equation. The simplest possible solution is known to be the Dirac equation.
At first sight Eq.~(\ref{eq_Heqom}) does not appear to be similar to the Dirac equation.
But with the help of the Fourier transform, the equivalence becomes obvious. The eigenvalues of ${\bf F}$
for a ``particle'' in a field-free region are given by $\pm\,i\,\sqrt{{\cal E}^2-\vec P^2}=\pm\,i\,m$.
Together with Eq.~(\ref{eq_QMop}) we can now flip the ``account'' of Eq.~(\ref{eq_Heqom}) by the
replacement of operators and eigenvalues (and vice versa):
\begary{rcl}
({\cal E}\,\y_0+P_x\,\y_1+P_y\,\y_2+P_z\,\y_3)\,\psi&=&\dot\psi={d\psi\over d\tau}\\
(i\,\d_t\,\y_0-i\,\d_x\,\y_1-i\,\d_y\,\y_2-i\,\d_z\,\y_3)\,\psi&=&\pm\,i\,m\,\psi\\
(\d_t\,\y_0-\d_x\,\y_1-\d_y\,\y_2-\d_z\,\y_3)\,\psi&=&\pm\,m\,\psi\,.
\label{eq_Dirac}
\endary
Apart from ``missing'' unit imaginary, which is ``hidden'' in the definition of the 
Dirac matrices $\y_\mu$, Eq.~\ref{eq_Dirac} is the Dirac equation. A unitary 
matrix which transforms between RDMs and the conventional Form of the Dirac matrices 
($\tilde\y_\mu$) is explicitly given in App.~\ref{sec_diracconv}. However - unitary 
transformations do not change the signs of the metric tensor. If we desire to have 
$\y_0^2={\bf 1}$, we still need to multiply the transformed matrices by the unit 
imaginary, so that:
\begeq
(-i\,\d_t\,\tilde\y_0+i\,\d_x\,\tilde\y_1+i\,\d_y\,\tilde\y_2+i\,\d_z\,\tilde\y_3)\,\psi=\pm\,m\,\psi\,.
\label{eq_DiracConv}
\endeq

\subsection{Lightlike Spinors and massive Multispinors}
\label{sec_mass}

So far we described spinors as a single list of fundamental variables of the 
form $\psi=(q_1,p_1,q_2,p_2)^T$ i.e. as a 4-dim. column vector and the matrix 
of second moments $\Sigma$ as $\Sigma=\psi\,\psi^T$ or explicitly 
$\Sigma_{ij}=\psi_i\,\psi_j$.
In this case the matrix $\Sigma$ as well as the matrix ${\bf
  S}=\Sigma\,\y_0$ have a vanishing determinant and the eigenfrequencies
(\ref{eq_eigenfreq}) are equally zero. The matrix ${\bf S}$ is a symplex
with the RDM-coefficients being second order monomials of the $\Gamma$ space 
coordinate $\psi$. In order to apply the EMEQ {\it and} to distinguish these 
productions from the ``external'' forces ${\bf F}$, we use different names 
(i.e. lower case letters and ${\cal U}$ instead of ${\cal E}$~\cite{Dirac}:
\begary{rcl}
{\cal U}&\propto&\frac{1}{2}\,(q_1^2+p_1^2+q_2^2+p_2^2)\\
p_x&\propto&\frac{1}{2}\,(-q_1^2+p_1^2+q_2^2-p_2^2)\\
p_y&\propto&(q_1\,q_2-p_1\,p_2)\\
p_z&\propto&(q_1\,p_1+q_2\,p_2)\\
e_x&\propto&(q_1\,q_2-p_1\,p_2)\\
e_y&\propto&(-q_1\,p_2-p_1\,q_2)\\
e_z&\propto&\frac{1}{2}\,( q_1^2-p_1^2+q_2^2-p_2^2)\\
b_x&\propto&(q_1\,q_2+p_1\,p_2)\\
b_y&\propto&\frac{1}{2}\,( q_1^2+p_1^2-q_2^2-p_2^2)\\
b_z&\propto&(p_2\,q_1-p_1\,q_2)\\
\label{eq_monomials2}
\endary
For dimensional reasons - Eqs.~\ref{eq_monomials2} represent {\it actions} -
we stay with proportionality. Evidently Eqs.~\ref{eq_monomials2} defines a bilinear 
mapping $\mathbb{R}^4\to\mathbb{R}^{10}$, i.e. for every single ``point'' in 4-D 
$\Gamma$ space there is a 3+1-dimensional structure including electromagnetic fields.
This mapping has the following general structural properties with respect to
every single phase space point:
\begary{rcl}
\vec p^2&=&\vec e^2=\vec b^2={\cal U}^2\\
0&=&\vec e^2-\vec b^2\\
{\cal U}^2&=&\frac{1}{2}\,(\vec e^2+\vec b^2)\\
{\cal U}\,\vec p&=&\vec e\times\vec b\\
{\cal U}^3&=&\vec p\cdot(\vec e\times\vec b)\\
m^2&\propto&{\cal U}^2-\vec p^2=0\\
\vec p\cdot\vec e&=&\vec e\cdot\vec b=\vec p\cdot\vec b=0\\
\label{eq_4thorder}
\endary
Since we made no other assumptions about the spinor, it follows
that single spinors are lightlike, i.e. have mass zero and the
vectors $\vec e$, $\vec b$ and $\vec p$ form a trihedron.

Multiple spinors are required to form the representation of a massive particle, 
more precisely: A massive object is composed of $\nu\ge 4$ linearly independent 
spinors. This raises questions about the 4-dimensional $\Gamma$-space - the
``phase space'' or ``spinor space'', for instance: How do we construct and represent 
these spinors? In a ``classical`` phase space we have two general methods to represent
an ensemble: The first is given by a classical density distribution $\rho(\psi)$ which describes 
the number of phase space points per unit volume and the second is sampling, i.e. the 
representation of the space space by $\nu$ ``samples'' that could for instance be 
represented by the columns of a $4\times \nu$-matrix $\Psi$. 
The single spinors are the columns of this matrix and the $\Sigma$-matrix is given 
by~\footnote{The root of the normalization factor 
$\frac{1}{\nu}$ could equally well be included into the definition of the spinor.}
\begeq
\Sigma=\frac{1}{\nu}\,\Psi\,\Psi^T\,.
\endeq 
Now consider that the density distribution $\rho(\psi)$ has an internal symmetry.
In the simplest case it might have a point-symmetric of the form:
\begeq
\rho(\psi)=\rho(-\psi)
\endeq
In this case we might replace $\rho$ by
\begeq
\rho\to\frac{1}{2}(\rho(\psi)+\rho(-\psi))\,.
\endeq
More generally, if the symmetry properties of the 4-dimensional phase space
can be expressed by a matrix $\eta$, then:
\begeq
\rho\to\frac{1}{2}(\rho(\psi)+\rho(\eta\,\psi))\,,
\endeq
or - if there are $N$ symmetries:
\begeq
\rho\to\frac{1}{N}\,\sum_k\,(\rho(\psi)+\rho(\eta_k\,\psi))\,.
\endeq
Instead of plugging the symmetry into the density, it could also be
expressed by the spinor, i.e. by replacing the single spinor $\psi$
by a sampling matrix of spinors:
\begeq
\psi\to (\psi,\eta_1\psi,\dots,\eta_N\,\psi)\,.
\endeq
In this form, it is still possible to multiply the matrix of second moments
by the density without any restriction to $\psi$. This is a mixed form that
allows to represent the phase space symmetry by the spinor and still write 
spinor and density as a product.
In this case we compute the matrix of second moments according to
\begeq
\Sigma=\int\,d\psi^{2n}\,\rho(\psi)\,\Psi\,\Psi^T\,.
\endeq
The expectation value of an operator ${\bf O}$ is then expressed by
\begeq
\langle{\bf O}\rangle=\int\,d\psi^{2n}\,\rho(\psi)\,\mathrm{Tr}(\Psi^T\,\Psi)\,.
\endeq

The highest degree of symmetry is given, if all single spinors 
(column vectors) that form the multispinor $\Psi$ are pairwise orthogonal. 
In this case the $\Sigma$-matrix is proportional to a unit matrix
\begeq
\Psi\,\Psi^T=(q_1^2+p_1^2+q_2^2+p_2^2)\,{\bf 1}\,,
\label{eq_sigortho}
\endeq 
and describes - according to the EMEQ - a massive particle in its rest frame.
If the single spinors of $\Psi$ (i.e. the columns of $\Psi$) - are $\psi_k$, 
then such a system can be constructed from a single spinor $\psi$ by the use of
linear orthogonal operators ${\bf R}_k$ according to
\begary{rcl}
\psi_k&=&{\bf R}_k\,\psi\\
\psi_j\cdot\psi_k&=&\psi_j^T\,\psi_k\\
&=&\psi^T\,{\bf R}_j^T\,{\bf R}_k\,\psi=0\,.
\endary
The orthogonality then requires that the product ${\bf R}_j^T\,{\bf R}_k$ must
be skew-symmetric:
\begary{rcl}
({\bf R}_j^T\,{\bf R}_k)^T&=&{\bf R}_k^T\,{\bf R}_j=-{\bf R}_j^T\,{\bf R}_k\\
\Rightarrow&&{\bf R}_k^T\,{\bf R}_j+{\bf R}_j^T\,{\bf R}_k=0\,.
\endary
Now it is known, that the $\y$-matrices are all orthogonal. If we replace
${\bf R}_k$ by $\y_k$, then we have the following condition:
\begeq
\y_k^T\,\y_j=-\y_j^T\,\y_k\,.
\endeq
If both operators $\y_j$ and $\y_k$ are symmetric or both skewsymmetric, then
the condition is equivalent to the requirement that they anticommute:
\begeq
\y_k\,\y_j+\y_j\,\y_k=0\,.
\endeq
If one is symmetric and the other one is skew-symmetric, then they have to commute:
\begeq
\y_k\,\y_j-\y_j\,\y_k=0\,.
\endeq
What is remarkable here is the fact that orthogonal operators {\it which are independent
of $\psi$} exist only in even-dimensional vector spaces. In odd dimensional spaces
such operators are impossible: Though it is always possible to define a linear
orthogonal operator ${\bf R}$ such that $({\bf R}\,{\bf x})\cdot{\bf x}=0$, 
only in even dimensional spaces the rotational operators ${\bf R}$ can be defined
in such a way that they are independent of ${\bf x}$, i.e. which rotate 
{\it any} spinor in such a way that $\psi^T\,{\bf R}\,\psi=0$. One such
example is $\y_0$. 

Without loss of generality we can restrict the first operator to equal the identity matrix 
${\bf R}_0={\bf 1}$. Then all ${\bf R}_k,\,\,k\in\,[1,2,\dots,2\,n-1]$ have to be 
skew-symmetric to make the product $\psi\,\psi_k=\psi\,{\bf R}_k\,\psi$ ($k=1,2,3$) 
equal to zero. In this case the product $\psi_j^T\,\psi_k=\psi^T\,{\bf R}_j^T\,{\bf R}_k\,\psi$
for $j\neq k$ must vanish and therefore the product of the two matrices must
also be skew-symmetric. In case of a $2\,n=4$-dimensional spinor phase space,
we can select some of the 15 traceless RDMs for the ${\bf R}_k$. In this case one 
needs to select 3 out of 6 skew-symmetric matrices. The 3 selected skew-symmetric 
matrices must pairwise anticommute.
There are (up to an orthogonal transformation and up to a signed permutation) 
only two sets that fulfill this conditions:
\begary{rcl}
\Psi_s&=&\frac{1}{2}\,(\psi,\y_7\,\psi,\y_{8}\,\psi,\y_{9}\,\psi)\\
\Psi_c&=&\frac{1}{2}\,(\psi,\y_0\,\psi,\y_{10}\,\psi,\y_{14}\,\psi)\,.
\label{eq_er_maps}
\endary
The two multispinors are then formally orthogonal matrices, i.e.
\begary{rcl}
\Sigma&=&\Psi_{s,c}\,\Psi_{s,c}^T=(q_1^2+p_1^2+q_2^2+p_2^2)\,{\bf 1}\\
{\bf S}&=&(q_1^2+p_1^2+q_2^2+p_2^2)\,\y_0\\
\endary
They can also be interpreted as geometric objects, i.e. as representations
of a ``phase space unit cell'' and/or as a certain symmetry pattern of a 
phase space distribution~\footnote{
Note that the 4 matrices ${\bf 1}, \y_7, \y_8, y_9$ as well as ${\bf 1}, \y_0, \y_{10}, \y_{14}$
are representations of the quaternions ${\bf 1,i,j,k}$.}. However the
possibility for such a $\psi$-independent definition of an orthogonal multispinor 
does not work in arbitrary even dimensional spaces: The next considered Clifford-algebra 
$Cl_{9,1}$ has a spinor of size $2\,n=32$. Hence we would need $31$ orthogonal
skew-symmetric and pairwise anticommuting matrices. If such a system of matrices
would exist, then it would be the basis of a $31$-dimensional Clifford 
algebra, which is obviously impossible. Therefore in 9+1 and higher dimensions
we cannot construct an orthogonal massive multispinor in the described way
simply by phase space symmetries. Neither a Hurwitz- nor a
Kustaanheimo-Stiefel transformation (see App.~\ref{sec_hurwitz}) exist for $32$ or more
variables. And therefore the wave function can not be written in a general way as a
product of density and $32$-component spinor: Only in $2$ and $4$ dimensions
the spinor space is separable from the phase space in the way described above.  
Hence the condition expressed by Eq.~\ref{eq_equaldim} excludes all but the 
$4$-dimensional phase space to be the simplest possible basis of ``objects'' of a 
physical world. 
If our arguments are cogent, then the corresponding {\it space of 
observables of any physical world} must have $3+1$ dimensions. 

We argued that the multispinor approach forces the phase space density $\rho(\psi)$ 
to have a specific symmetry. Then we should also have a method to suppress a
symmetry. This is technically difficult with a positive definite density $\rho$.
If the (multi-) spinors are multiplied with $\sqrt{\rho}$, then the option 
to have a negative $\sqrt{\rho}$ is mathematically not easily expressible. 
It is preferable to write the density as the square of a function $\phi$:
\begeq
\rho=\phi^2(\psi)\,,
\endeq
such that it is possible to express asymmetry without the use of negative
densities. If we then write
\begary{rcl}
\tilde\phi&=&\frac{1}{2}\,(\phi(\psi)+\phi({\bf X}\,\psi))\\
\rho&=&\tilde\phi^2=\frac{1}{4}\,(\phi^2(\psi)+\phi^2({\bf X}\,\psi)+2\,\phi(\psi)\phi({\bf X}\,\psi))\,.
\endary
then the density vanishes, where $\phi(\psi)=-\phi({\bf X}\,\psi)$, i.e.
where $\phi$ is purely skew-symmetric with respect to the transformation ${\bf X}$.

\subsection{Deformation of the Phase Space Unit Cell}
\label{sec_phasespace}

Now consider a deformed phase space ellipsoid. We give weights to
the phase space lattice as defined by the multispinors in Eq.~\ref{eq_er_maps}
\begary{rcl}
\Psi_s&=&(a\,\psi,b\,\y_7\,\psi,c\,\y_{8}\,\psi,d\,\y_{9}\,\psi)\\
\Psi_c&=&(a\,\psi,b\,\y_0\,\psi,c\,\y_{10}\,\psi,d\,\y_{14}\,\psi)\,,
\endary
Then the corresponding symplex $\Psi_c\Psi_c^T\y_0$ has the following form:
\begary{rcl}
\Psi_{c,s}\Psi_{c,s}^T\y_0&=&\psi^T\,\psi\,\mathrm{Diag}(a^2,b^2,c^2,d^2)\,\y_0\\
&=&\psi^T\,\psi\,\left((a^2+b^2+c^2+d^2)\,{\cal E}\right.\\
&+&(-a^2+b^2+c^2-d^2)\,P_x\\
&+&(a^2-b^2+c^2-d^2)\,E_z\\
&+&\left.(a^2+b^2-c^2-d^2)\,B_y)\right)/4\\
\endary
This corresponds to a decoupled oscillator matrix as described in
Ref.~\cite{geo_paper}. Any symplex with purely imaginary eigenvalues
is symplectically similar to this type of symplex, which represents a deformed 
phase space lattice. Note that the energy is represented by the squared diagonal 
of the phase space lattice while the mass is represented by a function
of the volume. In the next section we show how the phase space lattice is
related to CPT-Transformations.

\subsection{CPT-Transformations}
\label{sec_cpt}

The use of multispinors imprints a certain phase space symmetry. But this 
phase space symmetry is correlated with (or can be mapped to) specific 
``real world'' symmetries.

If we consider the difference between the multispinors
$\Psi_s$ and $\Psi_c$. The first thing we may note is that all single spinors
(columns) of $\Psi_s$ are constructed by symplectic transformations. In a
sufficiently ergodic system, they could - at least in principle be interpreted 
as a single phase space trajectory at different times, while $\Psi_c$ is
composed of (partially) disjunct components. It is one of the remarkable
features of the 4-dim. phase space that there are these disjunct areas.
A system with the phase space position $\psi$ at $t=0$ can by no means
be moved symplectically to position $\y_{14}\,\psi$ and vice versa.
The 4-dim. phase space is split into two symplectically disjunct regions.

In this context we have to take the effect of certain RDMs into account, which 
are used to model charge conjugation ($\y_{14}$), parity conjugation ($\y_0$) 
and time reversal $\y_{10}$:
\begary{rcl}
{\bf F}&=&{\cal E}\,\y_0+\vec P\,\cdot\,\vec\y+\y_0\,\vec E\,\cdot\,\vec\y+\y_{14}\,\y_0\,\vec B\,\cdot\,\vec\y\\
-\y_{0}\,{\bf F}\,\y_{0}&=&{\cal E}\,\y_0-\vec P\,\cdot\,\vec\y-\y_0\,\vec E\,\cdot\,\vec\y+\y_{14}\,\y_0\,\vec B\,\cdot\,\vec\y\\
\y_{10}\,{\bf F}\,\y_{10}&=&{\cal E}\,\y_0-\vec P\,\cdot\,\vec\y+\y_0\,\vec E\,\cdot\,\vec\y-\y_{14}\,\y_0\,\vec B\,\cdot\,\vec\y\\
\y_{14}\,{\bf F}\,\y_{14}&=&{\cal E}\,\y_0+\vec P\,\cdot\,\vec\y-\y_0\,\vec E\,\cdot\,\vec\y-\y_{14}\,\y_0\,\vec B\,\cdot\,\vec\y\\
\label{eq_cpt}
\endary 

The expectation values of an operator ${\bf F}$ using the multispinor $\Psi_c$ 
can also be written as
\begary{rcl}
\langle {\bf F}\rangle&=&\Psi_c^T\,\y_0\,{\bf F}\,\Psi_c={1\over 4}\,\sum\limits_{k=0}^3\,\psi_k^T\,\y_0\,{\bf F}\,\psi_k\\
&=&{1\over 4}\,(\psi^T\,\y_0\,{\bf F}\,\psi+\psi^T\,\y_0^T\,\y_0\,{\bf F}\,\y_0\,\psi\\
&+&\psi^T\,\y_{10}^T\,\y_0\,{\bf F}\,\y_{10}\,\psi+\psi^T\,\y_{14}^T\,\y_0\,{\bf F}\,\y_{14}\,\psi)\\
&=&{1\over 4}\,(\psi^T\,\y_0\,{\bf F}\,\psi-\psi^T\,\y_0\,\y_0\,{\bf F}\,\y_0\,\psi\\
&+&\psi^T\,\y_0\,\y_{10}\,{\bf F}\,\y_{10}\,\psi+\psi^T\,\y_0\,\y_{14}\,{\bf F}\,\y_{14}\,\psi)\\
&=&{1\over 4}\langle {\bf F}-\y_0\,{\bf F}\,\y_0+\y_{10}\,{\bf F}\,\y_{10}+\y_{14}\,{\bf F}\,\y_{14}\rangle
\endary
If we compare this to Eq.~\ref{eq_cpt}, then the result is clearly equivalent to a projection
resulting in $\langle {\bf F}\rangle={\cal E}$.

\subsection{Duality between Operators and Observables}
\label{sec_ops_obs}

It is well-known that observables have a dual role in both classical as
well as in quantum mechanics. Eq.~\ref{eq_env} describes the change of the
second moments of an ensemble of spinors due to the force matrix ${\bf F}$. 
In classical point mechanics we would have written ``external'' forces, since
self-interaction is almost always divergent in theories that assume space-time 
to be fundamental. In our approach the situation is different and in order to 
account for the dynamics of fundamental variables, self-interaction should be included.
This means that the force matrix ${\bf F}$ is the sum of self-forces ${\bf F}_{s}$ 
and ``external'' forces ${\bf F}_{x}$.
If no external forces are present, the spinor ensemble nevertheless ``oscillates'' 
with the frequency determined by the eigenvalues of ${\bf F}_{s}$. Since both, 
the self-force and the ${\bf S}$-matrix of the system are symplices, it is natural
to write the functional dependence of the self-force as a matrix-function of 
${\bf S}$: 
\begeq
{\bf F}_{s}=f({\bf S})\,.
\endeq
Since only odd powers of a symplex are again a symplex, the Taylor series of $f$ 
may contain odd powers only. It is obvious that the self-force and the 
${\bf S}$-matrix will commute so that the ${\bf S}$-matrix either depends 
explicitly on time (which implies external influences) or it can assumed to be 
static:
\begeq
\dot{\bf S}={\bf F}_s\,{\bf S}-{\bf S}\,{\bf F}_s=0\,.
\endeq
Without ``external'' influences the matrix of second moments remains static 
though the spinor itself oscillates - in accordance we the onto-logic of time.
Insofar we have to take care when interpreting Eq.~\ref{eq_lorentzforce}. 
``Classically'' the force matrix ${\bf F}$ contains only the external forces. 
But here we use the sum of external and the self-forces. In the absence of 
external interactions the oscillation is exclusively due to self-interaction.

\section{Anything Else?}
\label{sec_outlook}

\subsection{Once More: Cosymplices}
\label{sec_final1}

Much of what has been derived above is a consequence of the distinction between
symplices and cosymplices (or ``Hamiltonian'' and ``Skew-Hamiltonian'') matrices
and their expectation values. One central argument was that the generators of
the considered Clifford algebras have to be symplices {\it since only simplices
have non-vanishing expectation values}. Only symplices are generators of symplectic
transformations and hence only symplices represent forces with measurable effects.
One might argue that some entity might exist though its expectation value 
and its obvious consequences remain zero. Maybe it has hidden or indirect effects?
For instance the algebra of $2\,n\times 2\,n=8\times 8$ (co-) symplices represents 
a Clifford algebra $Cl_{3,3}$ that has - compared to the Dirac algebra - two additional 
cosymplices as time-like generators. Let the generators $\zeta_\mu$ with 
$\mu\in\,[0,\dots, 5]$ (with the real Pauli matrices $\eta_\nu$ defined in 
App.~(\ref{sec_rpm}) be defined by:
\begary{rcl}
\zeta_0&=&\eta_3\otimes\eta_3\otimes\eta_0\\
\zeta_1&=&\eta_3\otimes\eta_3\otimes\eta_1\\
\zeta_2&=&\eta_3\otimes\eta_1\otimes\eta_2\\
\zeta_3&=&\eta_3\otimes\eta_2\otimes\eta_2\\
\zeta_4&=&\eta_1\otimes\eta_0\otimes\eta_2\\
\zeta_5&=&\eta_2\otimes\eta_0\otimes\eta_2\,.
\endary
The first four elements correspond to (a variant of) the Dirac algebra,
and the last two represent ``hidden'' dimensions corresponding to additional 
energy contributions $\eps_1$ and $\eps_2$, but are cosymplices, if $\zeta_0$ 
is the symplectic unit matrix.
All $\zeta_\mu$ are pairwise anticommuting and the metric tensor
is given by $g_{\mu\nu}=\textrm{diag}(-1,1,1,1,-1,-1)$.
The commutator table of the first four generators then is (as in case
of the Dirac matrices) the electromagnetic field tensor~\cite{rdm_paper}. But since
(according to Eq.~\ref{eq_cosy_algebra}) the commutator of a symplex
and a cosymplex is a cosymplex, the mixed terms between symplices and
cosymplices have vanishing expectation values:
\begeq
\langle[\zeta_\mu,\zeta_\nu]\rangle=\bmtx{cccccc}
0   &E_x&E_y  & E_z &0&0\\
-E_x&0  &B_x  &-B_y &0&0\\
-E_y&-B_x& 0  &B_z  &0&0\\
-E_z& B_x&-B_y& 0   &0&0\\
0&0&0&0&0&\lambda\\
0&0&0&0&-\lambda&0\\
\emtx
\endeq
Hence the two additional elements are (in average) ``decoupled'' 
from the first four (and all other symplices). The commutator of the two 
extra cosymplices is a symplex and hence yields a non-zero field value $\lambda$, 
which acts only between the two additional time-like dimensions. 
Considerations like these are the background for the claim that time might indeed
be ``multidimensional'' - but it remains to be shown that we would be able to notice.

\subsection{Higher-Dimensional Spaces}
\label{sec_final2}

If spacetime is an emergent phenomenon, then the question arises
whether other (high-dimensional) spacetimes may not emerge ``in parallel''.
We have given a number of algebraic arguments why a $3+1$ dimensional space-time
is a very special case which can not be replaced easily by other dimensions.
Specifically we believe to have shown that it is not so much a question of
the properties of a ``real'' external world (which is ``independent'' of the observer)
which we experience and investigate. We should always keep in mind that our 
argumentation is founded simply on the possibility in principle of observation.
If our argumentation is sound, then the possibility in principle for the existence
of completely different universes with different dimensionality and different laws 
of nature is much smaller then considered elsewhere~\cite{mathworld}.

But let's ignore the above given arguments for $3+1$ dimensions for a moment: 
Indeed the strongest evidence for a 3+1-dimensional spacetime that we have 
at hand, is based on light and electrodynamics. Chemistry is almost exclusively 
based on electromagnetic interactions of electrons: we experience the world through 
electronic interaction. Our world is the world of electrons. What we see, is light 
emitted, reflected (or absorbed) by electrons. If we touch a solid object, then 
electrons are ``touching'' electrons. We can experimentally investigate weak and 
strong forces, gravitation and so on - but whatever we ``see'' with our own eyes, 
is to almost any degree of approximation based of the electromagnetic interaction 
of electrons. 

Assume that a $9+1$-dimensional systems exist in parallel {\it by
the properties of other (high-dimensional) fields} - then it still remains 
questionable, if and how observers would interpret these $9+1$-dimensional 
objects. Certainly we can not expect that $9+1$-dimensional entities are just
``3 times of the same'', since the algebraic features (for instance of rotations)
are different: An essential feature of 3 dimensions is that the spatial 
rotations do not commute. In more than 3 dimensions we can form commuting
rotators, say $\y_1\,\y_2$ and $\y_3\,\y_4$. Hence in 9 dimensions we
sort out 3 sets of 3 non-commuting rotators. Consider we use
\begary{rlll}
1) & \y_2\,\y_3,&\y_3\,\y_1,&\y_1\,\y_2\\
2) & \y_5\,\y_6,&\y_6\,\y_4,&\y_4\,\y_5\\
3) & \y_8\,\y_9,&\y_9\,\y_7,&\y_7\,\y_8\\
\endary
We then find that any two rotators from different groups commute~\footnote{
Dynamically emerging spacetimes have their own properties, which are
determined by the structure of the corresponding symplectic Clifford
algebra. One can not simply add another dimension as in arbitrary 
dimensional Euclidean spaces. Insofar {\it physical} spaces are much
more restricted than mathematical spaces.}.
On the level of observables we derive from Eq.~\ref{eq_eqom_meas} that
{\it if two observables commute they are decoupled}.
Presumably observers that are socialized in their perception to adapt to a 
3+1 dimensional world would not interprete the $9+1$ dimensional entity 
correctly. Instead of ``seeing'' a $9+1$-dimensional object ``parallel'' 
to our ``electronic'' $3+1$-dimensional spacetime, an observer might see 
$3$ objects located in $3+1$ dimensions. The ontologically motivated idea 
that our spacetime dimensionality must be a fundamental and a unique property 
of the world (and may not depend on the type of the interaction), might have 
guided physics into the wrong direction after all~\footnote{
There are however suggestions how to derive particle physics and even
dark matter from a {\it triplet algebra} of dimension $2^{12}=4096$ based
on a $64$-dimensional phase space~\cite{Sogami}.}.

Tab.~\ref{tab_obs} summarizes the theoretical number of (co-) symplices
that are to be expected in dynamically generated spacetimes based on
Clifford algebras. Obviously all algebraically possible spacetimes 
beyond $3+1$ dimensions include high-order symplices, i.e. penta-, 
hexa- and deca-vectors. However we would like to add a remark here:
{\it If} one would argue that only vectors and bi-vectors have non-negligible
effects, then we find that the $N$-dimensional symplex-algebra $Cl_{N-1,1}$
is composed of $N$ vectors, $N-1$ boosts and $(N-1)(N-2)/2$ rotations.
With $N=2\,n$ this sums up to an effective number of symplices $\nu_s^{eff}$ given by
\begary{rcl}
\nu_s^{eff}&=&N+N-1+{(N-1)(N-2)\over 2}={N\,(N+1)\over 2}\\
             &=&{2\,n\,(2\,n+1)\over 2}\,,
\endary
just as if the phase space dimension would be $N=2\,n$ and not $2^N$.
Since only the generators are observables, one might find arguments
for an effective dimensional reduction even of higher-dimensional
dynamical systems. The {\it effective} phase space dimension $n_{eff}=2\,n$ 
is then given by~\cite{rdm_paper}:
\begeq
n_{eff}=\sqrt{2\,\nu_s^{eff}+\frac{1}{4}}-\frac{1}{2}\,.
\endeq

\section{Summary and Conclusion}
\label{sec_summary}

Based on the identification of time with change, we introduced fundamental 
variables, which are defined exclusively by the property of variation in time. 
Measurability requires reference to invariant constant rulers. Since pure
constants are by definition not available on a fundamental level they have 
to emerge from dynamics in the form of constants of motion. We have shown
that this the simplest mathematical model that allows us to do so is given
by the Hamiltonian formalism. We found that the ostensible necessity of the 
use of complex numbers in quantum mechanics can be translated into the 
``classical'' finding that dynamical variables come as pairs.
The Hamiltonian ansatz resulted in an algebra of symplices and cosymplies. 
Reasonable assumptions as for instance a proper distinction 
of measurable and unmeasurable quantities guided us to Lie algebras that are
isomorphic to Clifford algebras $Cl(N-1,1)$. The most fundamental of these 
algebras is the Dirac algebra represented by real matrices.

Looking back we can identify pure variables with the components of the Dirac 
spinor, i.e. with the components of the quantum mechanical wavefunction. 
If we include a density function, then the spinors can be interpreted as
``probability amplitudes''. We argued that the separation into a multispinor
and a pure phase space density function only works, if $2^N=(2N)^2$ holds,
i.e. if the dimension of the Clifford algebra and spinor algebra are the same.
This was the last missing argument to show that Hamiltonian dynamics has
special properties in 4-dimensional phase spaces.

One goal of our gedankenexperiment was to demonstrate the emergence of 
a Minkowski geometry (i.e. spacetime geometry) through symplectic dynamics.
Spacetime can indeed be a mere interpretation of dynamics if the Lie algebraic 
construction of geometry based on symplices is isomorphic to the representation
of a Clifford algebra. Most (if not all) arguments that we used in the
derivation are related to specific (algebraic) symmetries.
Apparently these symmetries are less obvious in the conventional form of
the Dirac matrices. Furthermore the use of complex numbers in the 
conventional form of the Dirac theory wrongly suggests that we are in a 
non-classical domain. We have shown that some of the apparent differences 
between quantum and classical mechanics are of ontological nature and loose
much of their significance in the light of the ontology of existence in time.
Non-classicality is in this scheme neither connected to $\hbar$ or the unit
imaginary but to the fact that all quantities of the fundamental level vary 
at all times and to the consequences thereof.

The idea that the elements of Clifford algebras are related to Minkowski
spacetime is well-known and has been described by D. Hestenes and others in 
various publications on {\it spacetime algebra}~\cite{Hestenes,STA}: 
``The Dirac matrices are no more and no less than matrix representations of 
an orthonormal frame of spacetime vectors and thereby they characterize 
spacetime geometry''. 
However (to our knowledge) Hestenes never discussed {\it why} Minkowski spacetime 
should have these properties. According to our interpretation Clifford algebras 
are the optimal mathematical representation of spacetime {\it because}
spacetime emerges by a pattern of special symmetry properties which can be 
expressed by the isomorphism of the Hamiltonian dynamical structure to 
Clifford algebras. The methods of GA are fascinating, but the formal elegance of
GA remains unexplained unless we understand that both space-time and
electromagnetism emerge from the structure of the Dirac algebra. 
It is this connection that can explain why GA is the (only) appropriate 
mathematical tool to represent spacetime~\cite{RV}. But this connection
unfolds its full explanatory power only in combination with fact that
the Dirac algebra is also the algebra of symplectic coupling.
This can be shown in the simplest and most obvious way by the use of the 
real Dirac algebra.

We also claimed that these insights are arguments for the (apparent) 
dimensionality of spacetime. This might appear unacceptable to physicists
who are committed to a realistic ontology. However, already the question
{\it why} space-time {\it should} be 3+1 dimensional, implies the 
{\it possibility in principle} of a mechanism which could explain
the dimensionality of space-time. If this question is accepted as legitimate, 
then it appears to the author that the only logically possible answer 
must be related to the structure of interaction as suggested by Einstein. 
We have given a mathematically simple and sound quasi-classical explanation.
Our derivation of relativity does not require the principles of relativity,
neither the constancy of the speed of light nor the principle that the
laws of physics must be the same in all inertial frames. The presented
approach allows to {\it derive} both principles from the conventional 
classical concept of canonical transformations. The latter is a result
of the fact that Lorentz transformations are structure preserving 
(i.e. symplectic) transformations of the spinor part of the wavefunction.
The former is a result of our derivation of Maxwell's equations. 
3 central arguments were given in preparation of the derivation of the 
Lorentz force and electrodynamics in form of MWEQs: The first argument is 
based on the difference between even and odd elements of the algebra, the 
second on the fact that the expectation values of cosymplices must vanish 
in symplectic dynamics for symmetry reasons and the last one on the 
transformation properties. The wave equations follow from MWEQs and the 
comparison of the Fourier transformed MWEQs with the structure of decoupling 
then lead us to the identification of momentum (energy) with space (time) 
derivatives and hence to the Dirac equation. 

It is often claimed that the way in which probabilities appear in quantum
mechanics is special and unusual in some unspecified way. This alienation
is often related to the ``complex probability amplitudes''. But if we reinterpret 
the formalism of quantum mechanics by its close relation to the matrix of 
second moments, then its form and postulates are quite familiar. 

We do not claim that our ``derivation'' is rigorous in a strict mathematical
sense, nor that it is complete. And it can not be: our abstract ansatz may
only include {\it abstract objects} and becomes a physical theory only with an 
appropriate interpretation. Interpretations can not be proven, they can only
be adequate and consistent or not.

Our central ``interpretation'' is the EMEQ which is induced by the isomorphism
of certain classical quantities (energy, momentum, electric and magnetic field
components) with the abstract observables of the 3+1 dimensional Clifford algebra.
The fact that we can derive the Lorentz force equations from it, is neither a physical
nor a mathematical ``proof'' of this interpretation, but it is a proof that our
interpretation is compatible with this fundamental dynamical law.
It appears to be remarkable to the author that a single 3-dimensional orbit is 
mathematically isomorphic to the average envelope of a 2-dimensional ensemble. 
In this sense we join the view of Mark Van Raamsdonk:
``Everything around us - the whole three-dimensional physical world - is an illusion 
born from information encoded elsewhere, on a two-dimensional chip''~\cite{MvR}.
Most authors that analyzed the structure of the symplectic group $Sp(4,R)$ interpreted 
this space to be 2+1 dimensional. However - as we argued above - the 
space of {\it observables} that corresponds to $Sp(4,R)$ is a $3+1$ dimensional 
Minkowski spacetime. Mathematically it is evident that this view implies 
restrictions on the $3+1$ dimensional space - for instance the 
``quantization'' of the angular momentum. 

But even if we could present a more rigorous derivation of the (fundamental) 
relativistic equation of motion of quantum mechanics - the Dirac equation -
it would not automatically imply a ``derivation'' of quantum mechanics as a whole.
Though it is rarely explicitly mentioned, it is known (though not well-known)
that the equations of motion used in quantum mechanics are (taken as such)
classical~\cite{RalstonQM}. With respect to this question, our presentation
is not new. What is new (to the knowledge of the author) is the algebraic
connection between the classical (co-)symplex algebra with the Clifford
algebra of the Dirac matrices. 

We started with the classical Hamilton function, and were guided to 
interpret the fundamental variables as the quantum mechanical wave function in 
momentum space. Mathematically the alleged antagonism between classical and 
quantum mechanics seems to evaporate in the light of what we described above.
At least we could show that it is not the {\it mathematical structure} of the 
equations of motion that accounts for the interpretational difficulties with
quantum mechanics.

On the basis of the presented ontology of time we found an explanation for why the 
quantum mechanical wavefunction has some features that seem to be {\it
  mysterious} or ``non-classical'': Impossible to be directly measured and
without a well-defined dimensional unit and physical meaning. However these
are exactly the properties of fundamental variables that we derived from the 
ontology of time and a proper logic of measurement. 
Supposing our ontology is ``correct'', it is impossible to know {\it what} the
wave function {\it is}. We think that we gave an explanation why this is so: 
The question has no meaningful physical answer - it can not. The consequences 
of the presented ontology are certainly difficult to accept: On the fundamental 
level, the world is apparently very different from our every day experience: 
loosely speaking it is ``flat''. The apparent 3+1 dimensional spacetime is 
``in reality'' a four-dimensional phase space. However there are interpretations 
of quantum mechanics (and general relativity) which are way more esoteric than
this. 

The picture that we painted is a generalized, abstracted review of 
algebraic methods for a low-dimensional $\Gamma$-phase-space of classical 
statistical mechanics based on arbitrary continuously varying fundamental
variables. This is a generalization and abstraction because we did not use 
a-priori assumptions on the number of space-phase points making up a ``something'', 
a piece of matter, nor did we make any a-priori assumptions about the type of
interaction other than arguments based on symmetry considerations.
It is the onto-logic of time that guided us to the idea that observables are
moments of fundamental variables and the phase space coordinates themselves
are not directly accessible. Here we discussed almost exclusively the second
moments. However it might be worth considering the role of 4-th order
and higher moments in more detail for both, the static case of ``quantum
systems'' in their eigenstates as well as the potential role that odd moments
might play in quantum jumps. There are strong arguments to assume that 
quantization cannot be finally understood as a linear theory. It might be 
enlightening to investigate the physical and algebraic meaning of higher order 
moments. We only shortly touched this area with Eq.~\ref{eq_monomials2} and 
Eq.~\ref{eq_4thorder} with respect to a single phase space points.

Following this logic we found strong arguments for the dimensionality of 
energy-momentum-space. Since the energy-momentum space is conjugate to 
space-time, these arguments directly concern the dimensionality of the latter.
Along the way we found an interpretation that describes fundamental laws of relativistic 
electrodynamics and quantum mechanics. We described the bridge leading to
covariance matrices and statistical moments, i.e. the bridge to probability
theory. Significant theories have been built on less solid grounds. Maybe the above
is also a contribution to the question of whether, and in what sense, quantum mechanics 
deserves to be regarded as the final theoretical framework. 

\begin{acknowledgments}
Mathematica\textsuperscript{\textregistered} has been used for part of the
symbolic calculations. Additional software has been written in ``C'' and compiled 
with the GNU\textsuperscript{\copyright}-C++ compilers on different Linux distributions.
XFig 3.2.4 has been used to generate the figures, different versions of \LaTeX and
GNU\textsuperscript{\copyright}-emacs for editing and layout.
\end{acknowledgments}

\begin{appendix}

\section{Hurwitz and Kustaanheimo-Stiefel Matrices}
\label{sec_hurwitz}

With a minor modification (without influence on observables) the matrix $\Psi_s$ 
is proportional to the Hurwitz transformation~\cite{Hurwitz}:
\begary{rcl}
\Psi_{hw}&=&\frac{1}{2}\,(\psi,-\y_8\,\psi,\y_{9}\,\psi,-\y_{7}\,\psi)\\
&=&\frac{1}{2}\,\bmtx{cccc}
q_1&-p_1&-q_2&-p_2\\
p_1& q_1&-p_2&q_2\\
q_2& p_2& q_1&-p_1\\
p_2&-q_2& p_1&q_1\\
\emtx\,,
\endary
or (with a flipped sign of the last column) to the so-called 
Kustaanheimo-Stiefel (KS-) transformation~\cite{KS}:
\begeq
\Psi_{ks}=\frac{1}{2}\,(\psi,-\y_8\,\psi,\y_{9}\,\psi,\y_{7}\,\psi)\,.
\endeq
The matrices (\ref{eq_er_maps}) are also used in the Euler-Rodrigues 
formulation of attitudes and represent left- and right- multiplicative 
isoclinic mappings~\cite{Shuster}.

Since the KS-transformation is designed to map the Kepler- (or Coulomb-)
problem to the harmonic oscillator, it is not unreasonable to speculate that 
the appearance of the $\frac{1}{r}$-coulomb-potential of massive (charged) 
particles can also be explained in this way. It has been shown elsewhere that
non-bijective quadratic transformations allow to map the hydrogen atom 
to the harmonic oscillator~\cite{KS3}.

\section{Arbitrary constant Hamiltonians}

Can the Hamiltonian equivalently be of higher order? We presumed that the
Hamiltonian is a constant function of the fundamental variables $\psi$, i.e.
a constant of motion, we may ask if our derivation holds also for higher order
functions. Given the Hamiltonian has the general form of
Eq.~\ref{eq_Hamiltonian}:
\begary{rcl}
{\cal H}(\psi)&=&{\cal H}_0+\eps^T\,\psi+\frac{1}{2}\,\psi^T\,{\bf
  A}\,\psi+\frac{1}{3}\,{\bf B}_{ijk}\,\psi_i\,\psi_j\,\psi_k+\\
&+&\frac{1}{4}\,{\bf C}_{ijkl}\,\psi_i\,\psi_j\,\psi_k\,\psi_l+\dots\,,
\label{eq_h_general}
\endary
where all coefficients $\eps_i$, ${\bf A}_{ij}$, ${\bf B}_{ijk}$, ${\bf
  C}_{ijkl}$ are constants. Then we still assume that not only the matrix
${\bf A}$ but also the higher order ``tensors'' are symmetric in all indices.
The term ${\cal H}_0$ is a constant anyway and does not contribute and again
we choose the zeropoint of $\psi$ such that the linear term vanishes.
Then the time derivative of this Hamiltonian reads:
\begary{rcl}
\dot{\cal H}&=&\frac{1}{2}\,(\psi\,{\bf A}\,\dot\psi+\dot\psi\,{\bf A}\,\psi)+\\
&+&\frac{1}{3}\,({\bf B}_{ijk}\,\dot\psi_i\,\psi_j\,\psi_k+{\bf
  B}_{ijk}\,\psi_i\,\dot\psi_j\,\psi_k+{\bf
  B}_{ijk}\,\psi_i\,\psi_j\,\dot\psi_k)+\\
&+&\dots\\
\endary
Due to the symmetry in all indices, this symplifies to:
\begary{rcl}
\dot{\cal H}&=&(\psi_k\,{\bf A}_{ki}+{\bf B}_{jki}\,\psi_j\,\psi_k+\dots)\,\dot\psi_i\\
&=&(\nabla_\psi{\cal H})_i\,\dot\psi_i\\
\endary
with the gradient given by
\begary{rcl}
(\nabla_\psi{\cal H})_i&=&({\bf A}_{ik}+{\bf B}_{ijk}\,\psi_j+{\bf
  C}_{ijlk}\,\psi_j\,\psi_l+\dots)\,\psi_k\\
\y_0\,(\nabla_\psi{\cal H})&=&{\bf F}(\psi)\,\psi\\
\endary
where ${\bf F}(\psi)$ is the product of $\y_0$ with a {\it symmetric} matrix
${\bf\tilde A}$ - and therefore a symplex:
\begeq
{\bf\tilde A}_{ik}={\bf A}_{ik}+{\bf B}_{ijk}\,\psi_j+{\bf
  C}_{ijlk}\,\psi_j\,\psi_l+\dots\,,
\endeq
so that the solution still is:
\begary{rcl}
\dot\psi&=&\y_0\,(\nabla_\psi{\cal H})\\
\endary
We look at the second moments:
\begary{rcl}
{d\over d\tau}\,(\psi\,\psi^T)&=&\dot\psi\,\psi^T+\psi\,\dot\psi^T\\
&=&\y_0\,(\nabla_\psi{\cal H})\,\psi^T+\psi\,(\nabla_\psi{\cal H})^T\,\y_0^T\\
&=&{\bf F}(\psi)\,\psi\,\psi^T+\psi\,\psi^T\,{\bf F}^T(\psi)\\
\endary
Multiplication from the right with $\y_0$ and replacing ${\bf F}^T(\psi)$ with
 $\y_0\,{\bf F}(\psi)\,\y_0$ again yields:
\begary{rcl}
{d\over d\tau}\,(\psi\,\psi^T\y_0)&=&{\bf
  F}(\psi)\,(\psi\,\psi^T\,\y_0)+(\psi\,\psi^T\,\y_0)\,{\bf
  F}(\psi)\,\y_0^2\,,
\label{eq_nonlin1}
\endary
so that with $\y_0^2=-{\bf 1}$ and ${\bf S}\equiv\psi\psi^T\,\y_0$ we have the Lax pair:
\begary{rcl}
{\bf\dot S}&=&{\bf F}(\psi)\,{\bf S}-{\bf S}\,{\bf F}(\psi)\,.
\label{eq_nonlin2}
\endary
Note however, that the step from Eq.~\ref{eq_nonlin1} to
Eq.~\ref{eq_nonlin2} is in the general case only valid for single spinors.

\section{(Co-) Symplices for higher-dimensional Spacetimes}

\subsection{Which $k$-Vectors are (Co-) Symplices}
\label{app_highdim}

Given that we have a set of $N$ pairwise anticommuting (co-) symplices ${\bf S}_i$ (${\bf C}_i$),
which can be regarded as generators of real Clifford algebras, then the question if the $k$-products 
(i.e. bivectors, trivectors etc.) are symplices or cosymplices, depends on a sign and can be calculated 
as follows: 
\begary{rcl}
({\bf S}_1\,{\bf S}_2\,{\bf S}_3\,\dots\,{\bf S}_k)^T&=&{\bf S}_k^T\,{\bf S}_{k-1}^T\,{\bf S}_{k-2}^T\,\dots\,{\bf S}_1^T\\
&=&(-1)^{k-1}\,\y_0\,{\bf S}_k\,{\bf S}_{k-1}\,{\bf S}_{k-2}\,\dots\,{\bf S}_1\,\y_0\\
\endary
The number of permutations that is required to reverse the order of $k$ matrices is $k\,(k-1)/2$, so that
\begeq
({\bf S}_1\,{\bf S}_2\,{\bf S}_3\,\dots\,{\bf S}_k)^T=(-1)^s\,\y_0\,({\bf S}_1\,{\bf S}_2\,{\bf S}_3\,\dots\,{\bf S}_k)\,\y_0\,,
\endeq 
with $s$ given by:
\begeq
s=k-1+k\,(k-1)/2={k^2+k-2\over 2}\,.
\label{eq_symprod}
\endeq
If we consider $k$ cosymplices ${\bf C}_i$ instead, we obtain $k$ more sign reversals, so that 
\begeq
({\bf C}_1\,{\bf C}_2\,{\bf C}_3\,\dots\,{\bf C}_k)^T=(-1)^c\,\y_0\,({\bf C}_1\,{\bf C}_2\,{\bf C}_3\,\dots\,{\bf C}_k)\,\y_0\,,
\endeq 
with $c$ given by:
\begeq
c=2\,k-1+k\,(k-1)/2={k^2+3\,k-2\over 2}\,.
\label{eq_cosprod}
\endeq
If $s$ and $c$ are even (odd), respectively, then the products are (co-) symplices.
\begin{table}
\begin{tabular}{|c||c|c|c|c|c|c|c|c|c|c|c|c|}\hline
k        & 1 & 2 & 3 & 4  & 5  &  6 &  7 &  8 & 9  & 10 & 11 & 12\\\hline\hline
s        & 0 & 2 & 5 & 9  & 14 & 20 & 27 & 35 & 44 & 54 & 65 & 77\\\hline
$(-1)^s$ & + & + & - & -  & +  & +  & -  & -  & +  & + & - & - \\\hline
c        & 1 & 4 & 8 & 13 & 19 & 26 & 34 & 43 & 53 & 64& 76 & 89 \\\hline
$(-1)^c$ & - & + & + & -  & -  & +  & +  & -  & -  & + & +  & - \\\hline
\end{tabular}
\caption[]{Signs for products of $k$ anti-commuting (co-) symplices according
to Eq.~(\ref{eq_symprod}) and (\ref{eq_cosprod}). The ``plus'' signs correspond
to symplices: Products of $2, 5, 6, 9, 10,\,\dots$ symplices are again symplices.
\label{tab_signs}
}
\end{table}
Tab.~\ref{tab_signs} lists the resulting signs for $k=2\dots 10$. Clifford algebras
with more than $4$ generators include penta- and hexavectors as ``observables''.  
Furthermore one may conclude from Tab.~\ref{tab_signs}:
In $9+1$-dimensional spacetime, the pseudoscalar is the product of all $10$
generators and therefore a symplex. This allows (in principle) for a scalar field.
Since spatial rotators are combinations of two spatial generators, in $9+1$ dimensional
spacetime there are $\left( 9\atop 2\right)=36$ spatial rotations. A derivation of the 
corresponding MWEQs for such higher-dimensional spaces (if possible or not), lies beyond 
the scope of this paper, but if all bivector boosts have a corresponding electric field 
component and all bivector rotations a corresponding magnetic
field component, then we should expect $9$ ``electric'' and $36$ ``magnetic'' field components
in $9+1$ dimensional spacetime.

\subsection{The real Pauli matrices and the structure of (co-) symplices for $n$ DOFs}
\label{sec_rpm}

For one DOF and an appropriate treatment of the $2\times 2$-blocks of the general 
case we introduce the real Pauli matrices (RPMs) according to
\begary{rclp{10mm}rcl}
\eta_0&=&\bmtx{cc}
0&1\\
-1&0\\
\emtx&&
\eta_1&=&\bmtx{cc}
1&0\\
0&-1\\
\emtx\\
\eta_2&=&\bmtx{cc}
0&1\\
1&0\\
\emtx&&
\eta_3&=&{\bf 1}=\bmtx{cc}
1&0\\
0&1\\
\emtx\\
\endary
$\eta_0$ is the symplectic unit matrix (i.e. corresponds to $\y_0$ of the general case).
$\eta_0$, $\eta_1$ and $\eta_2$ are symplices. If we consider general $2\,n\times 2\,n$
Hamiltonian matrix ${\bf F}$ composed of $n^2$ $2\times 2$ blocks ${\bf A}_{ij}$, then $\y_0$
has the form given in Eq.~(\ref{eq_gamma0}).
The $2\,n\times 2\,n$-symplex ${\bf F}$ that fulfills Eq.~\ref{eq_symplexdef} has the form
\begeq
{\bf F}=\bmtx{ccccc}
{\bf D}_1&{\bf A}_{12}&{\bf A}_{13}&\dots&{\bf A}_{1n}\\
-{\bf\tilde A}_{12}&{\bf D}_2&{\bf A}_{23}&\dots&{\bf A}_{2n}\\
-{\bf\tilde A}_{13}&-{\bf\tilde A}_{23}&{\bf D}_3&\dots&{\bf A}_{3n}\\
\vdots&&\vdots&&\vdots\\
-{\bf\tilde A}_{1n}&-{\bf\tilde A}_{2n}&-{\bf\tilde A}_{3n}&\dots&{\bf D}_n\\
\emtx
\endeq
where the $2\times 2$ matrices ${\bf D}_k$ on the diagonal must be $2\times 2$-symplices and can therefore be written as
\begeq
{\bf D}_k=d_0^{k}\,\eta_0+d_1^{k}\,\eta_1+d_2^{k}\,\eta_2\,.
\label{eq_symplex2}
\endeq
If the blocks ${\bf A}_{ij}$ above the diagonal have the general form 
\begeq
{\bf A}=\sum\limits_{k=0}^3\,a_k\,\eta_k\,,
\label{eq_rpmcoeffs}
\endeq
then Eq.~\ref{eq_symplexdef} fixes the form of the corresponding blocks below the diagonal ${\bf\tilde A}_{ij}$ to
\begary{rcl}
-{\bf\tilde A}_{ij}&=&\eta_0\,{\bf A}_{ij}^T\,\eta_0\\
&=&-\eta_0\,(a_{ij}^0\,\eta_0+a_{ij}^1\,\eta_1+a_{ij}^2\,\eta_2+a_{ij}^3\,\eta_3)^T\,\eta_0\\
&=&\eta_0\,(-a_{ij}^0\,\eta_0+a_{ij}^1\,\eta_1+a_{ij}^2\,\eta_2+a_{ij}^3\,\eta_3)\,\eta_0\\
&=&a_{ij}^0\,\eta_0+a_{ij}^1\,\eta_1+a_{ij}^2\,\eta_2-a_{ij}^3\,\eta_3\\
\endary
From this one can count that each subblock ${\bf A}_{ij}$ has two antisymmetric
and two symmetric coefficients. We count $n(n-1)/2$ sublocks ${\bf
  A}_{ij}$. Together with $n$ diagonal subblocks with each having one
antisymmetric coefficients, we should have
\begeq
\nu_s^a=n\,(n-1)+n=n^2
\label{eq_nu_sa}
\endeq 
antisymmetric symplices. The number of independent antisymmetic parameters is
$\nu^a=\nu_s^a+\nu_c^a=n\,(2\,n-1)=2\,n^2-n$.
Therefore we have  
\begeq
\nu_c^a=n\,(2\,n-1)-n^2=n^2-n
\label{eq_nu_ca}
\endeq 
antisymmetric cosymplices. Since we have as many cosymplices as we have
antisymmetric matrix elements $\nu_c=\nu^a$, the number of symmetic
cosymplices $\nu_c^s$ is:
\begeq
\nu_c^s=\nu_c-\nu_c^a=2\,n^2-n-(n^2-n)=n^2
\label{eq_nu_cs}
\endeq
The number $\nu_s^s$ of symmetric symplices is:
\begeq
\nu_s^s=\nu_s-\nu_s^a=2\,n\,(2\,n+1)/2-n^2=n^2+n\,.
\label{eq_nu_ss}
\endeq

${\bf\tilde A}_{ij}$ is called the {\it symplectic conjugate} of ${\bf A}_{ij}$. If ${\bf S}$ and ${\bf C}$ are
the symplex and cosymplex-part of a matrix ${\bf A}={\bf C}+{\bf S}$, then the {\it symplectic conjugate} is
\begeq
{\bf\tilde A}=-\y_0\,{\bf A}^T\,\y_0={\bf C}-{\bf S}\,.
\endeq
As well-known one can quickly derive that
\begeq
{\bf\tilde{A B}}={\bf\tilde B}\,{\bf\tilde A}\,.
\endeq
If ${\bf A}$ is written ``classically'' as
\begeq
{\bf A}=\bmtx{cc}
a_{11}&a_{12}\\
a_{21}&a_{22}\\
\emtx
\endeq
then $\eta_0\,{\bf A}^T\,\eta_0$ is given by:
\begeq
\eta_0\,{\bf A}^T\,\eta_0=\bmtx{cc}
-a_{22}&a_{12}\\
a_{21}&-a_{11}\\
\emtx\,,
\endeq
so that
\begeq
{\bf A}\,(\eta_0\,{\bf A}^T\,\eta_0)=(a_{12}\,a_{21}-a_{11}\,a_{22})\,{\bf 1}=-\mathrm{Det}({\bf A})\,{\bf 1}\,.
\endeq
A matrix ${\bf A}$ is symplectic, if ${\bf A}\,{\bf\tilde A}={\bf 1}$. But we can say in any case, that
${\bf A}\,{\bf\tilde A}$ is a co-symplex:
\begary{rcl}
{\bf A}\,{\bf\tilde A}&=&({\bf C}+{\bf S})\,({\bf C}-{\bf S})\\
                      &=&{\bf C}^2-{\bf S}^2+{\bf S}\,{\bf C}-{\bf C}\,{\bf S}\\
\endary
since squares of (co-) symplices as well as the commutator of symplex and cosymplex are cosymplices.

A $2\,n\times 2\,n$-cosymplex ${\bf C}$ has according to Eq.~\ref{eq_cosymplexdef} the form
\begeq
{\bf C}=\bmtx{ccccc}
{\bf E}_1&{\bf B}_{12}&{\bf B}_{13}&\dots&{\bf B}_{1n}\\
{\bf\tilde B}_{12}&{\bf E}_2&{\bf B}_{23}&\dots&{\bf B}_{2n}\\
{\bf\tilde B}_{13}&{\bf\tilde B}_{23}&{\bf E}_3&\dots&{\bf B}_{3n}\\
\vdots&&\vdots&&\vdots\\
{\bf\tilde B}_{1n}&{\bf\tilde B}_{2n}&{\bf\tilde B}_{3n}&\dots&{\bf E}_n\\
\emtx
\endeq
where the $2\times 2$ matrices ${\bf E}_k$ on the diagonal must be $2\times 2$-cosymplices and are hence
proportional to the unit matrix
\begeq
{\bf E}_k=e_{k}\,{\bf 1}\,.
\endeq

\subsection{Which $k$-Vectors of $Cl_{N-1,1}$ are (Anti-) Symmetric}

Possible dimensionalities for emergent spacetimes are given by Eq.~\ref{eq_dim}.
Since all but one generator of $Cl_{N-1,1}$ are symmetric, the analysis of how many $k$-vectors
are (anti-) symmetric, is reasonably simple. The total number of $k$-vectors
is given by $\left({N\atop k}\right)$, the number $\mu_k$ of $k$-vectors generated 
only with spatial elements, hence is $\left({N-1\atop k}\right)$. For $k$-vectors
that are generated exclusively from spatial basis vectors, one finds:
\begary{rcl}
({\bf S}_1\,{\bf S}_2\,{\bf S}_3\,\dots\,{\bf S}_k)^T&=&{\bf S}_k^T\,{\bf S}_{k-1}^T\,{\bf S}_{k-2}^T\,\dots\,{\bf S}_1^T\\
&=&{\bf S}_k\,{\bf S}_{k-1}\,{\bf S}_{k-2}\,\dots\,{\bf S}_1\\
&=&(-1)^a\,{\bf S}_1\,{\bf S}_2\,{\bf S}_3\,\dots\,{\bf S}_k\,,
\endary
where first step is possible as all ${\bf S}$ are spatial basis vectors and hence symmetric.
The second step reflects the number of permutations that are required to reverse the order
of $k$ anticommuting elements:
\begeq
a=k(k-1)/2\,.
\endeq
If one of the symplices equals $\y_0$ (i.e. is anti-symmetric), then we have
\begary{rcl}
({\bf S}_1\,{\bf S}_2\,{\bf S}_3\,\dots\,{\bf S}_k)^T&=&{\bf S}_k^T\,{\bf S}_{k-1}^T\,{\bf S}_{k-2}^T\,\dots\,{\bf S}_1^T\\
&=&-{\bf S}_k\,{\bf S}_{k-1}\,{\bf S}_{k-2}\,\dots\,{\bf S}_1\\
&=&(-1)^{a+1}\,{\bf S}_1\,{\bf S}_2\,{\bf S}_3\,\dots\,{\bf S}_k\\
\endary
Hence, depending on whether $a$ is even or odd, we count $\mu_k^{s,a}=\left({N-1\atop k}\right)$ symmetric and $\mu_k^{a,s}=\left({N\atop k}\right)-\left({N-1\atop k}\right)$ skewsymmetric matrices or vice versa. The result for the simplest Clifford algebras is given in Tab.~\ref{tab_obs}.
\begin{table}
\begin{tabular}{|l||l|l|l|l|}\hline
N=p+q          & 4=3+1   & 10=9+1      & 12=11+1  &   \\\hline\hline
$2^N$          & 16      & 1032        & 4096     &   \\
n              & 2       & 16          & 32       &   \\\hline
               &\multicolumn{3}{|l|}{$\mu_k=\mu_k^s+\mu_k^a$}&\\\hline
k=0            & 1=1+0   & 1  =1+0     & 1=1+0    & c \\
k=1            & 4=3+1   & 10 =9+1     & 12=11+1  & s \\
k=2            & 6=3+3   & 45 =9+36    & 66=11+55 & s \\
k=3            & 4=3+1   & 120=36+84   & 220=55+165  & c \\
k=4            & 1=0+1   & 210=126+84  & 495=330+165 & c \\
k=5            & -       & 252=126+126 & 792=462+330 & s \\
k=6            & -       & 210=126+84  & 924=462+462 & s \\
k=7            & -       & 120=84+36   & 792=462+330 & c \\
k=8            & -       & 45=9+36     & 495=165+330 & c \\
k=9            & -       & 10=1+9      & 220=55+165  & s \\
k=10           & -       & 1=1+0       & 66=55+11    & s \\
k=11           & -       & -           & 12=11+1     & c \\
k=12            & -      & -           & 1=0+1       & c \\\hline
$\nu_s^s=n^2+n$ & 6      & 272        & 1056 &   \\
$\nu_c^s=\nu_s^a=n^2$   & 4      & 256        & 1024 &   \\
$\nu_c^a=n^2-n$ & 2      & 240        &  992 &   \\
$\mu^s=2\,n^2+n$& 10     & 528        &  2080 &   \\
$\mu^a=2\,n^2-n$& 6      & 496        &  2016 &   \\\hline
\end{tabular}
\caption[]{Numbers $\mu_k=\mu_k^s+\mu_k^a$ of $k$-vectors for
  $N$-dimensional dynamically emergent spacetime Clifford-algebras $Cl_{N-1,1}$.
$\mu_k^s$ ($\mu_k^a$) is the number of (anti-) symmetric k-vectors.
The 0-vector is identified as the unit matrix which is (of course) symmetric.  
$\nu_x^y=n^2+n$ is the number of (anti-) symmmetric (co-) symplices, were
  $y={(a),s}$ denotes (anti-) symmetry and $x={(c),s}$ denotes (co-) symplices.
\label{tab_obs}
}
\end{table}

\section{The Kustaanheimo-Stiefel Transformation}
\label{sec_KS}

During the ``derivation'' of the Lorentz force equation in Sec.~\ref{sec_lorentzforce}, we 
associated elements of the matrix of second moments (or ${\bf S}=\Sigma\,\y_0$) with energy and 
momentum of a ``particle'', i.e. we assumed that:
\begary{rcl}
{\cal E}&=&-\bar\psi\,\y_0\psi\propto \psi_0^2+\psi_1^2+\psi_2^2+\psi_3^2\\
P_x     &=& \bar\psi\,\y_1\psi\propto -\psi_0^2+\psi_1^2+\psi_2^2-\psi_3^2\\
P_y     &=& \bar\psi\,\y_2\psi\propto 2\,(\psi_0\,\psi_2-\psi_1\,\psi_3)\\
P_z     &=& \bar\psi\,\y_3\psi\propto 2\,(\psi_0\,\psi_1+\psi_2\,\psi_3)\\
\label{eq_e_p}
\endary
Eq.~(\ref{eq_e_p}) is (up to factor) practically identical to the regularization transformation
of Kustaanheimo and Stiefel~\cite{KS}, (KST). At the same time we introduced a scaling factor $\y$ to 
transform between the {\it eigentime} $\tau$ and the time of an observer $t$ - which is again 
a similarity to the use of the KST in celestial mechanics, where a Sundman transformation 
\begeq
{dt\over d\tau}=f({\bf q},{\bf p})
\endeq
is used~\cite{KS1,KS2}. Despite these remarkable similarities, there are also significant differences,
since in contrast to the KST, we do {\it not increase} the number of variables and we do not
transform coordinates to coordinates. The ``spinor'' $\psi$ was introduced as two canonical pairs 
(i.e. 2 coordinates and the 2 canonical momenta) and is used in Eq.~(\ref{eq_e_p}) to parameterize
a 4-momentum vector, while the KST uses 4 ``fictious'' coordinates to parameterize 3 cartesian 
coordinates.

\section{Non-Symplectic Transformations}
\label{sec_nonsym}

\subsection{Simple LC-Circuit}

\begin{figure}[h]
\parbox{80mm}{
\parbox{40mm}{
 \includegraphics[width=4cm]{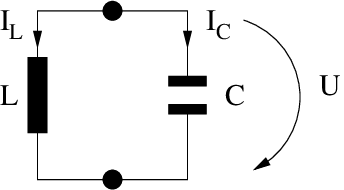}
}\hfill\parbox{30mm}{
\caption{Simple LC-circuit.
\label{fig_lc_simple}}}}
\end{figure}

The key concept of symplectic transformations was mentioned to be the {\it structure preservation}.
In the following we exemplify the meaning of {\it structure preservation} by giving examples for
non-symplectic transformations. Consider a simple LC-circuit as shown in Fig.~\ref{fig_lc_simple}.
Starting up naive, we take a Hamiltonian description using the total system energy, i.e.
we associate the potential energy with the energy stored in the capacitor $\frac{C}{2}\,U^2$
and the kinectic energy with $\frac{L}{2}\,I_L^2$:
\begeq
{\cal H}=\frac{C}{2}\,U^2+\frac{L}{2}\,I_L^2=\frac{C}{2}\,U^2+\frac{L}{2}\,I_C^2\,,
\endeq
and assume that $U$ is the canonical coordinate and $I_C$ is the canonical momentum.
Then the EQOM are:
\begary{rcl}
\dot U&=&{\d{\cal H}\over\d I_C}=L\,I_C\\
\dot I_C&=&-{\d{\cal H}\over\d U}=-C\,U\\
\endary
Quite obviously these equations fail to describe the correct relations for capacitors 
and inductors, which are:
\begary{rcl}
C\,\dot U&=&I_C\\
L\,\dot I_L&=&U=-L\,\dot I_C\\
\label{eq_LC}
\endary
We note that the mistake is a wrong scaling of the variables. We recall that the product of the
coordinate and the corresponding conjugate momentum should have the dimension of an action, 
while voltage times current results in a quantity with the dimension of power. 
Thus we introduce a scaling factor for the current and write:
\begeq
{\cal H}=\frac{C}{2}\,U^2+\frac{L}{2\,a^2}\,P^2\,,
\endeq
so that the canonical momentum is now $P=a\,I_C$ and the EQOM are:
\begary{rcl}
\dot U&=&{\d{\cal H}\over\d P}=\frac{L}{a^2}\,P=\frac{L}{a}\,I_C\\
\dot P&=&-{\d{\cal H}\over\d U}=-C\,U\\
\dot I_C&=&-\frac{C}{a}\,U\\
\endary
The comparison with Eq.~\ref{eq_LC} then yields $a=L\,C$. But note that the
scaling changes the product $p_i\,q_i$ and hence is not a symplectic
(structure preserving), but a non-symplectic (structure defining) transformation. 

\subsection{Coupled LC-Circuits}

\begin{figure}[h]
\parbox{80mm}{
 \includegraphics[width=7cm]{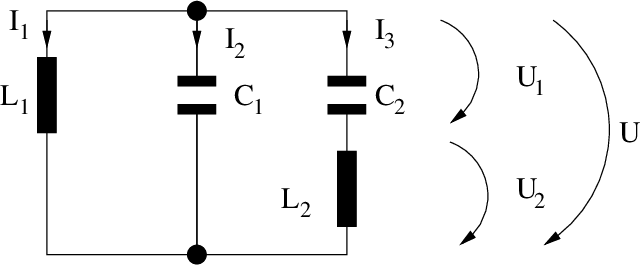}
}
\caption{Two coupled LC-circuits.
\label{fig_lc_coupled}}
\end{figure}

If we restrict ourselves to the simplest case (i.e. two capacitors, two inductors), the
structure of the only non-trivial way to couple two LC-circuits is shown if Fig.~\ref{fig_lc_coupled}.
The equations of motion can be derived directly from the drawing using the general relations
$\dot U_C=I_C/C$ for an ideal capacitor and $\dot I_L=U_L/L$ for the ideal inductor and Kirchhoff's rule:
\begary{rcl}
\dot I_1&=&U/L_1\\
\dot U&=&-(I_1+I_3)/C_1\\
\dot I_3&=&(U-U_1)/L_2\\
\dot U_1&=&I_3/C_2\\
0&=&I_1+I_2+I_3\\
U&=&U_1+U_2\\
\endary
The energy sum again yields a Hamiltonian of the diagonal form:
\begeq
{\cal H}=\frac{C_1}{2}\,U^2+\frac{L_1}{2}\,I_1^2+\frac{C_2}{2}\,U_1^2+\frac{L_2}{2}\,I_3^2\,.
\endeq
We use the voltages and currents of the Hamiltonian to define the state vector $\phi=(U,U_1,I_1,I_3)$ 
and we find for its derivative:
\begary{rcl}
\dot\phi&=&{\bf F}\,\phi\\
&=&\bmtx{cccc}
0 & 0 & -1/C_1 & -1/C_1\\
0 & 0 & 0 & 1/C_2\\
1/L_1 & 0 & 0 & 0\\
1/L_2 & -1/L_2 & 0 & 0\\
\emtx\,\phi\,.
\label{eq_LCcoupled}
\endary
The matrix ${\bf F}$ represents the structural properties of the LC-circuit. The matrix
is not a symplex and can not be transformed into a symplex by any symplectic transformation.
{\it This does not mean} that the dynamics of the system can not be derived from a 
Hamiltonian - it simply means that the transformation matrix which is required to map
the system to a Hamiltonian system, is not structure {\it preserving} but structure
{\it defining}.

In case of two coupled LC-circuits, it is likewise not sufficient to use scaling 
factors to obtain the canonical momenta and even if it was, the eigen-
frequencies of the circuit can not be guessed anymore. The square of the matrix from 
Eq.~\ref{eq_LCcoupled} is given by:
\begeq
\ddot\phi=\bmtx{cccc}
-{L_1+L_2\over C_1\,L_1\,L_2} & {1\over C_1\,L_2} & 0 &0\\
{1\over C_2\,L_2} & -{1\over C_2\,L_2} &0&0\\
0&0&-{1\over C_1\,L_1} & -{1\over C_1\,L_1}\\
0&0&-{1\over C_1\,L_2} & -{C_1+C_2\over C_1\,C_2\,L_2}\\
\emtx\,\phi\,.
\endeq

Obviously the state vector $\phi$ is not composed of canonical variables.
The Lagrangian ${\bf L}$ is given by 
\begeq
{\cal L}=\frac{L_1}{2}\,I_1^2+\frac{L_2}{2}\,I_3^2-\frac{C_1}{2}\,U^2-\frac{C_2}{2}\,U_1^2\,.
\endeq
Eq.~\ref{eq_LCcoupled} can be used to replace the currents by the derivatives of the voltages:
\begary{rcl}
I_1&=&-C_1\,\dot U - C_2\,\dot U_1\\
I_3&=&C_2\,\dot U_1\\
{\cal L}&=&\frac{L_1}{2}\,(C_1\,\dot U + C_2\,\dot U_1)^2+\frac{L_2}{2}\,C_2^2\,\dot U_1^2-\frac{C_1}{2}\,U^2-\frac{C_2}{2}\,U_1^2\,,
\endary
so that for the coordinates $q_1=U$ and $q_2=U_1$, the canonical momenta are:
\begary{rcl}
p_i&=&{\d{\cal L}\over\d\dot q_i}\\
p_1&=&{\d{\cal L}\over\d\dot U}=L_1\,C_1\,(C_1\,\dot U + C_2\,\dot U_1)\\
   &=&-L_1\,C_1\,I_1\\
p_2&=&{\d{\cal L}\over\d\dot U_1}=L_1\,C_2\,(C_1\,\dot U + C_2\,\dot U_1)+L_2\,C_2^2\,\dot U_1\\
   &=&-L_1\,C_2\,I_1+L_2\,C_2\,I_3\\
\endary
so that the transformation matrix from the state vector $\phi$ to the canonical variables $\psi$
is given by:
\begary{rcl}
\psi&=&{\bf T}\,\phi\\
\bmtx{c}q_1\\q_2\\p_1\\p_2\emtx&=&\bmtx{cccc}
1 & 0 & 0 & 0\\
0 & 1 & 0 & 0\\
0 & 0 & -L_1\,C_1 & 0\\
0 & 0 & -L_1\,C_2 & L_2\,C_2\\
\emtx\,\bmtx{c}U\\U_1\\I_1\\I_3\emtx
\endary
The transformed matrix ${\bf\tilde F}={\bf T}\,{\bf F}\,{\bf T}^{-1}$ then is:
\begary{rcl}
{\bf\tilde F}&=&\bmtx{cccc}
0    &    0 & {L_1+L_2\over C_1^2\,L_1\,L_2}&-{1\over C_1\,C_2\,L_2}\\
0    &    0 & -{1\over C_1\,C_2\,L_2}&{1\over C_2^2\,L_2}\\
-C_1 &    0 & 0 & 0\\
0    & -C_2 & 0 & 0\\ 
\emtx\\
\endary
The Hamiltonian can then be expressed in the canonical coordinates as:
\begary{rcl}
{\cal H}&=&\frac{1}{2}\,\psi^T\,{\bf A}\,\psi\\
{\bf A}&=&\bmtx{cccc}
C_1&0&0&0\\
0&C_2&0&0\\
0&0&{L_1+L_2\over C_1^2\,L_1\,L_2}&-{1\over C_1\,C_2\,L_2}\\
0&0&-{1\over C_1\,C_2\,L_2}&{1\over C_2^2\,L_2}\\
\emtx\,,
\endary
where ${\bf A}=\y_0\,{\bf\tilde F}$.
As was shown in Ref.~\cite{geo_paper}, the Hamiltonian formulation (for stable oscillating systems) 
is always ``similar'' to the case of completely decoupled oscillators, i.e. the non-symplectic
transformation maps the structure of Fig.~\ref{fig_lc_coupled} to two seperate systems as
shown in Fig.~\ref{fig_lc_simple}. The transformation ${\bf T}$ is not
symplectic and may serve as an example for a structure {\it defining}
transformation in contrast to symplectic structure {\it preserving} transformations.

\subsection{The Direction of Time}
\label{sec_vectortime}

If the matrix ${\bf Q}$ used in Eq.~\ref{eq_strucdef} is not only non-singular, but
orthogonal, then
\begeq
{\bf Q}^T\,{\bf S}\,{\bf Q}=\textrm{diag}(\lambda_0\,\eta_0,\lambda_1\,\eta_0,\lambda_2\,\eta_0,\dots\,,0,0,0)\,,
\label{eq_strucdef2}
\endeq
with some real coefficients $\lambda_k$~\cite{Horn}. In this case we would restrict to
$\lambda_k=1$ by the argument that the dynamics of the system has to be defined by the 
Hamilton function and not by the (otherwise arbitrary) skew-symmetric matrix
${\bf S}$. However this is true only for closed systems, i.e. systems in which
the matrix ${\bf A}$ is a function (of the constants of motion) of the variables
$\psi$ only. Since we interpreted $\y_0$ as the (unit vector in the) direction of
time, it appears to be wise to return to the possible algebraic and physical 
implications of this choice for the parameters $\lambda_k$. In
Ref.~\cite{geo_paper} we have demonstrated that a symplectic decoupling
analysis of any oscillatory symplex ${\bf F}$ leads to a normal form exactly
of the form on the right-hand side of Eq.~\ref{eq_strucdef2}. The only
difference is that the decoupling transformation is symplectic, while the
transformation Eq.~\ref{eq_strucdef2} is obviously not symplectic - and not
even generally orthogonal. We said that the system evolution in time is
symplectic. However the fact that the only restriction for the time-matrix
${\bf S}$ is its skew-symmetry calls for a deeper analysis of the possible
physical implications of the non-symplectic transformation by ${\bf Q}$.
Such a program extends the scope of this paper. It may concern the number, 
type, and possible interactions and transformations between the fermions.
Recall that the matrix ${\bf Q}$ can be used to transform ${\bf S}$ into
all possible $6$ already skew-symmetric matrices used to build the two
types of spinors in Eq.~\ref{eq_er_maps}. And it concerns the role that
the ``axial 4-vector''-components play in the transformations between 
these two spinor-types (or types of phase space symmetry).

\subsection{Nonsymplectic Transformations within the Dirac Algebra}
\label{sec_nsym_dirac}

In the following we present an orthogonal (non-symplectic) transformation that enables to
``rotate the time direction'', i.e. to transform from $\y_0$ (as we used it here) to other 
representations of $\y_0$. We define the (normalized) Hadamard-matrix $H_4$ according to
\begary{rcl}
H_4&=&{\y_2-\y_1-y_4+\y_5\over 2}={(\y_{15}+\y_0)\,(\y_2-\y_1)\over 2}\\
&=&{1\over 2}\,\bmtx{cccc}
 1 & 1 & 1 & 1 \\
 1 &-1 & 1 &-1 \\
 1 & 1 &-1 &-1 \\
 1 &-1 &-1 & 1 \\
\emtx\\
H_4&=&H_4^T\\
H_4^2&=&{\bf 1}\\
\y_0\,H_4\,\y_0&=&H_4\\
H_4\,\y_0\,H_4&=&-\y_0\\
\endary
$H_4$ is a symplex and it is antisymplectic.
Furthermore we define a ``shifter'' matrix $X$ according to
\begeq
X={-\y_0+\y_2+\y_6+\y_7\over 2}
 =\bmtx{cccc}
 0 & 0 & 0 & 1 \\
 1 & 0 & 0 & 0 \\
 0 & 1 & 0 & 0 \\
 0 & 0 & 1 & 0 \\
\emtx
\endeq
Some properties of the $X$-matrix are 
\begary{rclp{10mm}rcl}
\y_0\,X\,\y_0&=&X^T &&
X^T\,X&=&{\bf 1}\\
X^T\,\y_0\,X&=&-\y_7 &&
X^T\,\y_7\,X&=&-\y_0\\
X^T\,\y_8\,X&=&\y_{14} &&
X^T\,\y_{14}\,X&=&-\y_{8}\\
X^T\,\y_9\,X&=&-\y_{10} &&
X^T\,\y_{10}\,X&=&\y_{9}\,.
\endary
Note that $X$ is a symplex and hence could be used as a force matrix.
In this case we have a 4-th order differential equation of the form:
\begary{rcl}
\psi'&=&X\,\psi\\
\psi''''&=&\psi\,,
\endary
since one finds $X^4={\bf 1}$. 
The product of these two matrices $R_6=H_4\,X$ is an orthogonal matrix that transforms cyclic through 
all possible basis systems:
\begeq
R_6=H_4\,X={1\over 2}\,\bmtx{cccc}
 1 & 1 & 1 & 1 \\
-1 & 1 &-1 & 1 \\
 1 &-1 &-1 & 1 \\
-1 &-1 & 1 & 1 \\
\emtx\,.
\endeq
The most relevant properties are
\begary{rclp{10mm}rcl}
R_6^T\,R_6&=&R_6\,R_6^T={\bf 1}  &&
R_6^T\,\y_0\,R_6&=&\y_7\\
R_6^T\,\y_{10}\,R_6&=&-\y_8 &&
R_6^T\,\y_{14}\,R_6&=&\y_9\\
R_6^T\,\y_7\,R_6&=&-\y_{14} &&
R_6^T\,\y_8\,R_6&=&\y_0\\
R_6^T\,\y_9\,R_6&=&\y_{10}\,,
\endary
so that $R_6$ can be iteratively used to switch through all possible systems:
\begary{rcl}
R_6^T\,\y_0\,R_6&=&\y_7\\
(R_6^T)^2\,\y_0\,(R_6)^2&=&-\y_{14}\\
(R_6^T)^3\,\y_0\,(R_6)^3&=&-\y_{9}\\
(R_6^T)^4\,\y_0\,(R_6)^4&=&-\y_{10}\\
(R_6^T)^5\,\y_0\,(R_6)^5&=&\y_{8}\\
(R_6^T)^6\,\y_0\,(R_6)^6&=&\y_0\\
\endary
Finally one finds that $(R_6)^6={\bf 1}$. But $R_6$ is neither symplectic nor is it a symplex. 
Expressed by the $\y$-matrices, $R_6$ is given by
\begary{rcl}
4\,R_6&=&{\bf 1}+\y_{0}+\y_{1}-\y_{2}+\y_{3}-\y_{4}+\y_{5}+\y_{6}\\
      &+&\y_{7}+\y_{8}-\y_{9}-\y_{10}-\y_{11}-\y_{12}-\y_{13}-\y_{14}\,.
\endary
Another matrix with equivalent properties is given by
\begeq
\tilde R_6=H_4\,X^T\,.
\endeq
These transformations exemplifies part of the claim (i.e. the use of a specific 
(though arbitrary) form for $\y_0$) in Eq.~(\ref {eq_strucdef}).

\subsection{Transformation to the Conventional Dirac Algebra}
\label{sec_diracconv}

According to the fundamental theorem of the Dirac matrices~\cite{AJM},
any set of Dirac matrices is similar to any other set up to a unitary 
transformation. However a change in the sign of the metric tensor requires
in addition a multiplication with the unit imaginary.

Let ${\bf U}$ be the following unitary matrix
\begeq
{\bf U}=\frac{1}{2}\,\bmtx{cccc}
1&i&i&-1\\
-i&1&-1&-i\\
-i&-1&1&-i\\
1&-i&-i&-1\\
\emtx
\endeq
then it is quickly verified that
\begeq
\tilde\y_\mu=i\,{\bf U}\,\y_\mu\,{\bf U}^\dagger\,,
\endeq
where $\mu\,\in\,[0\dots 3]$ and $\tilde\y_\mu$ are the conventional Dirac
matrices~footnote{The explicit form of the real Dirac matrices is given
for instance in Refs.~(\cite{rdm_paper,geo_paper})}. Using $\tilde\y_\mu$, 
the other matrices of the Clifford algebra are quickly constructed. 
However, since we multiplied by the unit imaginary,
it is clear that the conventional Dirac algebra is a rep of $Cl_{1,3}$.

\section{Graphical Representation of Dirac Matrices}
\label{sec_geom_rdm}

\begin{figure}[h]
\parbox{70mm}{
\includegraphics[width=7cm]{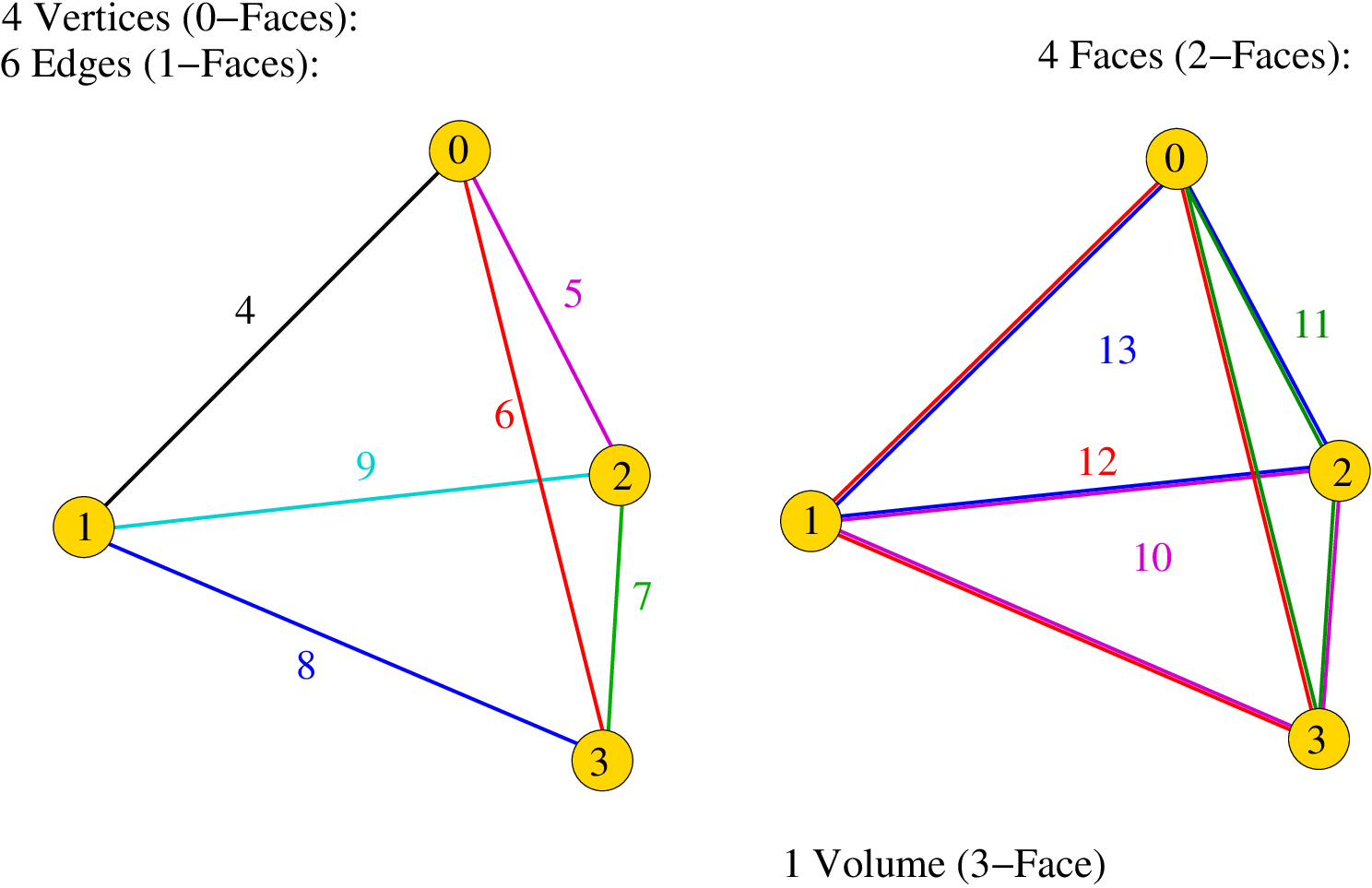}
}
\caption{The group structure of the RDMs (i.e. the Clifford algebra $Cl_{3,1}$) can be represented 
by a tetrahedron: The vertices represent the ``basic'' symplices $\y_0,\dots,\y_3$, the
edges the bi-vectors $\y_4,\dots,y_9$, the surfaces the components of the axial vector
$\y_{10},\dots,\y_{13}$ and the volume the pseudoscalar $\y_{14}$. Another graphical 
representation has been given by Goodmanson~\cite{Goodmanson}.
\label{fig_geom}}
\end{figure}

Fig.~\ref{fig_geom} illustrates the geometric interpretation of the RDMs. 

\end{appendix}


\begin{thebibliography}{9}
\bibitem{napier} This definition holds only for a {\it flat spacetime}.
    In curved spacetime, this definition would have to be modified; in spherical
    trigonometry the theorem of Pythagoras also has to be replaced by the rules of John Napier.
\end{thebibliography}
\end{document}